\documentclass{jfm}

\RequirePackage{packages}
\RequirePackage{newcommands}

\newcommand{\arev}[1]{#1}
\ifCUPmtlplainloaded \else
\checkfont{eurm10}
\iffontfound
\IfFileExists{upmath.sty}
{\typeout{^^JFound AMS Euler Roman fonts on the system,
		using the 'upmath' package.^^J}%
	\usepackage{upmath}}
{\typeout{^^JFound AMS Euler Roman fonts on the system, but you
		dont seem to have the}%
	\typeout{'upmath' package installed. JFM.cls can take advantage
		of these fonts,^^Jif you use 'upmath' package.^^J}%
	\providecommand\upi{\upi}%
}
\else
\providecommand\upi{\upi}%
\fi
\fi

\ifCUPmtlplainloaded \else
\checkfont{msam10}
\iffontfound
\IfFileExists{amssymb.sty}
{\typeout{^^JFound AMS Symbol fonts on the system, using the
		'amssymb' package.^^J}%
	\usepackage{amssymb}%
	  \let\leq=\leqslant
	  
}{}
\fi
\fi
\ifCUPmtlplainloaded \else
\IfFileExists{amsbsy.sty}
{\typeout{^^JFound the 'amsbsy' package on the system, using it.^^J}%
	\usepackage{amsbsy}}
{\providecommand\boldsymbol[1]{\mbox{\boldmath $##1$}}}
\fi

\title[Weakly non-linear transient waves on a shear current]{
  Weakly non-linear transient waves on a shear current: Ring waves and skewed Langmuir rolls
}
\author[A.~H. Akselsen and S.~\AA. Ellingsen]{Andreas H.\ Akselsen$^1$\thanks{Email address for correspondence: andreas.h.akselsen@ntnu.no} and Simen \AA. Ellingsen$^1$}
\affiliation{$^1$Department of Energy and Process Engineering, Norwegian University of Science and Technology, N-7491 Trondheim, Norway}
\date{\today}           %% By default, LaTeX uses the current date

\begin{document}

\maketitle

\begin{abstract}

We investigate the weakly nonlinear dynamics of   transient gravity 
waves at infinite depth under the influence of a shear current varying linearly with depth. The shear field makes this problem three-dimensional and rotational in nature, but an analytical solution is permitted via integration of the Euler equations. Although similar problems were 
investigated in the 1960's and 70's for special cases of resonance, this is to our knowledge the first general wave interaction (mode coupling) solution derived to second order with a shear current present. Wave interactions are 
integrable in a spectral convolution to yield the second order dynamics of 
initial value problems.

To second order, irrotational wave dynamics interacts 
with the background vorticity field in a way that creates new vortex structures. 
A notable example is the large parallel vortices which drive Langmuir circulation as oblique plane waves interact with an ocean current. We also investigate the effect on wave pairs which are misaligned with the shear current to find that similar, but skewed, vortex structures are generated in every case except when the mean wave direction is precisely perpendicular to the direction of the 
current. This is in contrast to a conjecture by \cite{leibovich83}.
Similar nonlinear wave-shear interactions are found to also generate near-field vortex structures in the Cauchy-Poisson problem 
with an initial surface elevation. These interactions create further 
groups of dispersive ring waves in addition to those present in linear theory.

The second order solution is derived in a general manner which accommodates any initial condition through mode coupling over a continuous wave spectrum. It is therefore applicable to a range of problems including special cases of resonance. 
As a by--product of the general theory, a simple expression for the Stokes drift due to a monochromatic wave propagating at oblique angle with a current of uniform vorticity is derived, for the first time to our knowledge.

\end{abstract}

%%%%%%%%%%%%%%%%%%%%%%%%%%%%%%%%%%%%%%%%%%%%%%%%%%%%%%	
%%%%%%%%%%%%%%%%% S E C T I O N %%%%%%%%%%%%%%%%%%%%%%	
%%%%%%%%%%%%%%%%%%%%%%%%%%%%%%%%%%%%%%%%%%%%%%%%%%%%%%	
\section{Introduction}
\label{sec:introduction}

\arev{	Ring waves are a canonical wave pattern studied scientifically for two centuries pioneered by \citet{cauchy1816theorie} and \citet{poisson1818m}. Even in the simplest situation, however --- deep water gravity waves --- the effect of weakly nonlinear wave steepness on these patterns has not been previously studied, and we find herein that they are profound and non-trivial. The addition of a  shear current, as recently studied to first order \citep{ellingsen_IVP}, makes the behaviour even richer.
}

%\adl{Surface waves contain a richness of nonlinear dynamics \ldots\ldots}

%\acommm{from here ...}
Traditionally, the analysis of a deterministic field of waves have been separated into two categories --- ray theory and mode coupling theory dealing with wave interactions in Fourier space 
\citep{wehausen60,whitham74}.
This text is concerned with the latter category.
	
Within mode coupling theory an extensive family of methods exists for problems where the wave distribution is narrowly centred around some specific wave vector.	Many of these narrow band approximation methods revolve around the application of multiple scale analysis.
An example is the nonlinear Schr{\"o}dinger equation, which can be obtained for the evolution of narrow banded wave packets \citep{stewartson_stuart_1971_Nonlinear_Schroedinger_formal}. Developed from the same principle, but with a different approach, is the equation derived by \citet{Zakharov_1968} during the same period.
 A number of years later this equation gained broader recognition  through a more extensive paper by \citet{Crawford_1981_about_Zakharov_eq}.	Multiple scale analysis can also be applied in systems oriented about the dispersive wave packets to yield equations of Korteweg-de Vries (KdV) type. 	This type of method has been used to study the far field of ring waves  \citep{Johnson_1990_KdV_ringwave_JFM, Karima_2016_KdV_stratified_ringwave_JFM, Karima_2016_KdV_ringwave}. 	Models such as those just mentioned have proven adapt at 	predicting	phenomena such as the formation of rough waves \citep{Karif_2003_rough_wave_many_models} 	and suitable for investigating wave packet stability in the presence of shear 	
\citep{Thomas_nonlinear_schroedinger_with_shear,francius17}. 
%\srev{
The fundamental reason for the divers wave phenomena occuring in the presence of vorticity is the chance in dispersion properties due to the interaction between waves
and current; see e.g.\ \citet{ellingsen_IVP} and the classic review by \citet{peregrine76}.
%}
%\acomm{Not clear what is meant by "the interaction" I think.}

The present paper revolves around mode coupling techniques for weakly nonlinear boundary value problems. A formalism is used similar to that presented by \cite{Phillips_1960_mode_coupling} and \cite{Longuet-Higgins_1961_four_wave_resonance} for selective wave components, and by \cite{Hasselmann_1962_mode_coupling_Taylordowntoz0,Hasselmann_part_2}, \cite{Benney_1962_mode_coupling_frequency_perturbation} and \cite{Holliday_1977_mode_coupling_compare_Taylor} for spectra.
Resonance and energy transfer between modes are 
the main foci in these references.
	
With the inclusion of a shear field new mechanisms for instability arise. Notable situations of wave--shear resonance include boundary layer transitions \citep{Benney_1960_first_Langmuirish,Benney_1960_O2_critical_layer_etc,Benney_1964_O2_BL_with_Lommel} and wind-driven mixing phenomena in the upper ocean layer \citep{Craik_1970_Langmuir_myidea,Craik_Leibovich_1976_Langmuir_averageNS}. Other interesting resonance effects involve resonant triads in the presence of strongly sheared flows, possible because the shear distorts the otherwise monotonic shape of the dispersion curve \citep{Craik_1968_resonance_critical_layer_Lommel}. The viscous region of the critical layer appears to be a vital mechanism for energy transfer in this case. Interactions between surface and `vorticity waves' in stratified flows is another notable example of resonance \citep{Drivas_2016_resonance_gravity_and_vorticity_waves}. 
Much useful insight on mode interaction phenomena in surface waves is gathered in the monographs by 
\citet{West_1981_lecture_notes_weak_interaction} and \citet{craik_1986_book_wave_interactions}.
Recent studies have demonstrated that striking and nontrivial phenomena occur in 3D wave--shear current systems also for linear waves 
\citep{ellingsen_IVP,ellingsen14a}.

\arev{
\cite{zakharov1990_nth_order} provides a foundation for solving the full three dimensional sheared Cauchy problem.
%The focus in this inspiring work is directed towards spectral evolution of ocean waves. 
Their focus is spectral evolution of ocean waves. 
It is shown that, in addition to the kinematic evolution attributable to resonant four-wave interactions \citep{Hasselmann_1962_mode_coupling_Taylordowntoz0}, a scattering process takes place from the resonant interaction of difference harmonics with the shear current via critical layers.
The formalism applied opens for an arbitrary but weak shear current through use of perturbation techniques for both current and nonliearities. 
\cite{shrira1993_shear_current_to_nth_order} has since proven convergence of current perturbation series in the linearised system whenever the characteristic perturbation ratio $U''(z)/\omega k$ is less than unity.

It is from a similar vantage point that we shall conduct our present study, for a strong%
\footnote{
\arev{
in the sense that wave-projected shear intensity 
$S (k_x/k) \sim \sqrt{g k} $.
}
}%
, albeit linear shear current. 
As opposed to the aforementioned authors, mainly active the 60's and 70's, we aim for a full spectral solution of the sheared Cauchy boundary value problem to second order in wave steepness.
Compared to boundary value problems, it is particularly the way in which the initial conditions are related to wave--shear interaction kinematics that furnish this Cauchy problem with new features.
Although endeavours such as these quickly become overwhelming in terms of complexity (three-dimensionally perturbed systems where potential theory is inapplicable) they are today more feasible through the available software for symbolic mathematics.
}

%Where the aforementioned authors explored significant territory with the mode coupling approach\arev{, mainly} in the 60's and 70's, they 
%all stopped short of a full spectral solution of the rotational \arev{Cauchy} problem. Solution complexity quickly becomes overwhelming when applying a Stokes expansion to a sheared wave spectrum 
%for which, being a 3D system, potential theory is not  applicable \citep{Ellingsen_vorticity_paradox}.
%This is perhaps why only  special cases of symmetry and resonance were considered. 
%In the present paper we seek full weakly nonlinear solutions to the sheared Cauchy boundary value problem to second order
%in wave steepness.

\arev{
A further motivation for our work is its relevance to climate research. Langmuir circulation, or it’s less sibling, Langmuir Turbulence, occurring in the presence of a spectrum of waves, is found to be the chief mechanism by which waves contribute to mixing of warm and cold water in the upper oceans. Existing ocean models account for this effect poorly or not at all \citep{dasaro2014quantifying,li2017statistical}, something that is believed to be a key reason for systematic mispredictions in fully coupled climate models \citep{belcher2012global}. The classical theories of Langmuir circulation originating from wave-current interactions have dealt exclusively with the situation where waves and currents are aligned with each other \citep{leibovich83}. In the oceans, the situation is often that waves (driven by wind) and currents make an oblique angle with each other. Simple parameterisations of this situation has recently been implemented in operational ocean models \citep{li2017statistical}, but these rely on only two practically unvalidated and mutually dependent large-eddy simulation studies \citep{vanRoekel_2012_Langmuir_LES_of_CL_eqs,mcwilliams2014langmuir}. Understanding the fundamental mechanism is therefore of the essence, and our work contributes the first steps in this direction, studying one of the two ways in which Langmuir turbulence might be created.
}

We use our theory to 2nd order to 
generalise the theory of \citet{Craik_1970_Langmuir_myidea} (developed further by \citet{Craik_Leibovich_1976_Langmuir_averageNS}) of a mechanism for Langmuir circulation, to more general and realistic situations. In particular we show that Langmuir rolls can be formed by second order wave--shear current interactions also when waves and current are not aligned but meet at oblique angles. In his famous review \citet{leibovich83} remarked ``No work has yet been done using the [Craik--Leibovich] theories when $\mathbf{u}_s$ is not parallel to the horizontally averaged current $\bU$. Heuristic considerations ... suggest that instability could occur whenever $\mathbf{u}_s\cdot\bU\gtrsim 0$, although there is no longer any reason to believe that rolls would be favoured.'' ($\bu_s$ is the Stokes drift velocity).
A large-eddy simulation by \cite{vanRoekel_2012_Langmuir_LES_of_CL_eqs}, however, indicates that Langmuir-like structures should occur in such a misaligned situation.
 We demonstrate here that the Craik--Leibovich `direct--drive' mechanism (called `CL1' by \citet{faller78}) will create (distorted) roll structures not only for oblique wave-current incidence angles, but for \emph{all} angles except $\bu_s\cdot\bU=0$. (Note, however, that the mechanism `CL2', if present, will have a stabilising effect when $\bu_s\cdot\bU<0$ \citep{leibovich83}; a stability analysis of such a situation is outside the scope of this study, but would be an important topic for the future).

%\acommm{Remove the next paragraph?}
%
%We also investigate oblique, nonlinear wave--shear current interaction phenomena in the Cauchy--Poisson (ring wave) problem. The transient behaviour of such a general initial value system sheds new light on mode coupling phenomena, particle motions and Stokes drift-like transport for a continuous spectrum of waves. 

The text is structured as follows: 
%\arev{
the problem is stated in Sec~\ref{sec:problem} and its perturbation series solution constructed in Sec~\ref{sec:solution_construction} for prescribable initial conditions.
Evaluating the solution integrals, explicit flow field and surface elevation expressions are derived to second order in Sec.~\ref{sec:O2_solution}, where we also discuss solution properties with regard to critical layers and dispersive and advective resonance.
%}
Expressions for approximating fluid particle trajectories are derived in Sec.~\ref{sec:particle_trajectories}, 
were we also present second order expressions for particle trajectories in the presence of a uniform shear current for monochromatic waves.
Numerical examples include 
obliquely interacting wave trains, both parallel and misaligned to the current,
as well as two-dimensional particle trajectories and
three-dimensional flow fields for the shared Cauchy-Poisson initial value problem.
These are found in Sec.~\ref{sec:results}, followed by a summary in Sec.~\ref{sec:conclusions}.
Bulky expressions of the internal flow field are relegated to Appendix~\ref{sec:uvw}--\ref{sec:special_cases}.
In Appendix~\ref{sec:statphase} the method of stationary phase (in 2D) is applied for an asymptotic approximation the present problem.

%%%%%%%%%%%%%%%%%%%%%%%%%%%%%%%%%%%%%%%%%%%%%%%%%%%%%%	
%%%%%%%%%%%%%%%%% S E C T I O N %%%%%%%%%%%%%%%%%%%%%%	
%%%%%%%%%%%%%%%%%%%%%%%%%%%%%%%%%%%%%%%%%%%%%%%%%%%%%%	
\section{Statement of the Problem}
\label{sec:problem}

\arev{
We examine flow consisting of a strong unidirectional shear current, aligned with the $x$-axis, perturbed by waves of moderate but finite steepness.
The Euler equations with accompanying free surface boundary conditions for this velocity field $U(z)\bm e_x + \p\bu(x,y,z,t)$ are}
%
%We seek an approximate solution to the Euler equations with free surface boundary conditions
\begin{subequations}
	\begin{alignat}{2}
		\left.
		\begin{aligned}
				 \pdtbar \p\bu + U' \p w \bm e_x + \p\nabla \p p  &= - ( \p\bu \!\cdot\! \p\nabla)\p \bu-  g \bm e_z\\
				\p\nabla \!\cdot\! \p \bu &=0
		\end{aligned}
		\right\};&
		\quad& z&\leq\p \zeta, \label{eq:problem:Euler}\\
		\left.
		\begin{aligned}
		\pp_t\p \zeta + \p \bu \!\cdot\! \p\nabla \p \zeta &= \p w \\
		\p p 
		&= 0%\p \Pa
		\end{aligned}\right\};
		& & z &= \p \zeta, \label{eq:problem:BC_free_surface}
		\\
		\quad \p w = 0\;; && z&\rightarrow-\infty.\label{eq:problem:BC_bed}
	\end{alignat}
	\label{eq:problem}%
	\end{subequations}%
	%
%for waves of moderate but finite steepness. 
Hatted symbols refer to real--space (as opposed to Fourier space) \arev{and density has been absorbed into the pressure. 
Here, $\p\nabla = (\pp_x,\pp_y,\pp_z)$. 
The problem (2.1) has been considered in a similar fashion by 
\citet{zakharov1990_nth_order} and \citet{shrira1993_shear_current_to_nth_order}, to whom we refer for further details.}
%An externally applied interfacial pressure $ \p \Pa\of\bmr$ has been included which can be taken to equal zero on a  free surface of uniform pressure. 
%
\arev{
Pressure and the horizontal velocities can be eliminated form the Euler equations to yield a Rayleigh 
(inviscid Orr--Sommerfeld) 
equation on the form
\begin{align}
    \pdtbar  \p \nabla^2 \p w - U''\pp_x \p w &= \pRay;
	&
	\pRay &= \pp_z \p\nabla\_h \!\cdot\! [(\p\bu\!\cdot\!\p\nabla)\p\bu\_h] - \p \nabla\_h^2  [(\p\bu\!\cdot\!\p\nabla)\p w]
   \label{eq:hRayleigh}
\end{align}
with $ \pdtbar=\pp_t+U(z) \pp_x$, $\p\nabla\_h=(\pp_x,\pp_y)$ and $\p \bu\_h=(\p u,\p v)$.

The surface boundary conditions are highly implicit as they are defined upon the same free surface which they describe. 
A common way of expressing these boundary conditions explicitly is by Taylor expanding them down to the reference plane $z=0$.
To further reveal the nature of these boundary conditions
%Upon doing 
one may
integrate the $z$-momentum equation, take the horizontal  Laplacian and use \eqref{eq:problem:BC_free_surface} and \eqref{eq:hRayleigh} to derive
 \begin{align}
	 \pdtbar^2(\pp_z\p w) -  \pdtbar (U' \pp_z \p w) - g \p \nabla\_h^2\p w &= 
 g \p \nabla\_h^2 \big[(\p L-1)(\p w- U \pp_x \p \zeta)-\p L(\p \bu\_h\!\cdot\!\p \nabla\_h\p\zeta )  \big]
\nonumber\\&
+  \pdtbar\big\{
\p \nabla\_h^2 \big[ \p \zeta \p L_1( \pdtbar \p w-\p \bu\!\cdot\! \p\nabla \p w)\big]
-\p\nabla\_h\! \cdot\! (\p\bu\!\cdot\!\p\nabla)\p\bu\_h  
 \big\}
   \label{eq:hBC}
\end{align}
where $\p L = 1+\p\zeta \pp_z+\tfrac12 \p\zeta^2 \pp_z^2+\ldots $ and 
$\p L_1 = 1+\tfrac12\p\zeta \pp_z+\tfrac16 \p\zeta^2 \pp_z^2+\ldots\ $.
A perturbation solution is admissible from the above system, both for dealing with nonlinearities and the shear current \citep{zakharov1990_nth_order,shrira1993_shear_current_to_nth_order}.
We will here adopt the Stokes perturbation for nonlinearities but will for the current instead consider and arbitrarily strong (zeroth order) linear  profile with a uniform shear strength $S$;
\begin{equation}
U(z)= S z.
%\label{eq:}
\end{equation}
}

\section{Constructing the Solution}
%\section{Second order solution}
%\label{sec:O2_solution_new}
\label{sec:solution_construction}
\arev{
The standard approach for solving the above nonlinear system is to seek
for each physical perturbation quantity $\p\psi$
 a Stokes perturbation solution
%\begin{equation}
$
	\p\psi = {\sum}_{n=0}^\infty \epsilon^n\,\psi\n,
	$
%\label{eq:Stokes_expansion}
%\end{equation}
where the smallness parameter $\epsilon$  represents the wave steepness ($\epsilon = k \eta$ where $k$ and $\eta$ is a characteristic wave vector and amplitude, respectively.)
Our problem is then reduced to a cascade of linearised problems for each perturbation order.
Each linearised problem has the same repeating structure; only the known lower order interaction terms (right-hand, inhomogeneous terms) differ, rapidly increasing in complexity at higher orders.
}

\arev{
%After substituting in the perturbation series 
We proceed by taking the Fourier transform in the horizontal plane of the first order perturbation components
\begin{equation}
\psi\oo = \Fourier_\bk \p \psi \oo,
\label{eq:Fourier_transform_O1}
\end{equation}
$\bk = (k_x,k_y)$.
%All higher order physical components generated form the cascade of linearised systems are then seen to be represented as nested inverse transforms
All higher order components generated by the system are then given in physical space by nested inverse transforms
\begin{equation}
\p\psi\ot = \invFourier_{\bk_1} \invFourier_{\bk_2}  \psi \ot,
\qquad
\p\psi^{(3)} = \invFourier_{\bk_1} \invFourier_{\bk_2}  \invFourier_{\bk_3}  \psi ^{(3)}, \quad \text{etc\ldots,}
\label{eq:Fourier_transform_nested}
\end{equation}
that is, convolutions of the lower order harmonics.
}
\arev{
%All gradient and Laplace operators simplify in Fourier space to
In Fourier space, the key operators become
\begin{align}
 \pdtbar&\rightarrow \dtbar = \pp_t+\rmi k_x U(z), 
&
\p\nabla &\rightarrow \nabla = (\rmi k_x, \rmi k_y, \pp_z), 
&
\p\nabla^2 &\rightarrow \nabla^2 =  \pp_z^2- k^2,
\end{align}
where $\bk$ is the sum of the convolution wave vectors $\bk_1$, $\bk_2$ etc.\ at each order and $k = |\bk|$.
We will in what follows restrict ourselves to solving the above system to second order in $\epsilon$ 
%(higher order expressions in shared three-dimensional flow turn very bulky indeed)
 and so $\bk = \bk_1+\bk_2$ whenever working with second order expressions.
}
%
%We impose a zeroth order background flow field with a uniform shear flow aligned with the $x$-axis 
%and choose the system of reference following the surface; $\bu^{(0)}(z)=Sz\bm e_x$, where $S$ is the vorticity of the unisturbed flow.
%This problem permits a Stokes perturbation solution on the form	
%%
%\begin{equation}
	%\up = [S z \bm e_x, -\rho g z,0]+ \sum_{n=1}^\infty \epsilon^n\, [ \p \bu, \p p, \p \zeta]\n,
%\label{eq:Stokes_expansion}
%\end{equation}
%%
%where $\epsilon = k \h$ is a measure of the wave steepness, $k$ being the wave vector and $\h$ the amplitude, and $(n)$ indicates the perturbation order in $\epsilon$. 
%Taking the Fourier transform of a first order perturbation quantity 
%$\p\psi\oo$ the increasing orders will contain nested convolutions of the preceding ones. We write
%\begin{align}
  %\p\psi\n(\bmr,z,t) &= \idotsint \frac{\dd\bk_1\dots\dd\bk_n}{(2\pi)^{2n}} 	\f\psi\n\of{\bk_1, \dots, \bk_n;z,t} \rme ^{\rmi \bk\cdot\bm r},
  %\label{eq:Fourier_transform}
%\end{align}
%where 
%$\bmr=(x,y,0)$, $\bk_m=(k_{mx},k_{my},0)^T$
%and $  \bk = \textstyle\sum_{m=1}^n\bk_m $. 
%Dependence on arguments $\bk_1,\dots,\bk_n$ is implicit in all that follows.
%We impose that any set of perturbations up to order $n$ must solve system \eqref{eq:problem} in Fourier space to $n$-th order accuracy in $\epsilon$. Recursively increasing $n$, this yields a linearised set equations for each order, parametric in wave vector.
	%
	%
	%
%%%%%%%%%%%%%%%%%%% S E C T I O N %%%%%%%%%%%%%%%%%%%%%%	
%\subsection{Internal flow field}
%\label{sec:internal_flow_field}
\arev{From here on we suppress the perturbation order superscript $(n)$ and drop the wave vectors in the argument lists.}
\renewcommand{\n}{}
\renewcommand{\nn}{}

Solving the Rayleigh equation \arev{\eqref{eq:hRayleigh}, now in Fourier space,} yields
\begin{equation}
   w\n\of{z,t} =  A\n\of{t}\rme ^{k z} + \tilde A\n\of{t}\rme ^{-k z}
	\arev{+ \int \dd z' C(z') \,\rme ^{-\rmi k_x S z' t} \sinh k (z-z')}
	+  w_\tpart\nn\of{z,t},
  \label{eq:w_sol}
\end{equation}	
\arev{where $  w_\tpart\nn\of{z,t}$ is the particular solution
generated by cross terms of oblique rotational wave interactions.
}
%comes from the cross terms of oblique rotational wave interactions. 
The integration coefficients $A\of t$, $\tilde A\of t$ and $C\of z$ are determined from the bottom and surface boundary conditions and from the initial condition, respectively. 
\arev{Making sure that we evaluate the integrals in forms that vanish as $z\rightarrow -\infty$ we set $\tilde A\of t\equiv 0$ in what follows.}
%As perturbations  of all orders must satisfy the infinite depth boundary condition, $\f w\n\rightarrow0$ as $z\rightarrow-\infty$, we have that $\mc R\nn$ vanishes in the same limit. Assuming the same property for $C(z)$, the infinite depth boundary condition then yields  $\tilde A\of t\equiv 0$. 

\arev{
An initial background velocity field can be contained in the integration coefficient $C\of z$. This field is then passively advected by the background shear flow via the kernel $\exp\br{-\rmi k_x S z t}$.
Such fields are usually not considered in analyses of, say, ocean waves, which have no particular origin in time.
One may then 
work without advected velocity fields
so that all frequencies are decoupled from the depth $z$.
Evaluating time derivatives or integrals is then simple. 
Ignoring advected fields in a Cauchy problem,
as \cite{ellingsen_IVP} incorrectly argued is necessary,
 implies that wave-generated vorticity is present at $t=0$.
%\cite{ellingsen_IVP} arguing incorrectly that doing so is necessary.
%\cite{ellingsen_IVP} ignored them all the same (initial rotation thus present being weak), arguing incorrectly that doing so is necessary.
%Such fields were ignored %also 
%in the first order treatment by \cite{ellingsen_IVP} who argued incorrectly that 
%doing so is necessary.
%Ignoring advected fields entails wave-generated vorticity being present at $t=0$.
%In the present work we 
%shall instead require an irrotational initial state (except for the vorticity $S$ in shear current) and use advective background vorticity to cancel that which is generated by the wave--shear interactions at $t=0$.
In the present work we shall at second order instead use advective background vorticity to cancel that which is generated by the wave--shear interactions at $t=0$ and thus generate an initial state which is irrotational (except for the vorticity $S$ in shear current).
%require 
%the solution to be initially irrotational except for the vorticity attributed to the shear current.
%%that the solution should initially contain no vorticity other than that of the unperturbed background flow.
%This means that the initial background vorticity field in $C(z)$ cancels the wave generated vorticity field at $t=0$. 
}
%
%The integration coefficient $C\of z$ determines the background vorticity present at $t=0$. This background vorticity is unrelated to the surface waves and is advected by the background shear flow, as can be seen in the kernel $\exp\br{-\rmi k_x S z t}$.
%Two choices for $C\of z$ are apparent:
%One may simply take $C\equiv0$ such that all dynamic vorticity is generated by the wave action. 
%This was done in the first order treatment by \citet{ellingsen_IVP}, who argued incorrectly that this choice is necessary. 
%In this case all frequencies are decoupled from the depth $z$ (at all orders $n$) and evaluating any time derivative or integral becomes trivial. 
%Wave vorticity is then initially present the solution.
%
%The other 
%possibility, for which we opt herein, 
%is to require $\f \f w_\tpart\nn=0$ at $t=0$, \ie, that the solution should 
%initially contain no vorticity other than that of the unperturbed background flow.
%This means that 
%$C\of z = - \textstyle\int\!\dd t \, \Ray\nn\of{z,t} \rme ^{\rmi k_x S z t}\big|_{t=0}$
%in order that the initial background vorticity field cancels the wave generated  vorticity field at $t=0$. 
\arev{
As we shall see later, this provides critical layers and advective resonances
 %appearing in our solution
 with an origin in time which also resolves ambiguity concerning integration paths around critical points.
}%

\arev{
With this choice we can absorb $C(z)$ into the particular solution $ w_\tpart\nn$ as a lower integration limit and write
}
\begin{equation}
 \f w_\tpart\nn \of{z,t} = 
\frac1k \int_{-\infty}^z \!\!\dd z'
\int_0^t \dd t' \Ray\nn\of{z',t'} \rme ^{-\rmi k_x S z' (t-t')
} \sinh k (z-z').
\label{eq:w_part}
\end{equation}	
%
%\arev{
%Because we seek so solve the full velocity field we proceed form here integrating out the pressure from the linearised $z$-momentum equation of \eqref{eq:problem:Euler}, using this in turn to evaluate the surface boundary conditions \eqref{eq:BC_system} and finally extracting the horizontal velocity components, also from \eqref{eq:problem:Euler}. 
%This turns out to be roughly as labour intensive as proceeding via \eqref{eq:huh}, \eqref{eq:hBC}.
%}
\arev{
%Because we seek so solve the full velocity field we proceed through the original system \eqref{eq:problem} as this is roughly as labour intensive as using \eqref{eq:huh}, \eqref{eq:hBC}.
An integration constant will appear also in the horizontal velocity components. These are chosen in a similar manner (see Appendix~\ref{sec:uvw}.) 
%
%Equivalent to working with \eqref{eq:hBC}, 
%%As a matter of preference,
 %we proceed by integrating out the pressure from the $z$-momentum equation of \eqref{eq:problem:Euler} 
%and evaluate the original boundary conditions \eqref{eq:problem:BC_free_surface}
%
%Doing so and proceeding with original boundary conditions \eqref{eq:problem:BC_free_surface} is roughly as labour intensive as working with \eqref{eq:hBC}.
%
We proceed by integrating out the pressure from the $z$-momentum equation of \eqref{eq:problem:Euler} and find
\begin{equation}
\f p\n\of{z,t} =
-\frac1k\br{\dtbar-\rmi S \frac{k_x}{k}}
 A(t) \rme ^{k z} 
+ p_\tpart(z,t).
%- \int_{-\infty}^z \!\! \dd z' \sqbrac{\br{  \f\bu\cdot\f\nabla\f w +
%\dtbar \f w_\tpart}\nn}' 
\label{eq:p_of_z}
\end{equation}%
$p_\tpart$ at second order is given in \eqref{eq:p_cross}.
}%

%
%\subsection{Boundary conditions}
\arev{
Equivalent to \eqref{eq:hBC},
the original boundary conditions \eqref{eq:problem:BC_free_surface}, Taylor expanded down to the reference plane $z=0$, yields 
\begin{align}
\pp_{t}^2 \f \zeta\n -  \rmi S \frac{k_x}k\pp_t\f \zeta\n + g k \f \zeta\n &=\D
  ,\label{eq:disp_eq_zeta_RHS}
		\\
		A\n(t) &=\pp_t \f \zeta\n -\Dkin,	\label{eq:A_n}
\end{align}
upon inserting \eqref{eq:p_of_z}.
Here, $\D = k \Ddyn + \br{\pp_t - \rmi S {k_x}/{k}}\Dkin$, 
$\Ddyn$ and $\Dkin$ containing the lower order interaction terms of the Taylor expanded dynamic and kinematic boundary conditions, respectively.
}
They are presented to second order in Appendix~\ref{sec:uvw}.
The solution of \eqref{eq:disp_eq_zeta_RHS} is
\begin{equation}
%\adl{
%\f \zeta\n\of t = 
	%\sumpm \f \zeta_\pms\n \rme ^{-\rmi \ompm  t}
	%+ \f\zeta_\bound\n\of t 	- 
	%\f \Pa\n/\rho g
	%}
\f \zeta\n\of t = 
	\sumpm \sqbrac{ \f \zeta_{\free,\pms}\n \rme^{-\rmi \ompm  t}
	+ \f\zeta_{\bound,\pms}\n\of t }	
\label{eq:zeta_of_t}
\end{equation}
where $\zeta_{\free,\pm}$ is independent of time, 
\begin{equation}
 %\f\zeta_\bound \n\of t = 
	%\sumpm \pms \int^t\! \dd t'\, \frac{\rmi  \D \of{t'}}{\om_+-\om_-} \rme ^{\rmi \ompm (t'-t) }
	 \f\zeta_{\bound,\pm} \n\of t = 
	\pm \int^t\! \dd t'\, \frac{\rmi  \D \of{t'}}{\om_+-\om_-} \rme ^{\rmi \om_\pm (t'-t) }
\label{eq:zeta_part}
\end{equation}
and 
$
\om_\pm = \Om_\pm\of\bk.
$
Eigenfrequencies of the homogeneous part of \eqref{eq:disp_eq_zeta_RHS} are 
	\begin{equation}
	\Om_\pm\of\bk =- \frac S2 \frac{k_x}{k} \pm \sqrt{\br{\frac S2 \frac{k_x}{k}}^2 + gk}.
	\label{eq:omega_pm}
	\end{equation}

The homogeneous components $\f\zeta_{\free,\pm}\n\rme ^{-\rmi \om_\pm  t}$
of \eqref{eq:zeta_of_t} are in what follows termed `free waves' as they propagate according to the dispersion relation \eqref{eq:omega_pm}. 
We are free to choose their amplitudes $\f\zeta_{\free,\pm}\n$ such that initial surface conditions are satisfied.
The modes of the particular solution $\f\zeta_{\bound,\pm}\n$ are in the following termed `bound modes'. 
Their frequency does not obey the dispersion relation as they arise as a nonlinear correction to the lower order solution.

%The corresponding terms for $A\n$, which link the surface elevation to the internal flow field, are found directly from the kinematic boundary condition: 
%%
%\begin{align}%
%A\n\of t &= \frac{1}{k} \pp_t\zeta\n -\frac{1}{k} \Taylor{\p w - \p \bu \cdot \p\nabla \p \zeta}. 
%\label{eq:A_n}%
%\end{align}%
%%\\

\arev{
We mentioned that the method of Taylor expanding the boundary conditions down to a reference plane, although common, does place restrictions on the wave spectrum width and can lead to convergence issues when short crested wave ride atop long crested ones \citep{Holliday_1977_mode_coupling_compare_Taylor,rainey18}. 
Our interest lies mainly in narrow spectra and so we are content with the above procedure. 
Boundary techniques suited for wider spectra 
can be constructed at the expense of increased complexity (for potential flows see
\citet{Zakharov_1968,Watson_1975_mode_coupling_Taylor,West_1981_lecture_notes_weak_interaction,Dommermuth_mode_coupling_numerical}).
}
\\

%%%%%%%%%%%%%%%%%%% S E C T I O N %%%%%%%%%%%%%%%%%%%%%%	
%\subsection{Initial conditions}
%\label{sec:IC}

\renewcommand{\n}{^{(n)}}

\arev{The final stage of laying out a solution is 
 to impose on the flow 
some prescribed physical initial surface elevation $\p\zeta\IC\of{\bmr}$ and its time derivative $\dot{\p\zeta}\IC\of{\bmr}$.
A solution in the form of a Stokes perturbation series must match these conditions at $t=0$.
Initial conditions apply to the sum of Stokes terms, but does not determine the initial value of each term individually, and exactly how these conditions are satisfied becomes a matter of choice.
We make the assumption that higher order terms can be made not to contribute to $\p{\zeta}$ and $\pp_t\p{\zeta}$ at $t=0$, whence the initial conditions become
(briefly reintroducing the $(n)$ order notation)}
\begin{align*}
	\p\zeta\oo\of{\bmr,0} &= \p\zeta\IC\of{\bmr}
	& &\text{and}&
	\pp_t\p\zeta\oo\of{\bmr,t}\big|_{t=0} &= \dot{\p\zeta}\IC\of{\bmr}
\end{align*}
which combine to yield
\begin{equation}
		\f\zeta\oo_{\free,\pm} = \mp\frac{\om_\mp \f\zeta\IC - \rmi \, \f{\dot\zeta}\IC}{\om_+-\om_-}.
	\label{eq:zeta_IC}
\end{equation}
	Here, $\f\zeta\IC\of{\bm k}$ and $\f{\dot\zeta}\IC\of{\bm k}$ are the Fourier transforms of $\p\zeta\IC\of{\bmr}$ and $\dot{\p\zeta}\IC\of{\bmr}$, respectively.
The bound modes from the higher order terms alters the initial state of our solution, but we can choose 
%the
 higher order free modes to compensate for this. 
Imposing $\f\zeta\n\of{t} = \pp_t\f\zeta\n\of t = 0$ at $t=0$ 
for all nonlinear orders, we find from \eqref{eq:zeta_IC} that
\begin{equation}
	\f\zeta\n_{\free,\pm} =\left. \pm \frac{\br{\om_\mp - \rmi \pp_t}\f\zeta_\bound\n\of t}{\om_+-\om_-}\right|_{t=0}
	\label{eq:zeta_ICn}
\end{equation}
\begin{subequations}%
for $n>1$, where  $\f\zeta\_{\bound}\n = \f\zeta\_{\bound,+}\n+\f\zeta\_{\bound,-}\n$. %\sumpm\f\zeta\_{\bound,\pms}$. 
Taking the time derivative of \eqref{eq:zeta_part} one readily finds that $\f\zeta\_{\bound}$ obeys
$\pp_t \f\zeta\_{\bound}\n = 
 %-\rmi\sumpm   \om_\pms \f\zeta\_{\bound,\pms}\n
-   \rmi \,\om_+ \f\zeta\_{\bound,+}\n
-   \rmi \,\om_- \f\zeta\_{\bound,-}\n
$ 
at $t=0$
so that \eqref{eq:zeta_ICn} yields
\begin{equation}
\f\zeta\n_{\free,\pm} =  -	 \f\zeta_{\bound,\pm}\n\Big|_{t=0} %  (t=0)%  
\label{eq:zeta_ICn_evaluated:IVP}
\end{equation}
for $n>1$.
%
%\begin{equation}
%\f \zeta\n\of t = 
	%\sumpm \br{ 
	 %\f\zeta_{\bound,\pms}\n\of t - \f \zeta_{\bound,\pms}\n\of 0 \rme^{-\rmi \ompm  t}}	- \frac{\f \Pa\n}{\rho g}
%\label{eq:zeta_net}
%\end{equation}
\arev{
%We thus ensure that initial
 %conditions on the surface elevation 
%remain unaffected by the introduction of higher order corrections, 
%by matching single wave interaction harmonics with free harmonics with equal amplitude and opposite phase at $t=0$.
Thus, by matching single wave interaction harmonics at $t=0$ with free harmonics of equal amplitude and opposite phase,
we ensure that initial conditions on the surface elevation 
remain unaffected by the introduction of higher order corrections.
}
%\Q{Above intended as response to Ref. 3: `In section 2.3 $\zeta^{(n)}(0)=0$, does it mean that there is no wave interaction initially?' OK?}

In Sec.~\ref{sec:Craik_case} we shall also consider oblique interactions of two monochromatic waves. 
Here we do then not require any particular initial state but instead impose
 %for $n>1$
\begin{equation}
\f\zeta\n_{\free,\pm}=0.
\label{eq:zeta_ICn_evaluated:Langmuir}
\end{equation}
\label{eq:zeta_ICn_evaluated}
\end{subequations}

%%%%%%%%%%%%%%%%%%%%%%%%%%%%%%%%%%%%%%%%%%%%%%%%%%%%%%%%	
%%%%%%%%%%%%%%%%%%% S E C T I O N %%%%%%%%%%%%%%%%%%%%%%	
%%%%%%%%%%%%%%%%%%%%%%%%%%%%%%%%%%%%%%%%%%%%%%%%%%%%%%%%	
\section{Second order solution}
\label{sec:O2_solution}

\renewcommand{\ot}{}

Consider first the linear solution. 
$\Ray\nn=\D=0$ in the first order components so that all particular solution terms drop out of \eqref{eq:w_sol}
%, \eqref{eq:p_of_z}, \eqref{eq:uv} 
and \eqref{eq:zeta_of_t}, 
leaving only the homogeneous components \citep{Ellingsen_vorticity_paradox}:
\begin{subequations} 
\begin{align}
  %\f u\oo(z,t) &= \sumpm A\oo_\pms \br{  \rmi\,k_x - S k_y \kyk  \frac{\rme ^{\rmi\, \oms_\pms t}-1}{ \rmi\, \oms_\pms} }		\rme ^{k z - \rmi \ompm t}
	%\label{eq:O1_flowfeeld:u}\\
  %\f v\oo(z,t)&= \sumpm A\oo_\pms \br{ \rmi\, k_y +  S k_x \kyk  \frac{\rme ^{\rmi \oms_\pms t}-1}{\rmi\, \oms_\pms} } \rme ^{k z - \rmi\, \ompm t} 
	%\label{eq:O1_flowfeeld:v}\\
	\f \bu\_h\oo(z,t) &= \sumpm A\oo_\pms \br{  \rmi\,\frac{\bk}{k} + S \frac{ \bk\bb}{k} \kyk  \frac{\rme ^{\rmi\, \oms_\pms t}-1}{ \rmi\, \oms_\pms} }		\rme ^{k z - \rmi \ompm t}
	\label{eq:O1_flowfeeld:uv}\\	
  \f w\oo(z,t)&= \sumpm  A\oo_\pms\rme ^{k z - \rmi\, \ompm t}  \\
  \f p\oo(z,t)&= \sumpm \rmi\, A\oo_\pms \frac1k\br{\oms_\pms + S \kxk  } \rme ^{k z - \rmi\, \ompm t}\\
  \f \zeta\oo(t) &= 	\sumpm \f \zeta_\pms\oo \rme ^{-\rmi \om_\pms  t}
;%	- \frac{\f \Pa}{\rho g};
		\qquad
		A_\pm\oo = -\rmi \,\zeta_\pm\oo \om_\pm,
\end{align}
\label{eq:O1_flowfeeld}%
\end{subequations}%
where and $\bk\bb=(-k_y,k_x)$ and the Doppler shifted frequency $\oms_\pm=\om_\pm - k_x S z$ originating from the $\dtbar$ operator.
As usual a dependence on $\bk$ is understood.

\arev{
Integration coefficients appearing in the horizontal velocities
%, put to zero in \citet{ellingsen_IVP},
have 
%for illustration here
in the above equation
 been made to yield an irrotational initial state, as per the discussion in the previous section. 
This part of the velocities is identified by the $\propto\exp (\rmi\, \oms_\pm t)$ kernel and constitutes an advection process.
}
\arev{A surface initially at rest ($\f{\dot\zeta}\IC=0 $) will then produce a flow field which is initially quiescent apart from the background current \arev{$S z$}. }
At the critical depths $z=\om_\pm/(k_x S)$ the flow field perturbation becomes proportional to
\begin{equation}
  \lim_{\oms\to 0} \frac{\rme^{\rmi\, \oms t} -1}{\rmi \bar\omega} = t,
	\label{eq:limit}
\end{equation}
\arev{rather than becoming a diverging singularity as it would without the advective term.
\arev{One should however keep in mind that 
in many practical situations, including shear-layer flow in the upper oceans or river delta plume flow, the depths at which such first order critical layers reside will typically lie far beneath the shear penetration depth above which the uniform current model is meaningful.
This is not necessarily true for the higher order harmonics.}

Including advective terms 
in the solution \eqref{eq:O1_flowfeeld} increases the
%increases solution 
complexity significantly, yet, away from critical layers,  
\arev{their contribution at first order is only moderate undulations.
Neglecting advective terms causes at first order merely a small deviation from the irrotational initial state.}
We therefore proceed \textit{without} the first order advective terms  $\exp ( \rmi \oms_\pm t )$ in what follows, accepting this slight initial fluid motion beneath the surface. 
It is, as will be shown, in dealing the second and higher order harmonics that imposing initial irrotationality can be crucial;
such \arev{states give} advective resonances (Sec~\ref{sec:advective_resonance}) an origin in time.
}

\arev{
We now turn to the second order quantities.
$\Ray$, the right--hand side the Rayleigh equation \eqref{eq:hRayleigh}, consists at second order only of interactions between freely dispersing waves. 
Two frequency branches are present in the first order components at every wave vector.
The time dependency of $\Ray$ will at second order then be made up of four frequency branch combinations to be summed together. 
Keeping this in mind, we 
let $\omega$ represent any of the four different frequencies $\Omega_{\sigma_1}(\bk_1) + \Omega_{\sigma_2}(\bk_2)$ with $\sigma_{1,2}=\pm$, and write
\begin{equation}
%\omsum \leftarrow \Om_{\pm_1}\of{\bk_1}+\Om_{\pm_2}\of{\bk_2} ,
%\qquad
\mc R\ott\of {z,t} = \tR\of z \rme ^{-\rmi \omsum t},
\end{equation}
remembering to sum all four branch combinations in the end.
}
\arev{Using partial fractions, $\tR$ can be written}
\begin{equation}    %A_{\pm_1}\oo\of{\bk_1} A_{\pm_2}\oo\of{\bk_2}
\tR
(z)
= A_1\oo A_2\oo \frac{\ktimesk}{k_1 k_2} \sum_{i=1}^2\sum_{j=1}^3 \frac{a_{ij}}{(\xi_i-z)^j}\rme ^{\ks z}.
\label{eq:mcRot_pmpm}
\end{equation}
We have here introduced the shorthand notations $\ks=k_1+k_2$ and $\ktimesk=k_{1x}k_{2y}-k_{1y}k_{2x}$, and the critical depths 
\begin{equation}
\xi_1=\frac{\om_1 }{k_{1x}S}
,\qquad \xi_2=\frac{\om_2}{k_{2x}S}
,\qquad \xi_3 = \frac{\omsum}{k_x S}.
%\label{eq:}
\end{equation}
$\ks$ should not be confused the \arev{modulus} of the second order wave vector, $k=|\bk_1+\bk_2|$.
\arev{
Note that
$\xi_1$ and $\xi_2$ will for parameters typical for the ocean ($S\sim0.01$/s, phase speed $\sim$m/s) be of order hundreds of meters, where the uniform shear model is unlikely to be representative.  
$\xi_3$ can on the other hand 
take on values much closer to the surface in the form of difference harmonics.
%be anything for difference waves. , as Zakharov...
}
%(Remember, $\bk=\bk_1+\bk_2$). 
Assuming $\xi_1\neq\xi_2$ 
(see Eq.~\eqref{eq:a_coef_xi1eqxi2} for the case $\xi_1=\xi_2$), 
the coefficients in \eqref{eq:mcRot_pmpm} read
\begin{subequations}
%
%\begin{align}
  %a_{i1}&= (-1)^i   \sqbrac{ k_1 k_2 (k_1 k_2-\kdk) - \ks  \tanthetam \frac{\ktimesk}{ \xi_1-\xi_2} }\tanthetai,\\
  %a_{i2}&= (-1)^i \sqbrac{ k_m (k_1 k_2-\kdk) -  \tanthetam \frac{\ktimesk}{\xi_1-\xi_2}  }\tanthetai,\\
  %a_{i3}&= (-1)^i k_m^2\tanthetai.
%\end{align}
\begin{align}
  a_{i1}&= (-1)^i   \sqbrac{ k_1 k_2-\kdk - \frac{\ks}{\xi_1-\xi_2}  \frac{\ktimesk}{k_1 k_2}\tanthetam  }\tanthetai,\\
  a_{i2}&= (-1)^i \frac{1}{k_i} \sqbrac{ k_1 k_2-\kdk - \frac{k_i}{\xi_1-\xi_2}  \frac{\ktimesk}{k_1 k_2}\tanthetam  }\tanthetai,\\
  a_{i3}&= (-1)^i \frac{k_m}{k_i}\tanthetai.
\end{align}
\label{eq:a_coef}%
\end{subequations}%
Here $i,m\in\{1,2\}$ so that $i\neq m$%
, and $\tan\theta_i = k_{iy}/k_{ix}$.
Inserting this into \eqref{eq:w_part} we first obtain
\begin{equation}
 \f w_\tpart\ott (z,t)
 = 
\frac{1}{ k }  \sumpmpm \rme ^{- \rmi\ompmpm t}\int_{-\infty}^z \!\dd z' \:
\frac{\rme ^{\rmi \oms' t}-1}{ \rmi \oms'} 
 \tR\of{z'} \sinh k (z-z'),
\label{eq:w_part_int_t}
\end{equation}	
\arev{$\oms' = \omsum - \rmi k_x S z' = k_x S( \xi_3-z')$.}
Once more applying partial fractions we find
\begin{equation}
\frac{\tR\of z}{ \xi_3-z} = A_1 A_2
\frac{\ktimesk}{k_1 k_2}\sumij \frac{b_{ij}}{(\xi_i-z)^j}
\rme ^{\ks z}
\label{eq:mcRot_omssum}
\end{equation}
with 
\begin{equation}
  b_{ij} = \sum_{m=j}^{3} \frac{-a_{im}}{(\xi_i-\xi_3)^{m-j+1}}, i=1,2;
  \qquad b_{31} = -b_{11} -b_{21};
  \qquad b_{32} = b_{33} = 0.
\label{eq:bij}
\end{equation}
The integral in \eqref{eq:w_part_int_t} can now be expressed in terms of the scaled exponential integral function
\begin{equation}
\tilde \E_j\of{\mu} = \rme ^\mu \,\mu^{j-1}\int_{\mu}^{\infty} \!\!\dd \tau\,\frac{\rme ^{-\tau}}{\tau^j}
\label{eq:Ej}
\end{equation}
%
%\arev{
whose integration path is not allowed \arev{to} cross the negative real axis.
%}
We get
\begin{align}
 w\_{\tpart}\ott\of {z,t} &=
\sumpmpm \frac{ \rmi }{ 2 k } \frac{A_1\oo A_2\oo }{ k_x S} \frac{\ktimesk}{k_1 k_2}
 \sumij \sumpm \pms
\frac{b_{ij}}{\dxiz^{j-1} }
 \nonumber\\&\times
\wigbrac{
\tilde\E_j\sqbrac{\kpmst \dxiz}\rme ^{-\rmi k_x S z t}-
\tilde\E_j\sqbrac{\kpms \dxiz}\rme ^{-\rmi \ompmpm t} 
}
\rme ^{\ks z}.
\label{eq:w_part_evaluated}
\end{align}
with $\kpm = \ks\pm k$ and $\kpmt=\kpm-\rmi k_x S t$.
Here, the time--dependent term $\sim \exp(-\rmi \om t)$ 
is related to dispersion of the surface waves. 
Time dependency is here decoupled from depth
and easy to handle.
The other time--dependent term, $\sim\exp(-\rmi k_x S z t)$,
originates from the advected background vorticity, \ie, 
\arev{
the lower limit in the time integral \eqref{eq:w_part}.}
%our choice of integration constant $C$ in \eqref{eq:w_part_C}. 
With it comes the effect that the shear's advection of vorticity waves increases with depth, generating the depth--frequency coupling. % found in this term.

Finally, we seek to evaluate the surface elevation
\arev{from \eqref{eq:zeta_part}. Integrating dispersive (wave--induced) terms in time is trivial as these are not coupled with depth, whereas the advective terms are more involved.}
\arev{We proceed by once more inserting}
 the definition \eqref{eq:Ej} and perform partial integration on the outer integral. 
One of the partial integration kernels will then turn out as the 
an exponential integral of negative order, 
which can be evaluated explicitly, 
\[
	\tilde \E_{-j}\of\mu = \frac{\rme ^\mu}{\mu^{j+1}} \int_{\mu}^\infty \!\! \dd \tau\;\tau^j\rme ^{-\tau} 
	= \sum_{\m=0}^j \frac{j!}{\m!} \mu^{\m-j-1}, \qquad j=0,1,2\ldots
\]
Each term of the sum now generates new exponential integrals. 
We present the solution in terms of its wave dispersive and advective parts,
\begin{equation}
	\f\zeta_\bound\ot = \sumpm\br{\f\zeta\_{\bound,\wave,\pms}\ot + \f\zeta\_{\bound,\bg,\pms
	}\ot},
\label{eq:zeta_bound}
\end{equation}
in order to distinguish the physical behaviour of the two fundamentally different modes of vorticity transport. 
Explicitly,
\begin{align}
%\f\zeta\_{\bound,\wave}\ot
%= & 
%\sumpmpm \frac{-\tD_{\wave} %D\ott_{\wave,\pmpms}
%\rme ^{-\rmi \ompmpm t}}{(\om_+ - \ompmpm)(\om_- - \ompmpm)},
%\label{eq:zeta_bound_wave}\\
\f\zeta\_{\bound,\wave,\pm}\ot
= & 
\pm\frac{\tDdisp 
\rme^{-\rmi \ompmpm t}}{(\om_+ - \om_-)(\om_\pm - \ompmpm)},
\label{eq:zeta_bound_wave}
\end{align}
and 
\begin{align}
\f\zeta\_{\bound,\bg,\pm}\ot =&
	\mp \frac{1}{\om_+-\om_-} \frac1k \sumpmpm \frac{A_1 A_2}{k_x S}
	\frac{\ktimesk}{k_1 k_2} \sumij  \frac{b_{ij}}{\xi_i^{j-1} }
	\bigg[
	 \tilde\E_j\br{\kpt \xi_i}	 +\tilde\E_1\br{\kpt \xi_\pm}
	\nonumber\\&
	+\sum_{\m=0}^{j-1} \frac{(j-1)!}{\m!}
	\frac{\kpt \xi_i \tilde\E_{j-\m}\br{\kpt \xi_\pm} + \m \tilde\E_{j-\m+1}\br{\kpt \xi_\pm} 
		- \kpt \xi_\pm \tilde\E_j\br{\kpt \xi_i}}{[\kpt(\xi_\pm-\xi_i)]^{j-\m}}
	\bigg].
\label{eq:zeta_bound_background_pm}
\end{align}
These expressions, sequentially inserted into \eqref{eq:zeta_bound}, \eqref{eq:zeta_ICn_evaluated} and \eqref{eq:zeta_of_t}, give the full second order surface elevation.
%}
\arev{$\tDdisp$ and the remaining parts of the velocity field are presented in Appendix~\ref{sec:uvw}.}

We note that $\tilde\E_j(\mu)$ has a singularity at the origin and a branch cut discontinuity along the negative real axis, physically representing a critical layer. 
\arev{
A `rule for going around the singularity', which imposes a history to critical layers, is thus required for when the arguments of $\tilde E_j$ are real negative \citep{zakharov1990_nth_order,Benney_1960_O2_critical_layer_etc, lin1955_hydrodynamic_stability}. 
Physically, this treatment affect the presence of Reynolds stresses in the solution. 
}
%It is therefore not defined when their arguments encountered in our solution, such as $k_\pm \xi_i$ and $k_\pm(\xi_i-z)$, are real and negative.
We here require that we should not cross the branch cut as we go from $t=0$ to $t>0$.
This is 
%\srev{
essentially equivalent 
%}
to replacing $\ks$ with $\ks-\rmi k_x  S \epsilon$ where $\epsilon\rightarrow 0^+$\arev{, and to the rule of the singularity applied by \cite{zakharov1990_nth_order,shrira1993_shear_current_to_nth_order}}.
\arev{
(Viscous analysis, such as that provided by \cite{Benney_1960_O2_critical_layer_etc,Benney_1964_O2_BL_with_Lommel}, is required for a fully physical treatment of critical layers.}
%; causality then uniquely defines all $z$-integrals where inviscid theory is formally ambiguous. 
%Introducing viscosity, \cite{Benney_1960_O2_critical_layer_etc,Benney_1964_O2_BL_wit_Lommel} 
 %resolved the ambiguity for some particular wave interactions.
\cite{Craik_1968_resonance_critical_layer_Lommel,Craik_1971_resonant_BL} has shown that the viscous region of these critical layers give significant contributions in the energy transfer between resonant triads.)

\arev{
The final stage in obtaining a second order 
%for a three-dimensional problem 
solution is evaluating convolution integrals of the form \eqref{eq:Fourier_transform_nested}.
The discrete convolution domains considered in Sec.~\ref{sec:results:3D_ringwave} constitute, 
for the resolution presented,
%with the resolution required for the presented results,%
 large arrays of data.
}
In order to split the computation into parallel chunks and avoid the full 4-D array storage we 
substitute $\bk_1$ and $\bk_2$ with $\bk=\bk_1+\bk_2$ and $\bk'=\bk_1-\bk_2$.
 %rewrite the convolution integral \eqref{eq:Fourier_transform} to the form
\arev{
Written by way of Fourier integrals, 
}
\begin{equation}
%\p\psi\ot\of\bmr =
\arev{\invFourier_{\bk_1} \invFourier_{\bk_2}\f\psi}=
\int \! \frac{\dd\bk}{(2\pi)^2} \wigbrac{
\int\! \frac{\dd \bk'}{16\pi^2}\, \f\psi\ot\sqbrac{\tfrac12(\bk+\bk'),\tfrac12(\bk-\bk')}
}\rme ^{\rmi \bk\cdot\bmr}
\label{eq:convolution_int_O2_inner_form}
\end{equation}
and we evaluate the inner integral sequentially while also taking advantage of the symmetries
\begin{subequations}
\label{eq:psi_symmetry_prop}
\begin{align}
\Om_\pm\of{-\bk} &= -\Om_\mp\of{\bk} \label{eq:symmetry_prop_Omage}
\\
\psi_{\pm_1\pm_2}\of{\bk_1,\bk_2} &= \psi_{\pm_2\pm_1}\of{\bk_2,\bk_1}
= \big[\psi_{\mp_1\mp_2}\of{-\bk_1,-\bk_2}\big]^*. \label{eq:symmetry_prop_psi}
\end{align}
\end{subequations}
%
%to ease the computational load.
Sign subscripts here indicate the frequency branches of the two interacting waves and an asterisk denotes the complex conjugate.
\arev{
A real solution in physical space is assumed in \eqref{eq:symmetry_prop_psi}.
}
%These symmetries arise from the property $\Om_\pm\of{-\bk} = -\Om_\mp\of{\bk}$%
%\arev{
%and the facts that our solution is real in physical space with no uniform perturbation in the physical plane.
%}
  Bear in mind that all second order perturbation quantities are to be summed over the four frequency branch combinations.

%%%%%%%%%%%%%%%%%%% S E C T I O N %%%%%%%%%%%%%%%%%%%%%%	
%\subsection{Resonances and special limits}

%%%%%%%%%%%%%%%%%%% S E C T I O N %%%%%%%%%%%%%%%%%%%%%%	
\subsection{Dispersive resonance}
\label{sec:dispersive_resonance}

\arev{
It is worthwhile to identify and label the two main types of resonance possible in our system.
}

A pole appears from the denominator of \eqref{eq:zeta_bound_wave} if $\omsum = \om_+$ or $\omsum =\om_-$.
This denominator is the dispersion relation with $\omsum$ replacing $\om_\pm$ 
%\arev{and a pole thus signifies the state at which the propagation of the bound harmonic coincides with that of a free surface wave.
\arev{and a pole thus signifies the state wherein the bound harmonic and free surface waves travel at the same (phase) speed.
Energy exchange is then possible,} 
causing dispersion resonance \citep{Hasselmann_1962_mode_coupling_Taylordowntoz0}.
Note that the singularity due to the poles in \eqref{eq:zeta_bound_wave} is cancelled by the free wave \eqref{eq:zeta_ICn_evaluated:IVP} in the full  solution \eqref{eq:zeta_of_t}
 \arev{ resulting in linear growth in the manner \eqref{eq:limit},}%
%, in a similar manner to Eq.~\eqref{eq:limit}, resulting in amplitudes growing linearly with time. 
%Here the expression converges smoothly towards a resonant point of linear wave growth:
\begin{equation}
\lim_{\omsum\rightarrow\om_\pm}\f \zeta_{\wave}\ot\of t
=
\pm \frac{\tDdisp
}{\om_+ -\om_-}
\br{\frac{\rme ^{-\rmi \om_+ t}-\rme ^{-\rmi \om_- t}}{\om_+ -\om_-}
+\rmi t \,\rme ^{-\rmi \om_\pm t}}.
\label{eq:zeta_n_with_IC_limit}
\end{equation}
%
%where the limit is approached in $k$-space.
%The resonance is manifest in the last term, which grows linearly with time.
%
\arev{
Dispersive resonance occurs by way of self-interactions at odd orders in monochromatic wave trains, manifests as frequency perturbations \citep{Fenton_fifth_order_Stokes_waves}.
}
\cite{Phillips_1960_mode_coupling} and \cite{Lounguet_mode_coupling_freq_perturbation} show that similar frequency perturbations result due to resonance between separate wave trains, and 
\cite{Benney_1962_mode_coupling_frequency_perturbation} showed it in the case of discrete wave spectra.
%\\

\begin{figure}%
		\subfigure[$++$ branch combination, $S=-3.0\sqrt{g k_1}$, angle $\bk_1$: $\pi/2$.]{\includegraphics[width=.49\columnwidth]{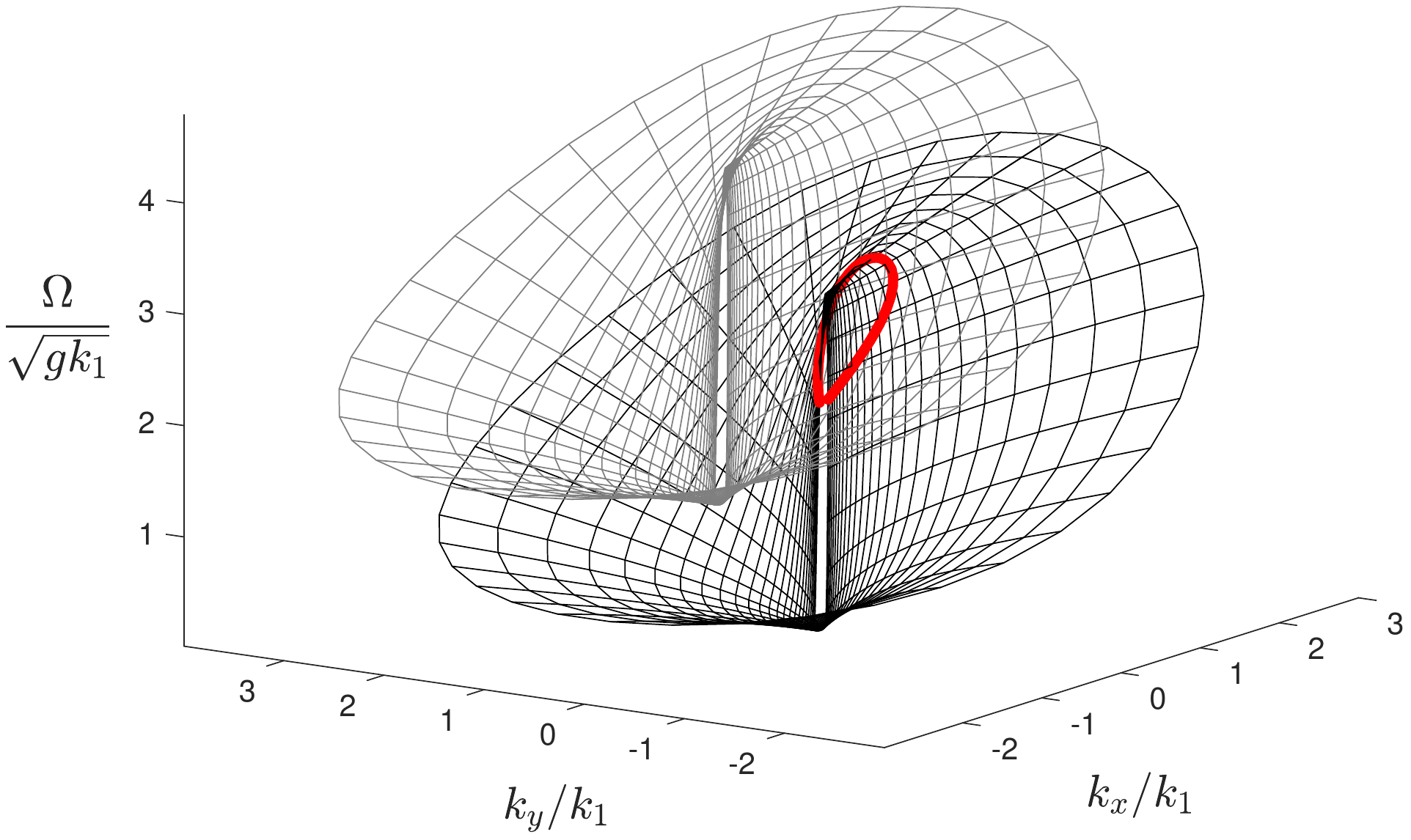} \label{fig:omega_resonance:pp} }
		\hfill
				\subfigure[$-+$ branch combination, $S=-3.75\sqrt{g k_1}$, angle $\bk_1$: $\pi$.]{\includegraphics[width=.49\columnwidth]{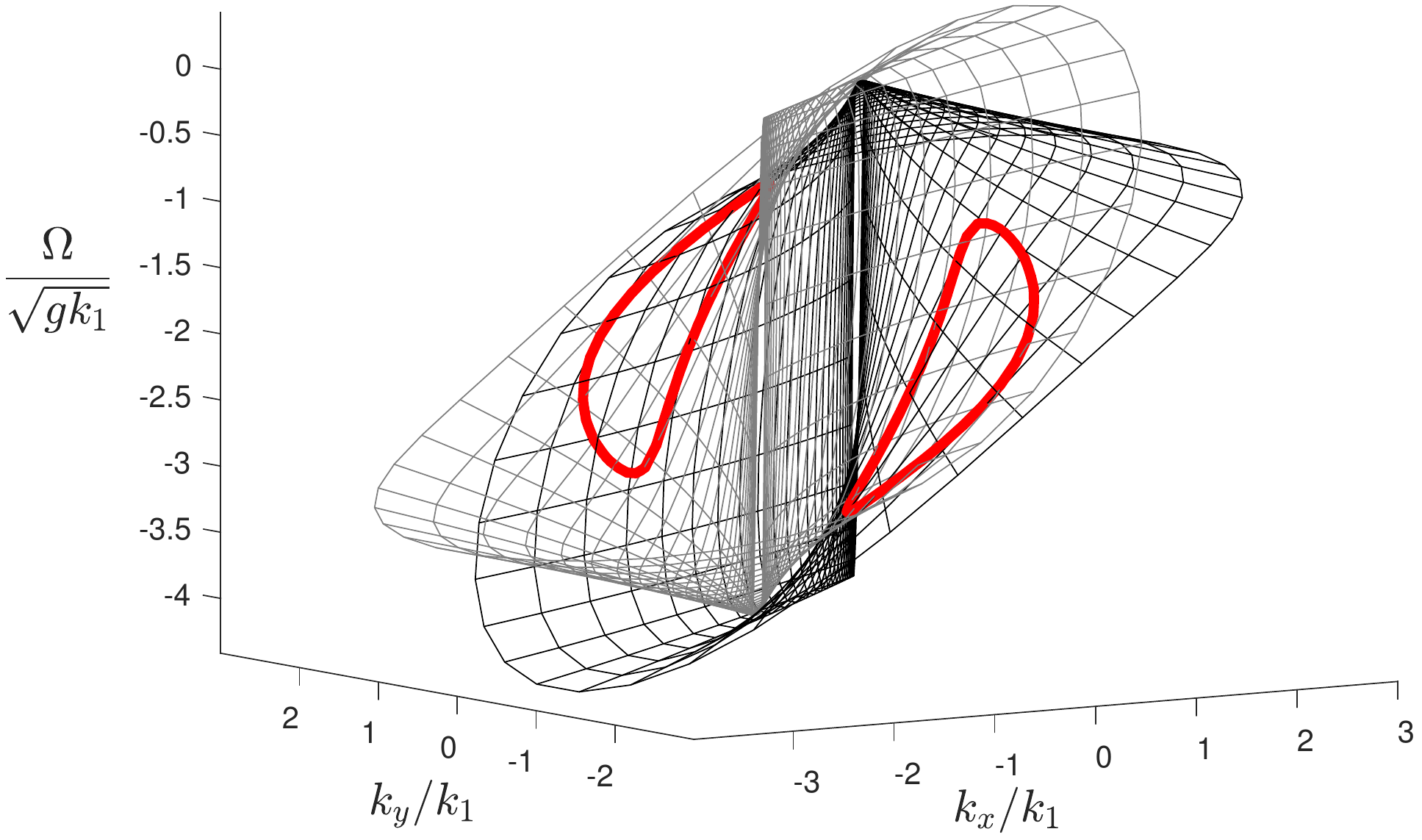} \label{fig:omega_resonance:mp}}
  \caption{
	\arev{
	Two dispersion relation curve branches in $\bk$-space are shown in each image, the origin of one surface shifted to a point $\bk_1$ atop the other.
	The shifted surface thus shows $\omsum=\Om_{\pm}\of{\bk_1}+\Om_+\of{\bk}$ for $\bk_1$ fixed.
	Triad resonance occurs if, for any  
	angle of $\bk_1$, there	exists an intersection cut between the two surfaces;
	%(possible in the strongly sheared right image but not in the left one); 
	along such a cut $\bk_{3}$ there exists pairs $\bk_1,\bk_2$ satisfying $\Om_\pm\of{\bk_{3}}=\Om_\pm\of{\bk_1}+\Om_+\of{\bk_2}$
	}
	%$\Om_+\of\bk$ and $\omsum=\Om_+\of{\bk_1}+\Om_+\of\bk$ at fixed $\bk_1$. Resonance occurs for values of $\bk_2$ where the two surfaces intersect. 
	}%
  \label{fig:omega_resonance}%
\end{figure}

\arev{
Gravity wave dispersion resonance is possible only at third order and above in irrotational flows \citep{Phillips_1960_mode_coupling,Kadomtsev_1971_mode_coupling_resonance}, yet it is important to remember that dispersion resonance  is possible at second order if the flow is three-dimensional and strongly sheared. 
}
\cite{Craik_1968_resonance_critical_layer_Lommel} found such resonance to be remarkably powerful.
\arev{To illustrate sheared triad resonance we visualise in Fig.~\ref{fig:omega_resonance} $\Om$-surfaces in $\bk$-space; }
%Fig.~\ref{fig:omega_resonance} shows $\omsum$ surfaces in $\bk$-space to illustrate the possibility of resonant triads in strongly sheared flow;
surface intersection, and thus resonance, is possible if the shear-induced curve stretching disrupts the otherwise monotone curvature sufficiently. 
\arev{
The shear strength (shear Froude number), the branch combination and the angle of the fixed wave vector $\bk_1$ are the only parameters in these figures.
 %with the presented 
%%nondimensionalization.   
%normalization.
%One can search for instability by looking for 
A simple study of graph topography reveals that resonance for the weakest shear strength 
%(irrespective of sign) 
 takes place as an
interaction of two `plus' or two `minus' branches (Fig~\ref{fig:omega_resonance:pp}), at $|S|= 2\sqrt{g k_1}$.
This minimal shear triad is 
%Resonant triads are seen to then be 
between two waves which are orthogonal to the shear and a third wave of large wavelength and high phase speed.%
\footnote{
\arev{
Note that such minima are sensitive to the scaling; 
\cite{Craik_1968_resonance_critical_layer_Lommel}, studying a special case, finds  a minimum to occur at an intermediate angle as he scales with $\sqrt{g k_{1x}}$ as opposed to $\sqrt{g k_{1}}$.
}
} 
%Resonance is then seen to be between waves which are orthogonal to the shear direction and which resonance waves of large wavelengths and high speeds.
Opposing branch resonance  (Fig~\ref{fig:omega_resonance:mp}) first occurs at
 %$|S|\approx 3.61 \sqrt{g k_1}$
 $|S|\approx 3.47 \sqrt{g k_1}$
and involves a triad where $\bk_1$ 
is near an angle $\pi/6$ to the current.
Resonance at $\bk_1$-angels lower than this appears with only slight increased  shear strength;
%and the resulting resonant wave ...
%%is parallel to the current.
%%, then involving waves which are parallel to the current.
%The resulting resonant wave is then oriented orthogonally to the current with roughly half the wavelength.
Fig~\ref{fig:omega_resonance:mp} shows a case of slightly stronger shear with $\bk_1$ parallel to the shear current.  
}

%%%%%%%%%%%%%%%%%%% S E C T I O N %%%%%%%%%%%%%%%%%%%%%%	
\subsection{Advective resonance}
\label{sec:advective_resonance}
%\scomm{New edits hence.}

A different type of resonance is possible from within the flow field.
Where the dispersion resonance originates from a wave-wave interaction at the surface, advective resonance comes about when energy is transferred to the wave field from the background shear current $U(z)$.
\arev{ %dvective resonance
It occurs where the phase velocity of a wave matches the shear advection velocity in the $x$-direction.}
%The phase velocity in the $x$-direction of a particular harmonic must then match the background shear advection velocity. % for this to happen.
This mode of resonance is seen 
%in expressions such as \eqref{eq:w_part_int_t} where 
\arev{repeatedly in the velocity field expressions, e.g. \eqref{eq:w_part_int_t}, 
in the form of factors such as Eq.~\eqref{eq:limit} resulting in terms diverging linearly with time.
}
%\[
  %\frac{\rme ^{\rmi \oms t}-1}{\rmi  \oms} \buildrel{\oms\to 0}\over{\longrightarrow} t.
%\]
%
It is further
auspicious  
to distinguish between two types of advective resonant behaviour. One appears 
in the form of 
critical layers at all depths $z=\xi_i$ 
\arev{
where $\om_i=k_{i,x} S z$. 
Vertical velocities remain continuous across the critical layers (due to the $\sinh$ in \eqref{eq:w_part_int_t}), 
although the vertical velocity pertaining to the disperive wave seen alone has a kink (discontinuous $z$-derivative) here.
The advective wave serves to cancel this kink initially so that the net vertical velocity is smooth across the critical layer but evolves into a kink as $t\rightarrow\infty$.
Horizontal velocities, presented in Appendix~\ref{sec:uvw}, will grow  like $\bu\_h\sim t\, \rme ^{-\rmi \omsum t}$ in a diverging manner near the critical layer and evolve towards a discontinuous singularity as $t\rightarrow\infty$.
}
The number of critical layers increases with the solution order. 
Just above or below a critical layer the flow remains periodic and bounded.

Another type of advective resonance occurs
 %if $\oms$ vanishes because
when $k_x$ and $\omsum$ each tends to zero independently. 
\arev{The resonance $\oms\to0$ then occurs globally.}
%Eq.~\eqref{eq:w_part} or the 
%$\oms\to0$ 
%limit of \eqref{eq:w_part_int_t} then shows that $w\sim t $ everywhere.
This type of resonance has been 
famously proposed 
to act as a key mechanism for the phenomenon of Langmuir circulation, 
in turn 
a vital  
mechanism for boundary layer mixing and the location and structure of the thermocline in oceans and lakes~\citep{leibovich83, craik_1986_book_wave_interactions}. 
We will return to this phenomenon later 
%when considering Langmuir circulation as a special case 
in Sec.~\ref{sec:Craik_case}.
\arev{
In the confines of the initial value problem 
%\acommm{`--Poisson'; I meant Cauchy problem in the sense of a Cauchy initial value problem. Isn't Cauchy--Poisson problems about ringwaves specifically?} 
 there is, for moderate times,  no ambiguity  
near a global advective resonance
-- in fact, this was the main motivation for demanding irrotational initial condition.
There is
%, in the Cauchy problem,
 therefore no need for excluding the parts of the interaction spectrum where $k_x=k_{1x}+k_{2x}$ is small, as done by \citet{zakharov1990_nth_order}. % with a cut-off wave vector. 
}

%\scomm{Why are the exponential integrals interesting here? Probably some reason i don't see.}
%\acomm{Believe I wanted to make the point that a) solutions becomes a lot simpler in 2D, and/or b) oblique interactions is an integral reason for the appearance of exponential integrals.}
Note that all the terms associated with nonlinear advective resonance 
 contain the factor $\ktimesk$, and so are artefacts of oblique wave interactions. 
\arev{
The same is true for the appearance of all exponential integrals.
%, whose computational load dominates.
}
\\

\arev{
Our scope is restricted to currents of linear depth dependence, hence we have excluded the possibility of a final type of resonance, which is due to the curvature of $U(z)$ \citep{drazinReid2004}.
Such resonances are possible in free surface flows even when there are no inflection points.
%Finally we mention resonance due to curvature on the shear current, possible in free surface flows even when there are no infliction points.
%Such resonances are excluded form the present study as we have restricted our scope to linear currents at infinite depths.
One may refer to 
\citet{shrira1993_shear_current_to_nth_order} who derived a perturbation series solution for linearised wave fields atop arbitrary, weak shear currents to any prescribed accuracy. 
%One may refer to 
%\citet{shrira1993_shear_current_to_nth_order}, whose perturbation technique can evaluate wave fields atop arbitrary, weak shear currents 
%%in linear wave fields 
%to any prescribed accuracy. 
Instabilities related to critical layers are found.
\cite{Drivas_2016_resonance_gravity_and_vorticity_waves} 
provide,
with their
study of a bilinear shear current of finite penetration depth,
 a related example of three-dimensional triad instability which is related to the current profile `kink'.
 %Here, the 'kink' of the shear current $U(z)$ can give rise to resonant triads.
\cite{carpenter2011} provide a review for two-dimensional   homogeneous and density stratified shear flows.
}

%%%%%%%%%%%%%%%%%%%%%%%%%%%%%%%%%%%%%%%%%%%%%%%%%%%%%%%%	
%%%%%%%%%%%%%%%%%%% S E C T I O N %%%%%%%%%%%%%%%%%%%%%%	
%%%%%%%%%%%%%%%%%%%%%%%%%%%%%%%%%%%%%%%%%%%%%%%%%%%%%%%%	

\section{Fluid particle trajectories}
\label{sec:particle_trajectories}

\renewcommand{\n}{}
\renewcommand{\ot}{^{(2)}}

As is well known, the second-order Stokes expansion of a steady periodic wave creates a net mass transport in the direction of propagation, referred to as Stokes drift. From a Lagrangian perspective, the trajectories of individual fluid particles are not closed but is slightly shifted for each cycle. Naturally a similar phenomenon will be present in the presence of a transient wave such as may be created by an initial disturbance.

Surprisingly to us,
\arev{
literature on fluid particle trajectory and Stokes drift in the presence of a uniform shear current have been found to be scarce;
\citet{kishida1988stokes_uniform_shear_O3} provide expressions for monochromatic two dimensional flow, but we are not aware of any literature for monochromatic waves propagating at oblique angles  with a sheared current,
nor of any reliable expressions for particle trajectories in sheared flow.
}
The following theory has therefore been kept rather general, allowing for both  monochromatic waves and discrete and continuous wave spectra.

We define a fluid particle trajectory $\p \bx\_p\of t=(\p x\_p\of t, \p y\_p\of t, \p z\_p\of t)$ as the parametric position of an imaginary particle whose velocity always coincides with the velocity filed at the immediate trajectory position. Precisely,
\begin{equation}
\ddiff{\p \bx\_p}{t} = \p \bu\of{\p \bx\_p\of t, t}
	\label{eq:particle_traj_def}
\end{equation}
where the initial particle position is close to a point $\bx_0 = ( x_0, y_0, z_0)$ at $t=0$.
\arev{
Assume that $\bx\_p$ revolves around an orbit centre point  $\p\bx\orb$
which
%Let $\p\bx\orb$ the centre of a moving orbit which slides 
slides in the horizontal plane with a constant velocity $\pbuo$.
}
\arev{
Taylor expanding $\p \bu$ about $\p\bx\orb$}
 generates, in index notation,
\begin{equation}
	 \p u_i\of{\p \bx\_p\of t, t} 
	\approx (\p u_{i})\orb 
	+ (\p\bx\_p - \p\bx\orb) (\p\nabla \p u_{i})\orb ^T
	+ \tfrac12 (\p\bx\_p - \p\bx\orb)	(\p\nabla \p\nabla\p u_{i})\orb	(\p\bx\_p-\p\bx\orb)^T 
	+\dots
	\label{eq:velocity_Taylor}
\end{equation}
%
%where the right-hand side is evaluated at $\bx=\p\bx\orb$.
where $\orbsymbol$-suffixes indicate evaluation at $\bx=\p\bx\orb$.
Inserting \eqref{eq:velocity_Taylor} into \eqref{eq:particle_traj_def} makes $\p \bx\_p$ too a Stokes series expansion with increasing 
number of nested convolutions
%convolution layers 
in the form of \eqref{eq:Fourier_transform_nested}.
We use the sliding orbit $\p\bx\orb$ to remove 
linear time dependencies in $\p\bx\_p$
%non-periodic functions in time
(from current advection and Stokes drift) from the right-hand side of \eqref{eq:velocity_Taylor}, 
%so that higher order polynomials in $t$ does not accumulate in the subsequent orders, leaving only periodic and linear terms.
denying $\p\bx\_p$ higher polynomials in time in the subsequent orders.
%allowing $\p\bx\_p$ to obtain terms which are only periodic or linear in time. 
Consequently, also $\p\bx\orb$ takes the form of a Stokes expansion.
The procedure is analogous to the well-known frequency perturbation of the Poincar{\'e}--Lindstedt method.
At zeroth order \eqref{eq:velocity_Taylor} then yields 
%This rule applies at zeroth order so that $\bx\_p\on=\f\bx\orb\on$ and \eqref{eq:velocity_Taylor} yields 
%
\begin{equation}
\f\bx\_p\on\of t= \f\bx\orb\on\of t = \bx_0+S z_0  \bm e_x t,
%\label{eq:}
\end{equation}
as expected.

Evaluation of the perturbation velocity field along the sliding orbit $\p\bx\orb$ is equivalent to a Galilean transformation and generates a Doppler shift
\begin{equation}
\p\bu\n\of{\p\bx\orb\of t,t}=\p\bu\orb\n\of t
\rightarrow \f\bu\orb\n\of t = \f\bu\n\of t \rme^{\rmi \bk\cdot\pbuo t}
%\label{eq:}
\end{equation}
in Fourier space.
$\bx_0$ is here the independent variable of the corresponding inverse Fourier transform.
A cascade of ordinary differential equations, 
\arev{
\begin{align}
	%\br{\ddiff{\f x\ppi}{t}}\n
	\ddiff{\f x\ppi}{t}\n
	- S \f z\_p \n \delta_{i,1} 
	&=	 \f u\iorb\n 
	\nonumber\\
	&+ 
	\Big[	
	\rmi  \omorb \f x\ppi +
	(\f\bx\_p - \f\bx\orb)	(\f\nabla \f u\iorb)^T
	+ \frac12 (\f\bx\_p -\f\bx\orb) (\f\nabla \f\nabla\f u\iorb )(\f\bx\_p-\f\bx\orb)^T+ %\ldots
	\Big]\nn ,
	\label{eq:part_pos_Fourier}
\end{align}
}%
results for the subsequent orders.
\arev{(Only interaction terms of the appropriate combined order enter among the inhomogeneous interaction terms in the square brackets on the right-hand side.)}
 %$n\geq1$.
%As before, ${(<\!n)}$ indicates that we in each term sum all variable combinations whose orders are less than $n$ and whose summed order equal $n$.
The 
\arev{first bracket term,}
%second right-hand term, 
where 
\begin{equation}
\omorb=\om-\bk\cdot\pbuo,
%\label{eq:}
\end{equation}
originates from incorporating the Stokes drift into particle orbit position and becomes active at third order. 

Non-zero contributions to the sliding orbit velocity $\buo$ are found at even orders where the time dependency in self-interacting waves cancel.
%$\f\bu\orb$ gets non-zero contributions from self-interacting waves at even orders where some right-hand terms in \eqref{eq:part_pos_Fourier} become time independent.
Physically, this signifies the phenomenon of Stokes drift. 
Such contributions appear as finite terms when working with monochromatic waves or discrete wave spectra.
In a continuous wave spectrum $\pbuo\n \rightarrow 0$ at all non-zero orders. 
Instead, Stokes drift then manifests as poles whose limits generate linear time dependency in the manner of \eqref{eq:limit}.
%These in turn affect $\bx\_p$ at higher odd orders 
%and are analogue to the well-known frequency perturbation of the Poincar{\'e}--Lindstedt method as our inverse Fourier transformation is to be evaluated about $\bx=\p\bx\orb$.
\\

To first order, \eqref{eq:part_pos_Fourier} yields
\begin{equation}
	\f \bx\_p\oo = 
	\frac{\rmi}{\omorb\on}\br{
	\f \bu\orb\oo+ \frac{\rmi}{\omorb\on}S\f w\orb\oo \bm e_x 
	}
	\label{eq:part_pos_O1}
\end{equation}
with $\omorb\on =  \om - k_x S z_0$. 
Summation over the two frequency branches \arev{of the first order wave field} is implied.
$\f \bx\_p\oo$ contains only terms which are periodic in time and so $\buo\oo=\bm 0$.

A simple study of \eqref{eq:part_pos_O1} reveals that the second term serves to make the first order trajectories elliptical as a result of the linearly differing current advection above and below the orbit centre.
We remark that this simple mechanism is not present in the solution presented by \cite{Hsu_2013_particle_traj_with_shear}%
. % which brings us to question that work.
Likewise, Stokes drift does not appear to be present in \citet{Umeyama_proc_with_traj_expressions}. 
%\deleted{and subsequent publications.}
%\acomm{Would be good if you'd check whether you agree about these papers. I find the details hard to follow.}
We further note that our particle trajectory approximation contains critical layers, even in the 2D case where the velocity field itself is overall smooth.
This is because particle oscillation ceases if the advection causes the wave crest to remain stationary relative to the particle orbit. 
Elliptical shear stretching thus increases near critical layers and our trajectory approximation ceases to be valid if the trajectory orbit and the critical layer are too close.

The spectral expressions for the second order trajectory are similar to \eqref{eq:part_pos_O1} in appearance but 
decidedly bulkier, and we do not quote them explicitly.
Instead, because they seem 
absent in present literature, we quote to second order our result for velocity field and fluid particle position in a monochromatic wave propagating at an oblique angle to a current of uniform vorticity at finite depth.
This has been calculated with the above procedure for validation purposes. 
\arev{Introducing dimensionless parameters
\begin{align}
\bK &= \frac{\bk}{k};
&
\Ru &= \sigu \frac{k_x}{k}\frac{S}{\oms};
&
\R &= \sig \frac{k_x}{k}\frac{S}{\om\orb\on};
&
\A\n&=\frac{k^2 A\n}{\om\orb\on} %,\;n=1,2.
\end{align}
and the orthogonal wave vector $\bk\bb = (-k_y,k_x)^T$,
}
the velocity field of a monochromatic wave with wave vector $\bk$ and frequency $\om$ reads
\begin{subequations}
\begin{align}
\p\bu\_h&=
S z \bm e_x + \frac{\epsilon A\oo}{\sqrt{1-\sigu^2}}\br{\bK + \bK\bb \frac{k_y}{k_x}\Ru} \cos(\bmr\cdot \bk-\om t)
\nonumber \\
& + \frac{\epsilon^2 }{1-\sigu^2}
\wigbrac{
 A\ot\sqbrac{ \bK(1+\sigu^2) + \bK\bb \frac{k_y}{k_x}\Ru } 
-\frac{\bK\bb}{4} \frac{k }{\oms} \br{A\oo}^2 \frac{k_y}{k_x}\Ru^2
}\cos2(\bmrdk-\om t)
\\
\p w&= 
\epsilon A\oo\frac{\sigu}{\sqrt{1-\sigu^2}}
\sin(\bk\!\cdot \!\bmr-\om t)
%\nonumber \\&
+\epsilon^2 A\ot\frac{2\sigu}{1-\sigu^2}\sin 2(\bk\!\cdot\!\bmr-\om t)
\end{align}
\label{eq:part_vel_monochromatic}
\end{subequations}
to second order, with
\begin{align}
A\oo &= \frac{\om}{k} \frac{\sqrt{1-\sigu^2}}{\sigu}\bigg|_{z=0},
&
A\ot &= \frac{\om}{k} \frac{1-\sigu^2}{4\sigu^4}
\sqbrac{3\br{1-\sigu^2+\Ru}+\Ru^2}\bigg|_{z=0},
\label{eq:A_monochromatic}
\end{align}
which agrees with the field quoted by \cite{Hsu_2D_O3_finite_depth_sol} in 2D.
This generates the fluid particle trajectories
\begin{subequations}
\begin{align}
\p{\bm r}\_p &=
\bmr_0+S z_0  \bm e_x t
-\frac{\epsilon \A\oo}{k} \frac{\bK }{\sqrt{1-\sig^2}}  (1+\R) \sin(\bk\!\cdot\!\bmr_0-\om\orb t)
\nonumber \\
&-\frac{\epsilon^2}{2k} \frac{\bK}{1-\sig^2} \sqbrac{
\A\ot(1+\sig^2+\R) - \frac12 \br{\A\oo}^2\br{1-\sig^2+\R+\frac{\R^2}2}
}\sin2(\bk\!\cdot\!\bmr_0-\om\orb t)
\nonumber \\
&+\frac{\epsilon^2}{2k}\frac{1}{1-\sig^2} \br{\A\oo}^2
\sqbrac{\bK (1+\sig^2+\R)+ 2\bK\bb \frac{k_y}{k_x} \R(1+\R)  } \om\orb\on t
\label{eq:part_pos_monochromatic:r}
\\
\p z\_p &=
z_0 +
\frac{\epsilon\A\oo}{k}\frac{\sig}{\sqrt{1-\sig^2}} \cos(\bk\!\cdot\!\bmr_0 -\om\orb t)
\nonumber \\
&+\frac{\epsilon^2}{k}\frac{1}{1-\sig^2}\sqbrac{\A\ot-\frac14 \br{\A\oo}^2\R}\cos 2(\bk\!\cdot\!\bmr_0-\om\orb t)
\label{eq:part_pos_monochromatic:z}
\end{align}
\label{eq:part_pos_monochromatic}%
\end{subequations}%
to second order. 
$\sigu=\tanh k(z+d)$ and $\sig=\tanh k(z_0+d)$ are here the bathymetry parameter for a depth $d$; at infinite depth we first insert \eqref{eq:A_monochromatic} and then take the limit $\sigu,\sig \rightarrow 1$, noting that $(1-\sigu^2|_{z=0})/(1-\sig^2) \rightarrow \exp 2 k z_0$.
The last expression in \eqref{eq:part_pos_monochromatic:r} adheres to the Stokes drift and is non-periodic. 
Following the described procedure we put $\bx\orb\ot$ equal to this expression when proceeding to third order, removing the quadratic time dependency otherwise appearing there. 
Again, our Stokes drift expression contains several shear-dependent terms not reported in \citep{Hsu_2013_particle_traj_with_shear}.
It has been verified numerically that the code used to compute the results of Sec.~\ref{sec:results:particle_trajectory} reproduces the above result at infinite depth 
when perturbing the bound frequency with an imaginary component $\rmi\varepsilon$ in the manner
\begin{equation}
\p\bmr\_p\ot\sim
	\lim_{\varepsilon\rightarrow0^+} 
 \br{\frac{\rme ^{-\rmi (\omorb-\rmi\,\varepsilon) t}}{\omorb -\rmi\, \varepsilon}
+\frac{\rmi  \varepsilon}{(\omorb-\rmi\,\varepsilon)^2}
}
		\xrightarrow{\omorb=0}
		\lim_{\varepsilon\rightarrow0^+} \frac{ \rme ^{-\varepsilon t}-1}{  \rmi \varepsilon}
		= -\rmi t.
\label{eq:Stokes_drift_frequency_perturbation}
\end{equation}
\arev{The term which is time independent 
is a sagaciously chosen integration constant which appears 
%we have added as an integration constant 
when evaluating \eqref{eq:part_pos_Fourier}; it appropriately balances the singularity while vanishing away from $\omorb=0$.}
\\

%\acomm{Added the following as an afterthought. Is it trivial? Also, check!}\\
Finally, we remark that a uniform current can easily be incorporated into these results by linearly shifting the frame of reference in the manner
\begin{subequations}
\begin{align}
 \p{\bm U}\of{\bx,t} &= \p\bu\of{\bx-\bm U_0 t,t} + \bm U_0,
\\
	\p{\bm X}\_p\of{t} &= \p\bx\_p\of{t} + \bm U_0t.
\end{align}
\end{subequations}
$\p{\bm U}$ and $\p{\bm X}\_p$ are now the velocity field and fluid particle trajectory in the presence of a uniform current with velocity $\bm U_0$, respectively;
choosing $\bm U_0=S d$ in \eqref{eq:part_vel_monochromatic}--\eqref{eq:part_pos_monochromatic} implies  zero current velocity at the bottom surface.
%}
%\\\acomm{It appears to be a whole lot of analyses, eg.\ Chen et.\ al.\ 2012, Zaman and Baddour 2010 etc., who write new papers augmenting the research `in the presence of a uniform current'.
%Their results (usually!) amounts to the above transformation.
 %}
The effect of uniform currents on the particle trajectory was investigated by \cite{Constantin_2001_Pressure_and_trajectories}.

%%%%%%%%%%%%%%%%%%%%%%%%%%%%%%%%%%%%%%%%%%%%%%%%%%%%%%%%	
%%%%%%%%%%%%%%%%%%% S E C T I O N %%%%%%%%%%%%%%%%%%%%%%	
%%%%%%%%%%%%%%%%%%%%%%%%%%%%%%%%%%%%%%%%%%%%%%%%%%%%%%%%	
\section{Numerical examples}
\label{sec:results}

%%%%%%%%%%%%%%%%%%% S E C T I O N %%%%%%%%%%%%%%%%%%%%%%	
\subsection{
Generalized Langmuir vortices from obliquely incident wave trains
}
\label{sec:Craik_case}

The resonance of \eqref{eq:w_part_int_t} as $\omsum=k_x=0$ was first proposed as a mechanism for Langmuir-type vortices in a model by \cite{Benney_1960_first_Langmuirish},  later investigated numerically by \cite{Antar_Collins_1975_numerical_viscous_Langmuir}. 
The presence of these large rollers can sometimes be observed in the form of parallel `windrows' forming in the wind direction on oceans and lakes due to the gathering of seaweed and flotsam in the downwelling regions they create.

\begin{figure}
  \centering\includegraphics[width=.75\columnwidth]{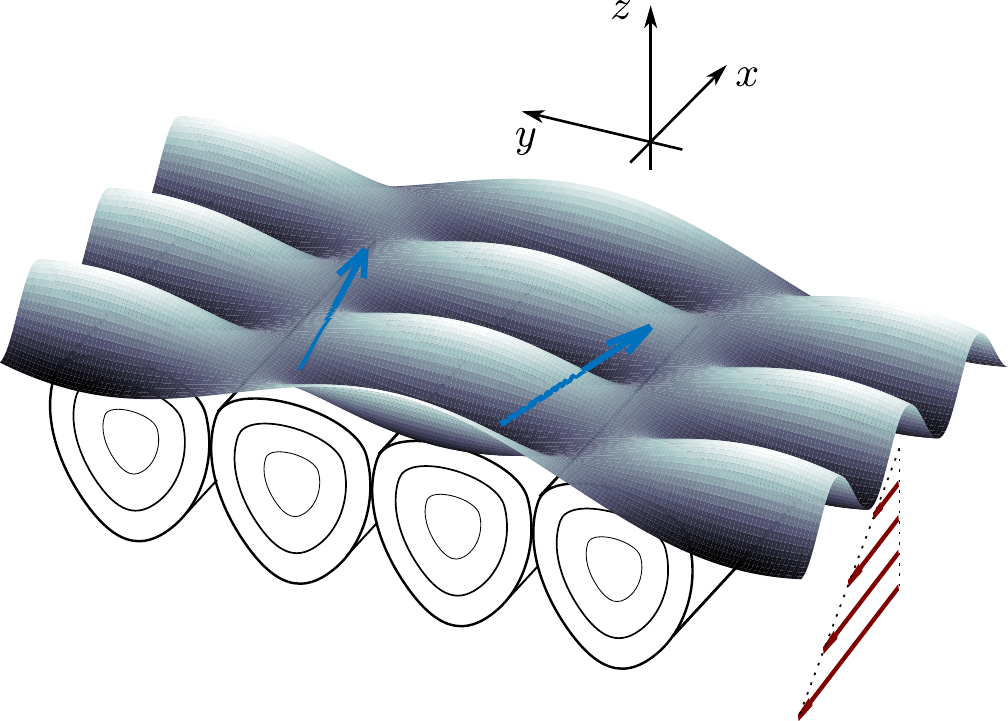}%
  \caption{
  Geometry of the classic set-up originally considered by \cite{Craik_1970_Langmuir_myidea}: Two monochroatic waves (wave--vectors illustrated with arrows) propagate symmetrically about the background shear flow. Langmuir vortices are created beneath the surface.  We begin by revisiting the set--up considered by \cite{Craik_1970_Langmuir_myidea} as a verification and benchmark, before considering more general, skewed configurations.
  }%
  \label{fig:Langmuir_setup}%
\end{figure}

Presently we consider an
 inviscid model proposed by \cite{Craik_1970_Langmuir_myidea} in which we consider a pair of plane waves 
whose 
wave vectors
 are oblique to, and symmetrical about, the 
$x$-axis;
a train moving in the direction $\bk_{\mr{L}+}=(k_{\mr{L}x},k_{\mr{L}y})^T$ and another in the direction $\bk_{\mr{L}-}=(k_{\mr{L}x},-k_{\mr{L}y})^T$.
%\srev{
The set-up is sketched in Fig.~\ref{fig:Langmuir_setup}. Remember that we work in a frame of reference moving with the water surface. In the lab frame this would most likely represent waves propagating \emph{along with} the surface flow, e.g.\ if the model should mimic a surface current created by the wind. 
%}

Such a case is represented with the first order `spectrum'
\begin{equation}
  \f\zeta_\pm\oo
  (\bk)
  = 2 \h \pi^2 \sumpm\delta\of{\bk \mp \bk_{L\pms}},
%\label{eq:}
\end{equation}
$\delta$ being the Dirac delta function.
The second order (bound) waves are then of the form
\begin{align}
\p\zeta\ot 
&= \sumpmpmnew \int \frac{\dd \bk_1 \dd \bk_2}{(2\pi)^4}
\f\zeta_{\pms_1}\oo\of{\bk_1} \f\zeta_{\pms_2}\oo\of{\bk_2} \big[\cdots\big] 
\rme ^{\rmi (\bk\cdot\bmr-\ompmpm t)}
\nonumber\\&=
\frac{\h^2}{4} \!\!\sum_{\pms_{1},\pms_{2},\pms'_{1},\pms'_{2}=\pm}\!\! \big[\cdots\big] \, 
\rme ^{\rmi (\bk\cdot\bmr-\ompmpm t)}\bigg|_{\scriptsize
\begin{array}{l}
	\bk_1 =\pms_1 \bk_{\mr{L}\pms_1'}
	\\
	\bk_2 =\pms_2 \bk_{\mr{L}\pms_2'}
\end{array}
}.
\end{align}
The sum runs over sixteen sign combinations.
Half of these are duplicates where the wave indices are swapped, leaving eight distinct kernels. 
These in turn form pairs of complex conjugates, leaving four physically distinct 
types of two--wave interactions as listed in Table \ref{tab:Craik_cases}.

\begin{table}
\begin{center}
\begin{tabular}{clcl}
Type & Distinction &Signs& Description\\
\\[-2ex]
A	&
$\bk_2 =  \bk_1$ &
${\begin{array}{c}	\pm_1=\pm_2\\\pm'_1=\pm'_2\end{array}}$ & 
\parbox{5cm}{Self interaction;\\ a 1D/2D phenomenon.}
\\[3ex]
B & 
$\bk_2 = -\bk_1$ &
${\begin{array}{c}	\pm_1=\mp_2\\\pm'_1=\pm'_2\end{array}}$ & 
\parbox{5cm}{Self cancellation;\\ a constant (zero) contribution. } 
\\[3ex]
C & 
$\bk_2 = (k_{1x},-k_{1y})^T$ &
${\begin{array}{c}	\pm_1=\pm_2\\\pm'_1=\mp'_2\end{array}}$ & 
\parbox{5cm}{Oblique wave interaction;\\ oscillatory and uniform in $y$.} 
\\[3ex]
D & $\bk_2 = (-k_{1x},k_{1y})^T$ &
${\begin{array}{c}	\pm_1=\mp_2\\\pm'_1=\mp'_2\end{array}}$ & 
\parbox{5cm}{Oblique wave interaction;\\ non-oscillatory and uniform in $x$.} 
\end{tabular}
\end{center}
\caption{Four distinct types of interactions between two waves $\bk_1, \bk_2$ of equal wavelength.
%\scomm{JFM has a very particular style of tables. A bit strange.}
}
\label{tab:Craik_cases}
\end{table}

\begin{figure}
\centering
\includegraphics[width=.9\columnwidth]{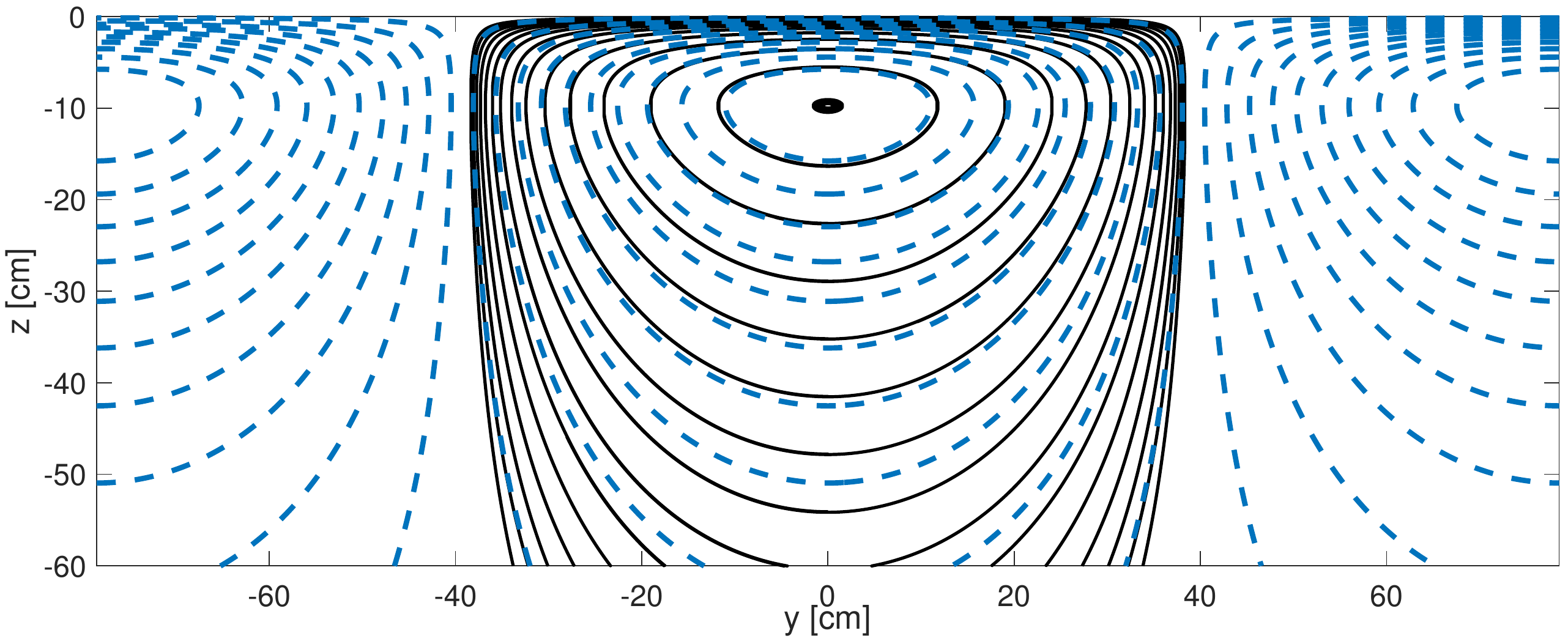}%
\caption{Langmuir vortex cells (case D). 
Blue, stippled: streamlines from stream function  \eqref{eq:Craik_stream_function} from \cite{Craik_1970_Langmuir_myidea}. 
Black, solid: streamlines numerically computed from velocity field presented in Sec.~\ref{sec:special_cases}. Flow state in given reference.
}%
\label{fig:Langmuir}%
\end{figure}

Types A and B do not entail any three-dimensional interaction and the oblique interaction term $w\_{\tpart}\ott$, given in \eqref{eq:w_part_evaluated}, disappears from \eqref{eq:w_sol} as 
$\ktimesk\equiv 0$ in these cases.
Type C is a wave interaction propagating in the $x$-direction with frequency $2\om_+\of{\bk_0}$.
Our main interest lies in Type D, whose particular interaction can set up vortical 
`roll' structures parallel to the $x$-axis,
a candidate mechanism for Langmuir circulation. 
For Type D, $\omsum$ and $k_x$ are both zero,
activating the resonance in \eqref{eq:w_part_int_t} at all depths --- the flow field solution component $w\_{\tpart}$ 
taking on a constant acceleration in inviscid theory.
Fig.~\ref{fig:Langmuir} shows the flow field from case D in the $yz$-plane. 
Streamline plots of the velocity field for the special case presented in 
Appendix~\ref{sec:special_cases} are here presented along with contours of the stream function solution \eqref{eq:Craik_stream_function} presented by \citet{Craik_1970_Langmuir_myidea}. 
We mention here that the solution presented in said reference contains an error which is rectified in 
Appendix~\ref{sec:special_cases}.

\begin{figure}%
\centering
\begin{tabular}{cl}
    \includegraphics[width=.7\columnwidth]{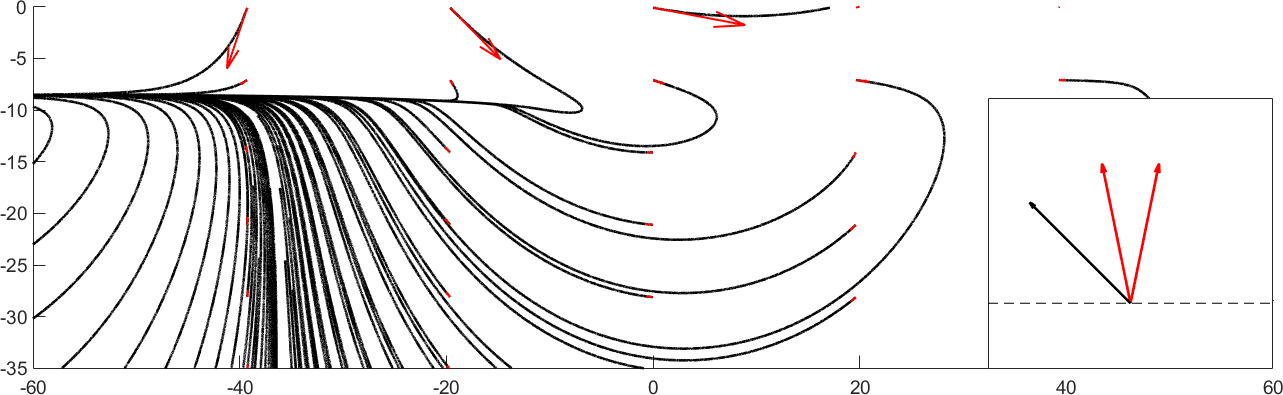} & (a) $T=5$ \\
    \includegraphics[width=.7\columnwidth]{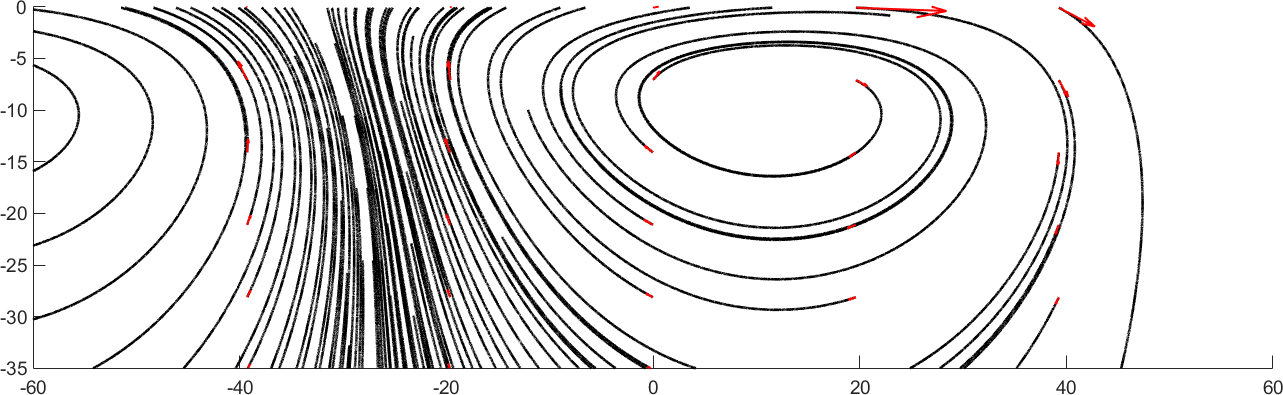} & (b) $T=10$ \\
    \includegraphics[width=.7\columnwidth]{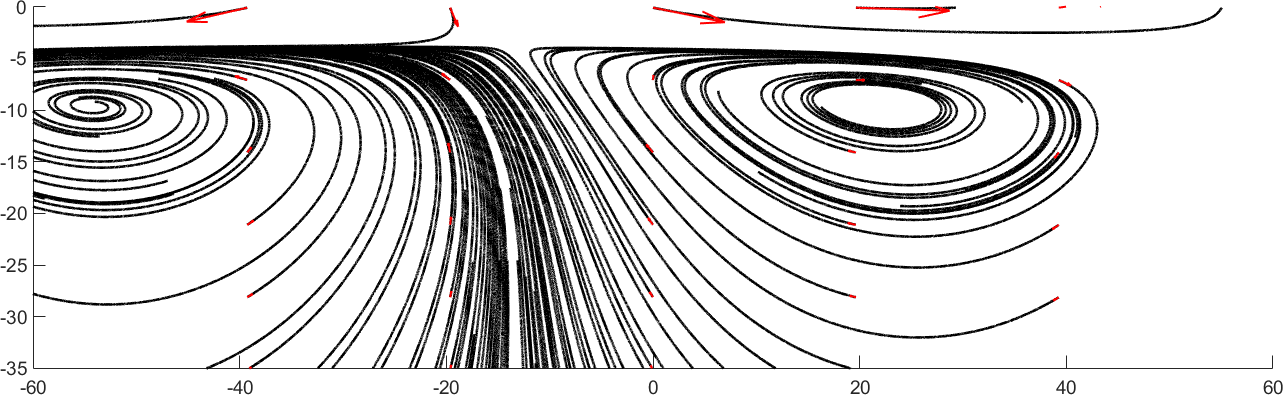} & (c) $T=20$ \\
    \includegraphics[width=.7\columnwidth]{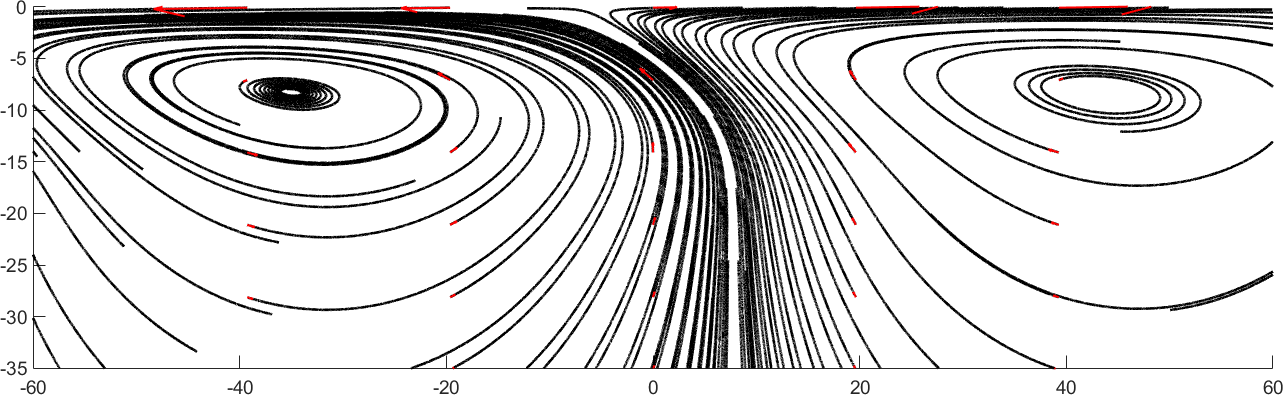} & (d) $T=40$\\
		    \includegraphics[width=.7\columnwidth]{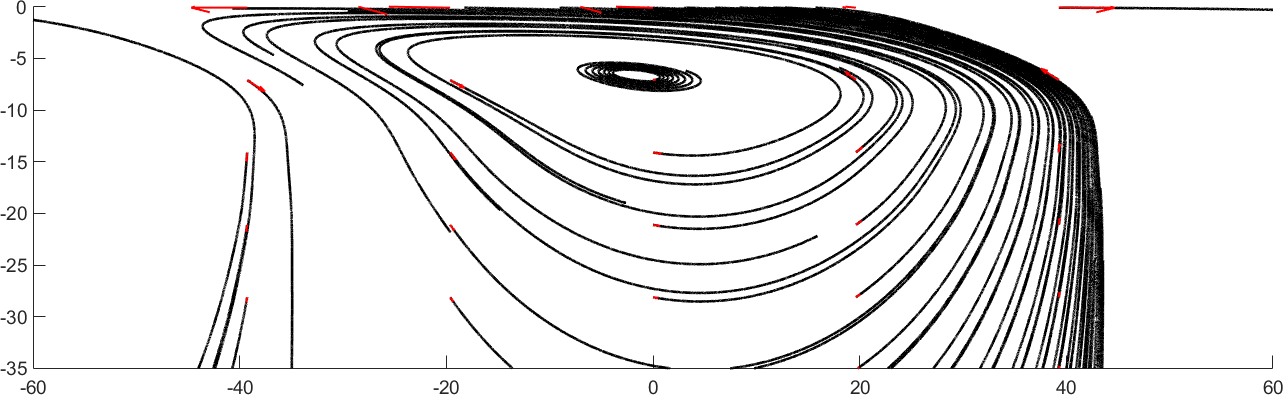} & (d) $T=80$\\
				    \includegraphics[width=.7\columnwidth]{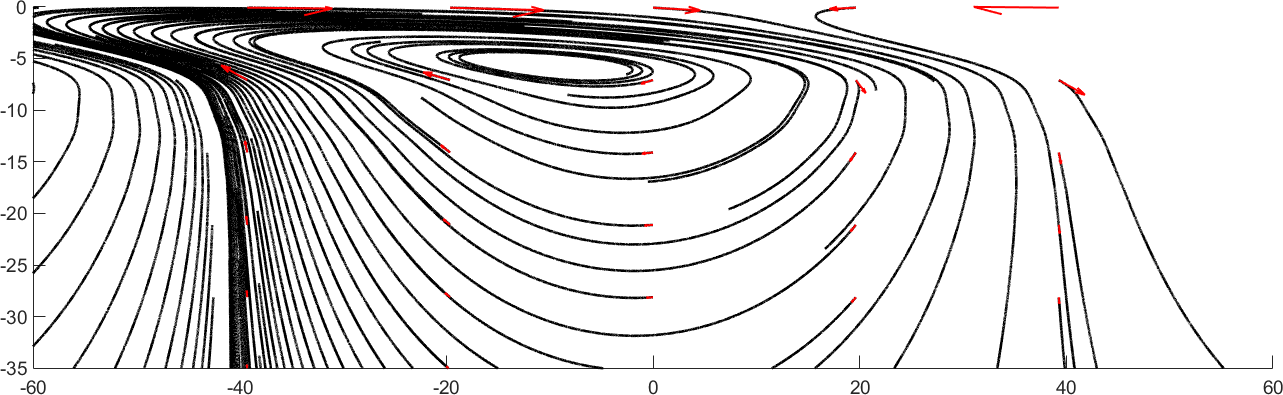} & (d) $T=160$
  \end{tabular}
\caption{
Slowly developing skewed Langmuir--like vortical structures due to a wave pair propagating, on average, at $45^\circ$ relative to the shear current.
%\arev{(wave pair and current directions indicated in top side panel)}.
\arev{Streamlines are shown in the plane orthogonal to the vortex rolls 
illustrated with a dashed line in the inset in the top panel, where also the wave pair and current directions are indicated.
}
(See movieFile1 and movieFile2.)
}%
\label{fig:Langmuir_angle:45}%
\end{figure}

\begin{figure}%
  \centering
	\begin{tabular}{cl}\centering
    \includegraphics[width=.7\columnwidth]{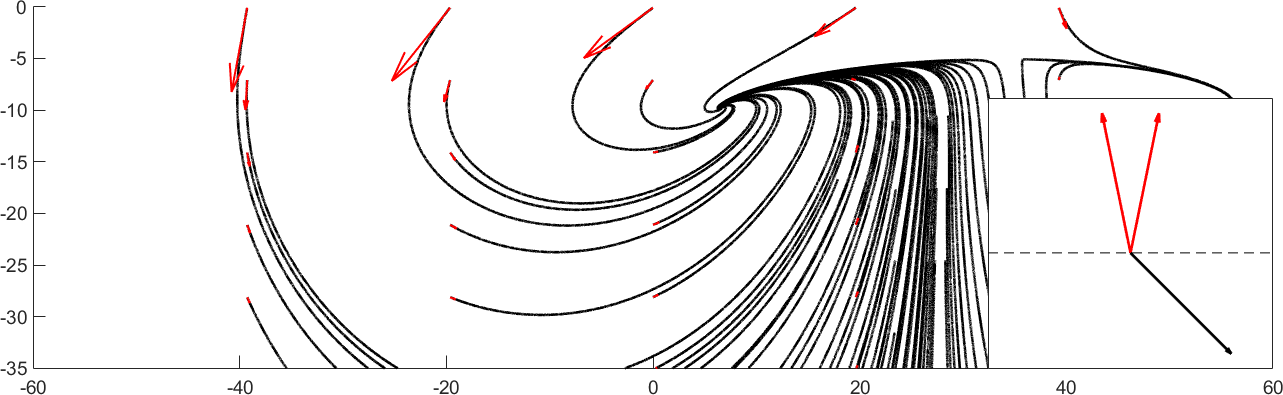} &  (a) $T=5$ \\
    \includegraphics[width=.7\columnwidth]{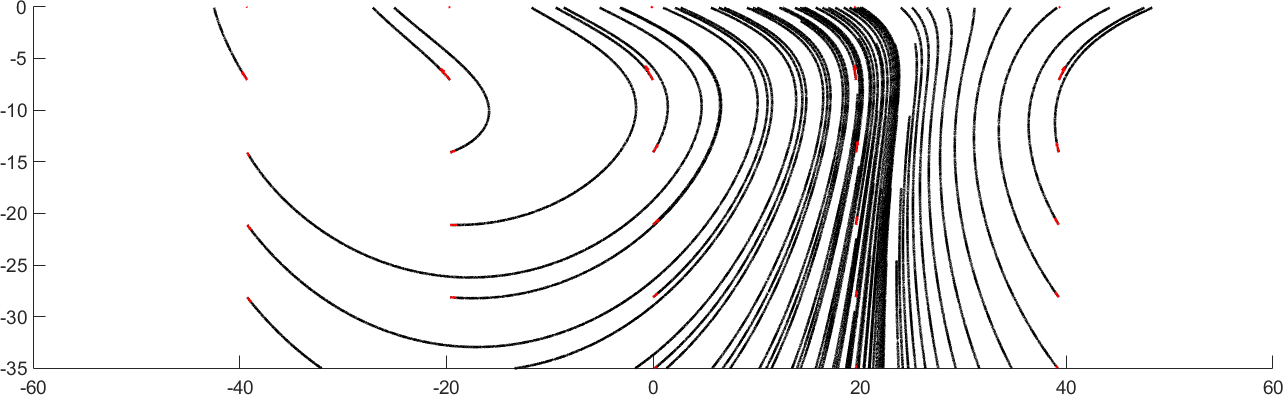} & (b) $T=10$ \\
    \includegraphics[width=.7\columnwidth]{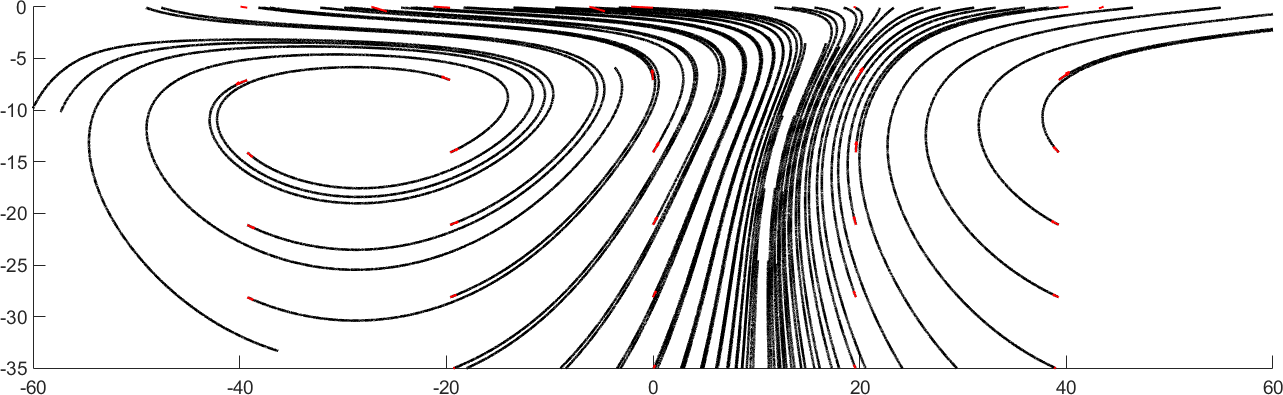} & (c) $T=20$ \\
    \includegraphics[width=.7\columnwidth]{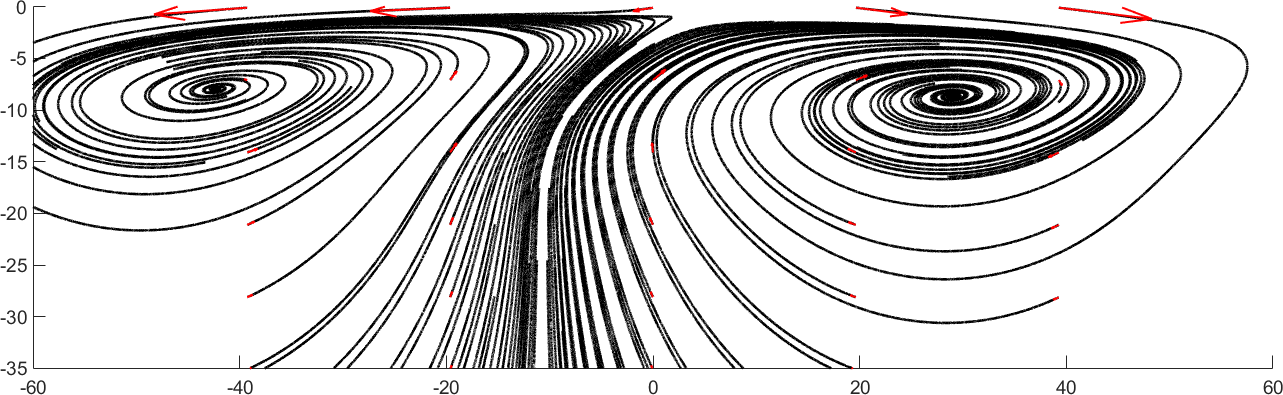} & (d) $T=40$\\
		    \includegraphics[width=.7\columnwidth]{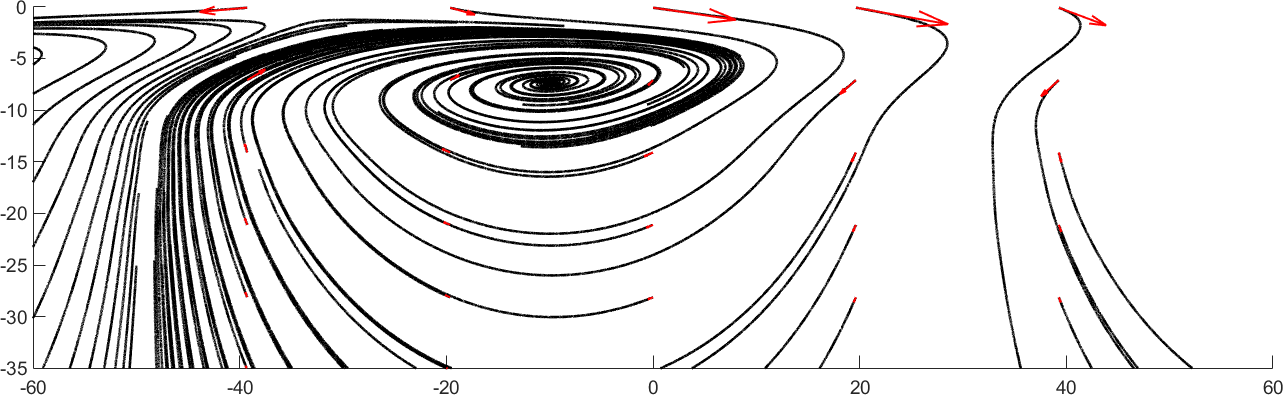} & (d) $T=80$\\
				    \includegraphics[width=.7\columnwidth]{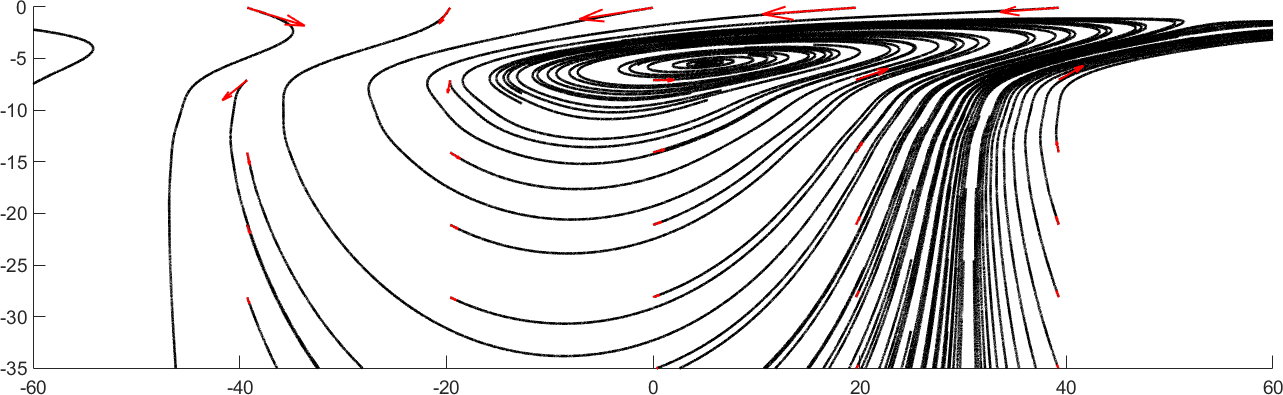} & (d) $T=160$
  \end{tabular}
\caption{
  Same as Fig.~\ref{fig:Langmuir_angle:45}, but 
  the angle between $\bk$ and shear current is now $-135^\circ$.
  }%
\label{fig:Langmuir_angle:m135}%
\end{figure}

\begin{figure}
  \centering 
		\includegraphics[width=.9\columnwidth]{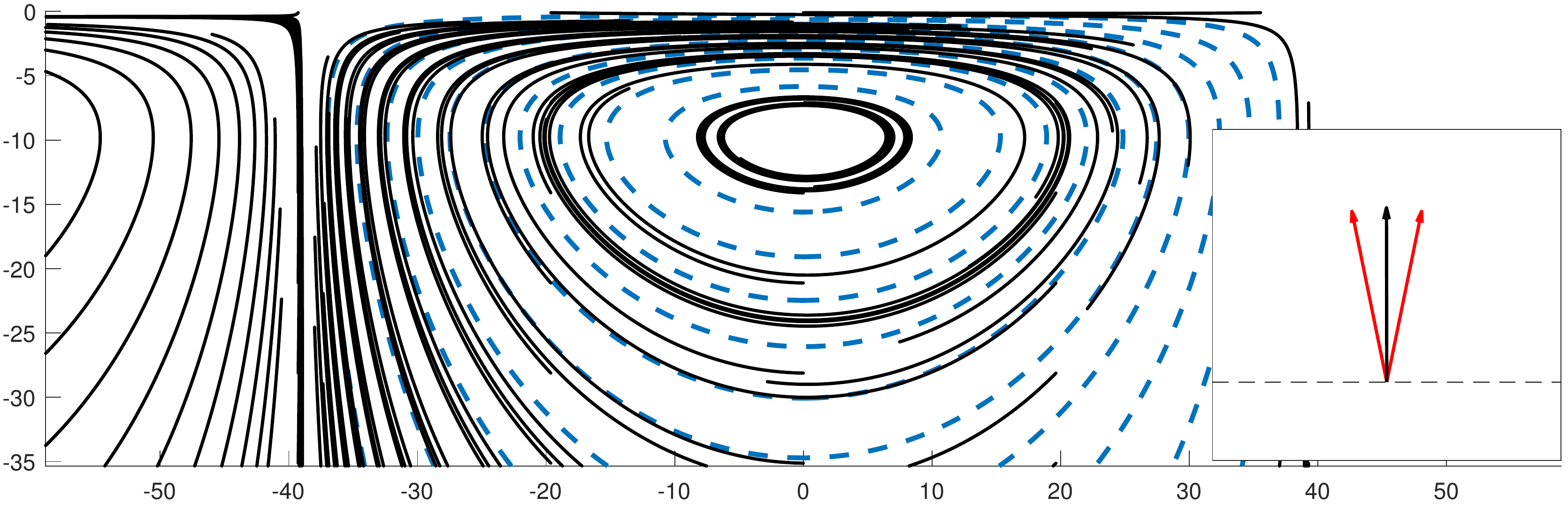}%
  \caption{Angle: $0.1^\circ$ between shear current and $\bk$, 
	%\adl{$T=1000$}\arev{
	$T=100$
	%}. 
  Dashed lines: contours of the 
  stream function \eqref{eq:Craik_stream_function}   for angle $0^\circ$ 
  from \citet{Craik_1970_Langmuir_myidea}.}
  \label{fig:Langmuir_angle:0p1}%
\end{figure}

With the more general solution one can investigate this kind of set--up further. 
In Figs.~\ref{fig:Langmuir_angle:45}--\ref{fig:Langmuir_angle:0p1} we 
rotate the direction of the shear current relative to the propagation direction of the wave pair. New dynamics are now observed.
As time progresses the vortices evolve towards a similar profile as in the symmetric case but then slowly begin to skew in the direction of the shear flow until a steady vortex shape is reached.  
%\scomm{By `rate of skewing' do you mean the rate in time? Or `amount' of skewing?}
%\acomm{In time. Isn't time implied in the word `rate'?}
The rate of skewing of the vortex increases with the angle to the shear direction, but the evolution is insensitive to whether the $k$-vectors point along or against the shear flow direction. 
This is shown in Fig.~\ref{fig:Langmuir_angle:45}--\ref{fig:Langmuir_angle:m135}.
\arev{
Some periodic wave motion is initially prominent near the surface in these figures. 
At larger times this motion is hidden by the increasing vortex intensity.
%Skews of the in the direction of the current also intensifies.
}
The dimensionless time is 
%\adl{$T=t \sqrt{L/g}$}\arev{
$T=t \sqrt{g/L}$
%}
 where $L=\pi/(2 k_{\mr Ly})$ is the symmetric vortex width.
%\arev{
Vortex structures are here aligned with $\bk=\bk_1+\bk_2$.
(Case D loses spatial dependence in $x$ in a coordinate system aligned with $\bk$.)
All rolls will however uniformly drift sideways with time in the direction of the current.
%}

Finally, Fig.~\ref{fig:Langmuir_angle:0p1}, where the angle between $\bk_1+\bk_2$ and the $x$-axis is $0.1^\circ$ and 
	$T=100$, demonstrates how the general solution converges towards Craik's symmetric case solution.

%\adl{
%The other wave interactions (case A--C) come in addition to the above plots, but these will be of a time--periodic nature. 
%}\arev{
The non-swirling interactions (cases A--C) contribute with a limited periodic disturbance to the vortex motion%
, which initially dominates the second order motion of
 Fig.~\ref{fig:Langmuir_angle:45}--\ref{fig:Langmuir_angle:0p1} but is then overwhelmed by the vortex motion whose period is considerably longer. 
In the symmetric case (Fig.~\ref{fig:Langmuir}), this periodic motion is in time completely wiped out by the unbounded vortex motion.
%}
%\acomm{I changed the figures \ref{fig:Langmuir_angle:45}--\ref{fig:Langmuir_angle:m135} with plots that do contain interaction A--C. (See previous draft for comparison.)}\\
%\arev{
The driving force for the Langmuir circulation reduces as the angle between shear and mean wave direction approaches $90$ degrees. At the same time the skewing mechanism intensifies. 
No Langmuir vortices are generated if the mean wave direction is perpendicular to the direction of the current. 
This is not dissimilar to the findings of \cite{vanRoekel_2012_Langmuir_LES_of_CL_eqs} who performed large eddy simulations of the Craik-Leibovich equations with misaligned Stokes drift and wind forcing;
their results show diminishing Langmuir turbulence as the angle between Stokes drift and Langmuir cell alignment is made to increase.
%}

%%%%%%%%%%%%%%%%%%% S E C T I O N %%%%%%%%%%%%%%%%%%%%%%	
\subsection{Fluid particle trajectory beneath a localized disturbance in 2D}
\label{sec:results:particle_trajectory}

%\scomm{The only problem with this section is that it doesn't have any conclusions other than maybe `we can do this'. }
%\acomm{Added minor stuff.}

As we proceed towards spectral computations we start by considering 
a two-dimensional Cauchy problem. 
This is similar to the problem studied by \cite{Abou-Dina_2001_integral_mode_coupling_topography} for cases of varying bathymetry. 
Two-dimensionality means that all wave vectors are parallel such that no oblique interaction between waves and shear can take place; $\ktimesk\equiv0$. 
As a result, all the rotational cross-terms and advective terms disappear. 
We look at this problem including fluid particle trajectories.

\begin{figure}
  \begin{center}
	   \includegraphics[width=\columnwidth]{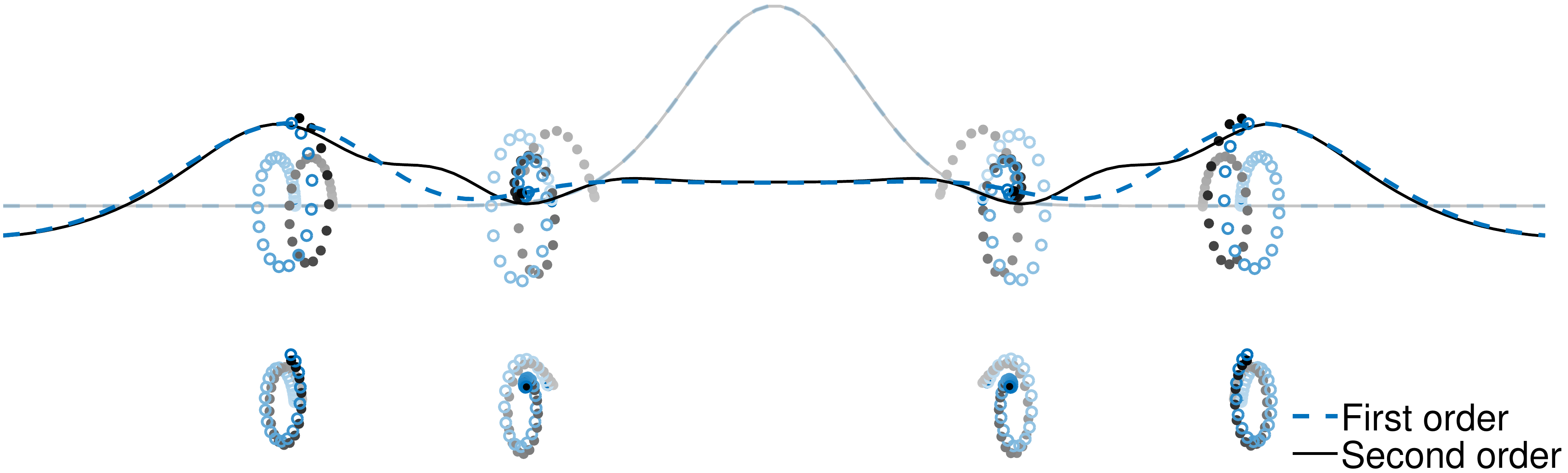} \\
	   \includegraphics[width=\columnwidth]{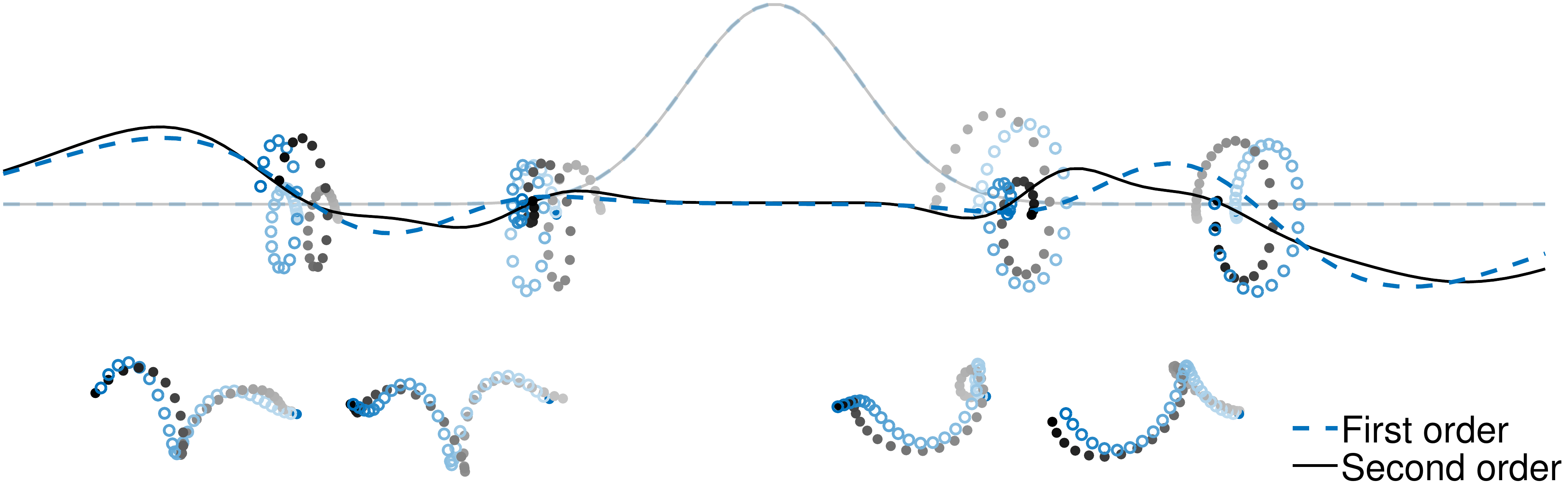}%
  	\tikz[overlay]{
     \node at (-1.6,1.7){		\includegraphics[width=.1\columnwidth]{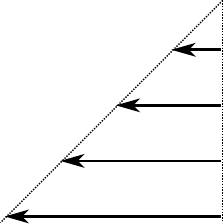}};	} 
  \end{center}
\caption{2D Cauchy--Poisson problem --- Gaussian initial elevation profile 
with `steepness' $H= 0.2$. 
Blue, dashed/open: first order solution; Black, solid/filled: second order solution. Top: $\FrS=0$; bottom: $\FrS=0.5$ 
%\adl{\srev{($b$ being the Gaussian distribution width parameter)}}. 
Surface elevation is plotted at $T=0$ and $T=6.0$, together with particle positions within this time interval.
%Decreasing transparency 
Increasing opacity indicates the evolution in time.
(See movieFile3 and movieFile4.)
}%
\label{fig:2D_ringwave}%
\end{figure}

Fig.~\ref{fig:2D_ringwave} shows the surface elevation and particle trajectories to second order.
The initial surface elevation is here a Gaussian \arev{profile} at rest,
\begin{align}
\p\zeta\IC &= \h 
  \rme^{-\pi^2 x^2/b^2}
%\exp\br{-\frac{\pi^2 x^2}{b^2}};
&
\f\zeta\IC &= 
  (\h b/\sqrt{\pi})
  \rme^{-b^2 k_x^2/(2\pi)^2}
%\frac{\h b}{\sqrt \pi}  \exp\br{-\frac{b^2 k_x^2}{(2\pi)^2}};
&
\dot{\p\zeta}\IC&=\dot{\f\zeta}\IC=0,
\end{align}
and the surface pressure is uniform.
$b$ is here the Gaussian distribution width parameter.
A respective dimentionless steepness, time and shear Frude number 
\begin{equation}
H = \h/b,
\qquad
T = t \sqrt{g/b},
\qquad
\FrS = S \sqrt{b/g},
%\label{eq:}
\end{equation}
have here been introduced.
Initial conditions have been chosen uniform in the $y$-direction, although no significant complexity is added by letting the waves disperse at an angle to the shear current
%\srev{
beyond the fact that also the vorticity field will be perturbed in this case \citep{Ellingsen_vorticity_paradox}.
%}
The procedure of perturbing the bound frequency, described in and around \eqref{eq:Stokes_drift_frequency_perturbation}, is employed to avoid 
pole singularities appearing with self-interacting wave components at second order. 
As opposed to the linear solution, 
%\adl{the orbits of the particle trajectories to second order are seen to shift}\arev{
this shifts the second order orbits of the particle trajectories
%}
as the main wave bulk flushes past. 
%\adl{, most perceptibly in the trajectories nearest to the surface and closest to the centre.}
This is the transient manifestation of the phenomenon of Stokes drift 
%\adl{(equation \eqref{eq:Stokes_drift})}
in the Cauchy--Poisson problem.

The Stokes drift mechanism is most perceptible in the trajectories nearest to the surface and closest to the origin of the ring wave.
This is because the immediate Stokes drift is proportional to the immediate wave steepness squared, which is greatest in the early stages, near the 
centre of the figure.
The drift follows the wave direction so that net mass transport due to Stokes drift is  away from the region of the initial Gaussian bell. 
In the two-dimensional infinite depth case the effect of the shear is inferred form \eqref{eq:part_pos_monochromatic:r} to generate a factor
\[
\times\frac{\om}{\om-S k_x z_0}\br{ 1 + \frac12\frac{S}{\om-S k_x z_0}}
\]
to the Stokes drift compared to non-sheared flow. 
At the surface the factor is simply $1+S/2\om$ below which the drift reduces exponentially. 
Assuming no dominating critical layer in the surface region we therefore conjecture/surmise that $S>0$ 
serves to strengthen the Stokes drift transport away from the centre towards the positive $x$-direction, conversely in the negative $x$-direction if $S<0$.
%.
%}

%%%%%%%%%%%%%%%%%%% S E C T I O N %%%%%%%%%%%%%%%%%%%%%%	
\subsection{A three-dimensional ring wave}
\label{sec:results:3D_ringwave}

Similar to the previous example, 
let the initial perturbation be a Gaussian at rest, but now in three dimensions, 
\begin{align}
\p\zeta\IC &= \h 
  \rme^{-\pi^2 r^2/b^2}
%\exp\br{-\frac{\pi^2 r^2}{b^2}};
&
\f\zeta\IC &= 
  (\h b^2/\pi)
  \rme^{-b^2 k^2/(2\pi)^2}
%\frac{\h b}{\sqrt \pi}  \exp\br{-\frac{b^2 k^2}{(2\pi)^2}};
&
\dot{\p\zeta}\IC&=\dot{\f\zeta}\IC=0.
\end{align}

\renewcommand{\tabcolsep}{1mm} %tabilar column width

\begin{figure}
\begin{center}
  \begin{tabular}{ccc}
   Dispersive & Advective \\
  \rotatebox{90}{\hspace{.15\columnwidth} Bound}  
   \includegraphics[width=.4\columnwidth]{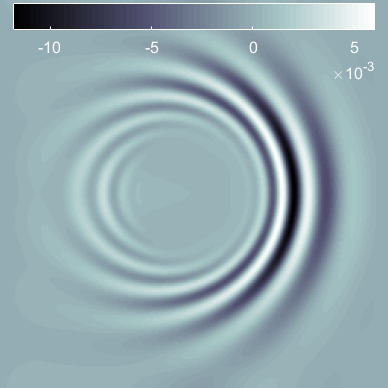}
  &  \includegraphics[width=.4\columnwidth]{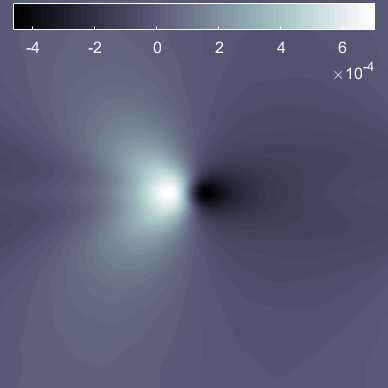} \\
   (a) & (b) \\
  \rotatebox{90}{\hspace{.15\columnwidth}Free}
   \includegraphics[width=.4\columnwidth]{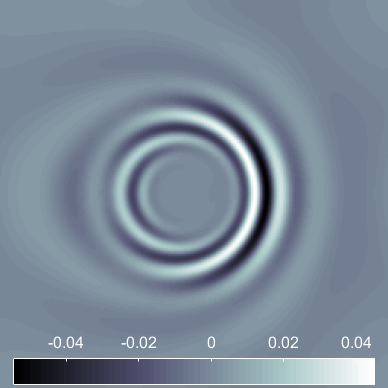} 
  & \includegraphics[width=.4\columnwidth]{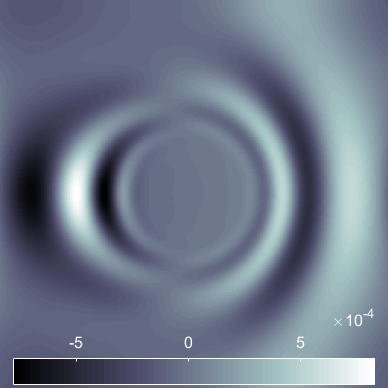} \\
   (c) & (d) \\
\end{tabular}
\begin{tabular}{ccc}
  \includegraphics[width=.32\columnwidth]{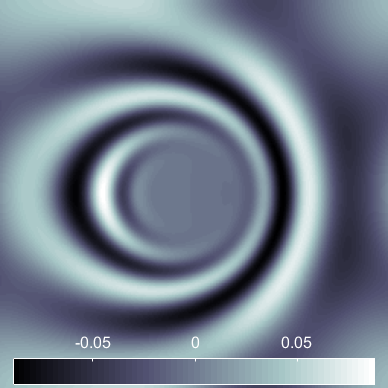}
  & \includegraphics[width=.32\columnwidth]{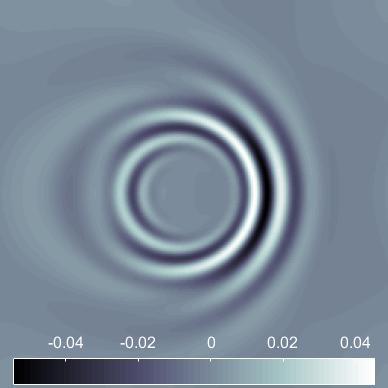}
  & \includegraphics[width=.32\columnwidth]{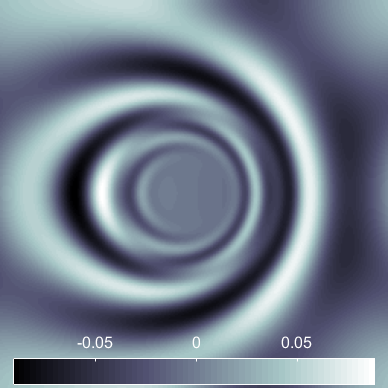} \\
  (e)&(f)&(g)
\end{tabular}
\caption{
   Second order surface elevation components of Cauchy problem --  Gaussian initial profile. 
   FFT from \eqref{eq:convolution_int_O2_inner_form}. Parameters: $\FrS=1.0$, $T =10$, $H=0.2$. Domain length: $8b$.
  %$128^4$ interaction modes.
  (a) $\p\zeta_{\bound,\wave}\ot/\h$; 
  (b) $\p\zeta_{\bound,\bg}\ot/\h$; 
  (c) $\p\zeta_{\free,\wave}\ot/\h$; 
  (d) $\p\zeta_{\free,\bg}\ot/\h$;
  (e) $\p\zeta\oo/\h$;
  (f) $\p\zeta\ot/\h$;
  (g) $(\p\zeta\oo+\p\zeta\ot)/\h$.
  %\scomm{The numbers on the colour bars are too small to be readable. Especially the (important!) power of $10$}
  }
  \label{fig:ringwave_3D_T1_Frs1}
\end{center}
\end{figure}

\begin{figure}
\begin{center}
  \begin{tabular}{ccc}
   Dispersive & Advective \\
  \rotatebox{90}{\hspace{.15\columnwidth} Bound}  
   \includegraphics[width=.4\columnwidth]{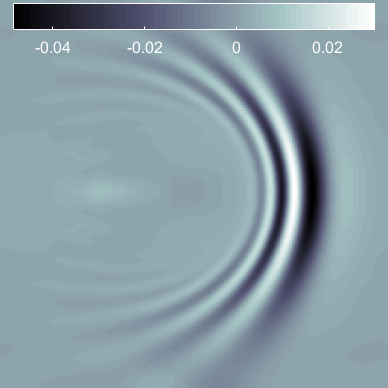}
  &\includegraphics[width=.4\columnwidth]{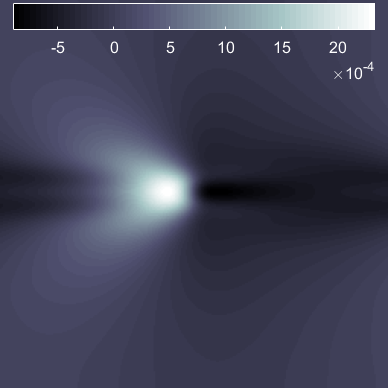} \\
   (a) & (b) \\
  \rotatebox{90}{\hspace{.15\columnwidth}Free}   
   \includegraphics[width=.4\columnwidth]{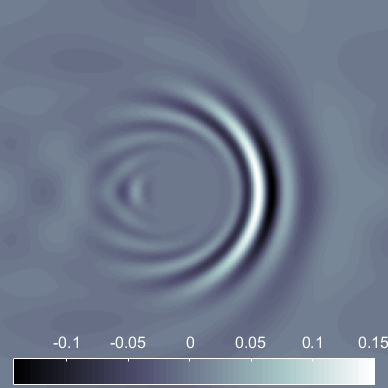}
  & \includegraphics[width=.4\columnwidth]{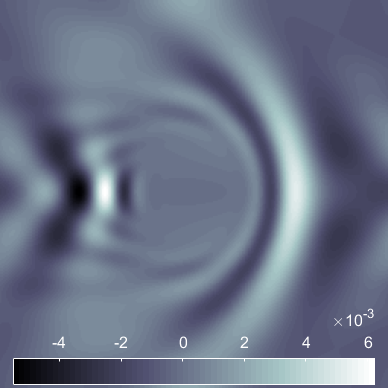}\\
   (c) & (d) 
\end{tabular}
\begin{tabular}{ccc}
  \includegraphics[width=.32\columnwidth]{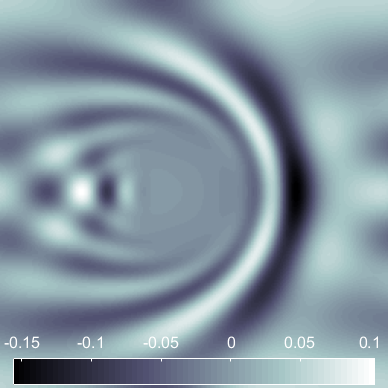}
  & \includegraphics[width=.32\columnwidth]{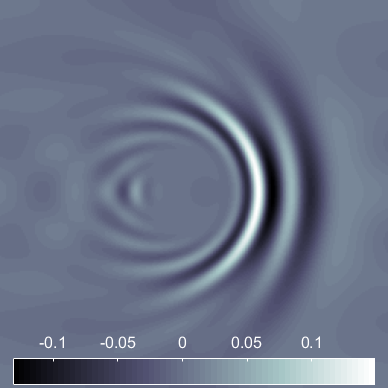}
  & \includegraphics[width=.32\columnwidth]{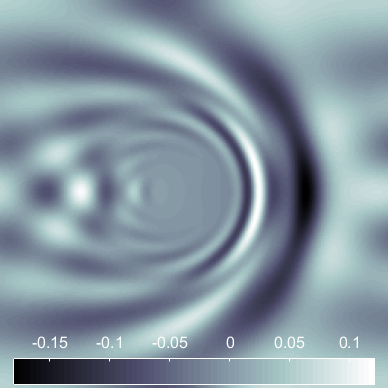}\\
  (e)&(f)&(g)
\end{tabular}
\caption{
   Second order surface elevation components of Cauchy problem --  Gaussian initial profile. 
   FFT from \eqref{eq:convolution_int_O2_inner_form}. Parameters: 
   $\FrS=2.5$, $T =10$, $H=0.2$. Domain length: $8b$.
  %$128^4$ interaction modes.
  (a) $\p\zeta_{\bound,\wave}\ot/\h$; 
  (b) $\p\zeta_{\bound,\bg}\ot/\h$; 
  (c) $\p\zeta_{\free,\wave}\ot/\h$; 
  (d) $\p\zeta_{\free,\bg}\ot/\h$;
  (e) $\p\zeta\oo/\h$;
  (f) $\p\zeta\ot/\h$;
  (g) $(\p\zeta\oo+\p\zeta\ot)/\h$.
  %\scomm{The numbers on the colour bars are too small to be readable. Especially the (important!) power of $10$}
  }
  \label{fig:ringwave_3D_T1_Frs2p5}
\end{center}
\end{figure}

Fig.~\ref{fig:ringwave_3D_T1_Frs1} and \ref{fig:ringwave_3D_T1_Frs2p5} show the surface elevation at a specific point in time for moderate and strong shear, respectively. 
Panels   (a)--(d) in each of the figures 
show the four wave components that make up the second order waves.
Below that, panels 
(e)--(g) 
show the net first, second and combined surface elevation.
These figures are computed with inverse fast Fourier transforms.

The upper left panels, \ref{fig:ringwave_3D_T1_Frs1}a and \ref{fig:ringwave_3D_T1_Frs2p5}a, show the profile of the second order bound dispersion wave $\p\zeta_{\bound,\wave}\ot$, computed from the convolution of \eqref{eq:zeta_bound_wave}.
This is the harmonic which appears as a second order correction for nonlinear convection and boundary 
condition terms.
Its dispersive rate is bound by two interacting first order waves to give a phase velocity 
 $\bm c\_p=(\bk/k)(\omsum/k)$ where $\omsum=\Om_{\pm_1}\of{\bk_1}+\Om_{\pm_2}\of{\bk_2}$ and $\bk=\bk_1+\bk_2$.

Advective waves are presented in panels  \ref{fig:ringwave_3D_T1_Frs1}c and \ref{fig:ringwave_3D_T1_Frs2p5}c. 
These are not directly artefacts of surface wave dynamics but of the internal flow field and its initial state.
The advective waves constitute a passively advected background flow whose purpose is to cancel the wave-induced internal rotational motion at $T=0$.
The intensity of these waves is centred around 
where the wave--shear rotational interaction was initially greatest.
Because this field is advected with the shear current, and our frame of reference follows the surface, these waves remain 
in the vicinity of the initial centre.
Advective waves do however diminish in amplitude with time as this wave field is smeared out and transported away by the shear current below the surface.
Note also that the magnitude of the advective waves is small compared to that of the dispersive ones and does not 
 affect the net surface elevation visibly.

Free wave dispersion 
(see Eqs.~\eqref{eq:zeta_of_t} and 
\eqref{eq:zeta_ICn_evaluated:IVP}
) is shown in 
panels b and d in Figs.~\ref{fig:ringwave_3D_T1_Frs1} and \ref{fig:ringwave_3D_T1_Frs2p5}.
%}
These waves are again not directly related to the nonlinear interactions but manifest as corrections to the first order modes and abide by the linear 
dispersion relation.
Their function is to cancel whatever contribution the two bound waves impose on the surface at $T=0$ 
to ensure that our solution is in 
accordance with the prescribed initial state.

To second order, the ring wave problem is seen to exhibit five distinct `chunks' of dispersion as there are five dominant group velocities present. 
One is the first order free dispersion. 
Another pertains to the second order bound dispersion, whose frequency and wave vector are roughly twice those of the free first--order harmonic.
A third group is the free second--order dispersive wave. 
This consists of the same wavelengths as the bound dispersive wave (in precise antiphase at $T=0$), but disperses more slowly. 
Finally there are two advective--type waves, one remaining fairly stationary
with respect to the surface and one dispersing freely. 
%\scomm{Not sure whether I think `fairly stationary' is too vague.}
These have small amplitude compared to the other groups. 
The free dispersive wave quickly dominates among the second order waves because it decays more slowly with time than its bound counterpart.
%, as we showed in the asymptotic analysis in Sec.~\ref{sec:asymptotic}.
%
\arev{This is apparent from the asymptotic analysis given in Appendix~\ref{sec:statphase};
comparing the bound dispersive asymptotic expression \eqref{eq:asymptote:bound_wave_O2} with 
%that for free waves in \eqref{eq:asymptote:free_wave_O2},
the free wave asymptotics (combining \eqref{eq:convolution_int_O2_inner_form} and \eqref{eq:asymptote:O1})
 one sees that the former 
vanishes as $t^{-2}$ while the \arev{latter} does so as $t^{-1}$.
As these waves are initially of the same order of magnitude we can expect the free waves to dominate among the second order waves in the far field. 
}

\arev{
We remark that the second order effects we have studied constitute a significant correction to the first order ring wave when reasonable steepness of the initial perturbation is assumed. 
The concept of wave steepness is however not directly translatable to the initial spectrum
 of the initial value problem as steepness is intended as a measure of the perturbation magnitudes;
velocities are initially zero in the present problem and the surface elevation decays rapidly.  
Rather, the kinematic history of the solution ought to be considered.
In the two depicted cases of Figs.~\ref{fig:ringwave_3D_T1_Frs1} and \ref{fig:ringwave_3D_T1_Frs2p5} the maximal steepness observed it physical space is roughly 
$0.25$, $0.17$, $0.13$ and $0.10$ at the times $T=2.5$, $5.0$, $7.5$ and $10$, respectively.
%$0.54$, $0.17$ and $0.10$ at the times $T=0$, $5$ and $10$, respectively.
We therefore conclude that the presented problems are appropriately within  the weakly nonlinear regime and note that second order effects are then highly conspicuous  in the full solution (panels g of said figures). 
%Although we rely on the conventional concept of wave steepness of monochromatic waves for distinguishing orders of nonlinearity, this notion is not directly translatable to the initial spectrum
 %of the initial value problem. 
%Rather, the kinematic history of the solution ought to be considered.
%In the two depicted cases of Figs.~8 and 9 the maximal steepness observed it physical space is roughly $0.54$, $0.17$ and $0.10$ at the times $T=0$, $5$ and $10$, respectively.
%One must here keep in mind that steepness is intended as a measure of perturbation magnitudes and that velocities are zero at $T=0$. 
%We therefore conclude that the presented problems are appropriately within  the weakly nonlinear regime and note that second order effects are then highly conspicuous  in the full solution (panels g of said figures). 
}

\arev{
In order to clearly visualise the interaction of shear and second order wave modes, the
 shear strengths here presented are strong, far greater than what one would expect from ocean wind generated shear currents, but feasible in 
other types of flows such as river shallows, discharge plumes and surface jets.
(Dimensionally, wavelengths on the order $\sim\!10\,$m will have shear strength $S\sim\FrS$/s.
The vorticity of the tidal current in the Columbia River mouth has been reported at around $0.4$/s in the top $5\,$m of the water column \citep{dong2012_ColumbiaRiver}.
) 
Weaker shear results in even weaker advective waves
%With weak (absent) shear the advective waves are even weaker (absent) 
while dispersive waves remain the same order of magnitude, though closer to being cylindrically symmetric.
The image of the net wave, with grouped dispersion of bound and free first and second order waves, remain the same.
}

\begin{figure}
  \begin{tabular*}{\columnwidth}{cc}
    \includegraphics[width=.48\columnwidth]{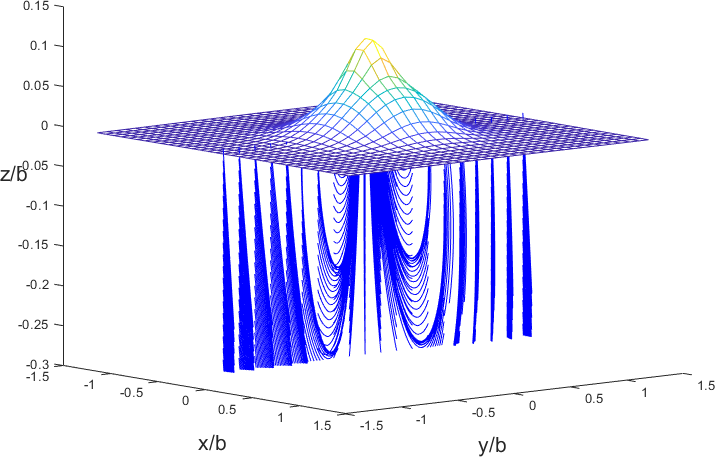}
    &\includegraphics[width=.48\columnwidth]{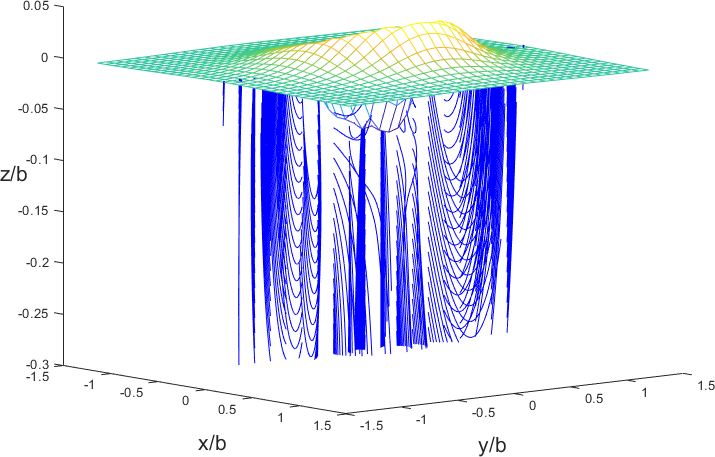} \\
    $T=0.5$ & $T=1.0$ \\
    \includegraphics[width=.48\columnwidth]{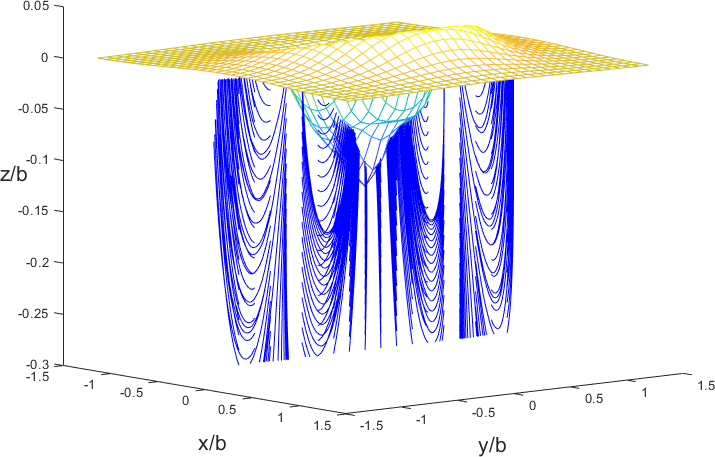}
    &\includegraphics[width=.48\columnwidth]{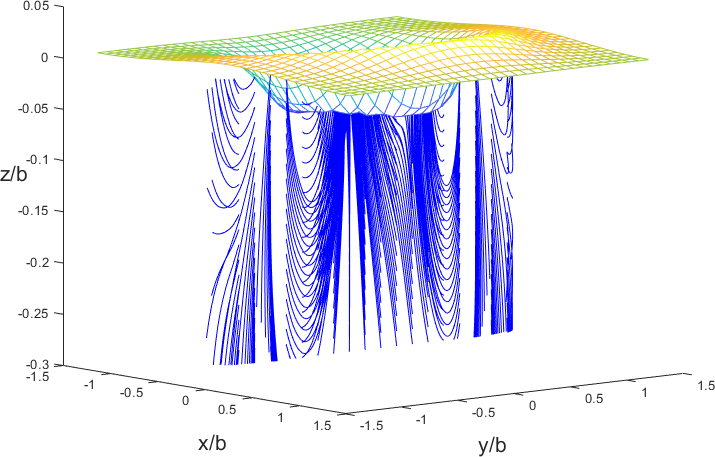} \\
    $T=1.5$ & $T=2.0$
  \end{tabular*}
\caption{
Streamlines showing the motion beneath the surface of a ring wave on a shear current.
\arev{(Only streamlines originating at points in the plane $x=0$ are shown.)}
Only the second order homogeneous wave motion is shown (originating from the $A\ot$-terms first appearing in \eqref{eq:w_sol}). 
This, together with the motion from oblique interactions shown in Fig.~\ref{fig:ringwave_vel:cross}, comprise the net second order motion.
\arev{Surfaces show the net second order elevation $(\p\zeta\oo+\p\zeta\ot)/b$. }
Initial parameters are 
$\FrS=1.0$ and $H=0.2$, as in Fig.~\ref{fig:ringwave_3D_T1_Frs1}.
%\Q{Referee 1 listed this and the following figure as being poorly described, but I can't see what needs clarification.}
}
\label{fig:ringwave_vel:A}
\end{figure}

\begin{figure}
\begin{tabular*}{\columnwidth}{cc}
  \includegraphics[width=.48\columnwidth]{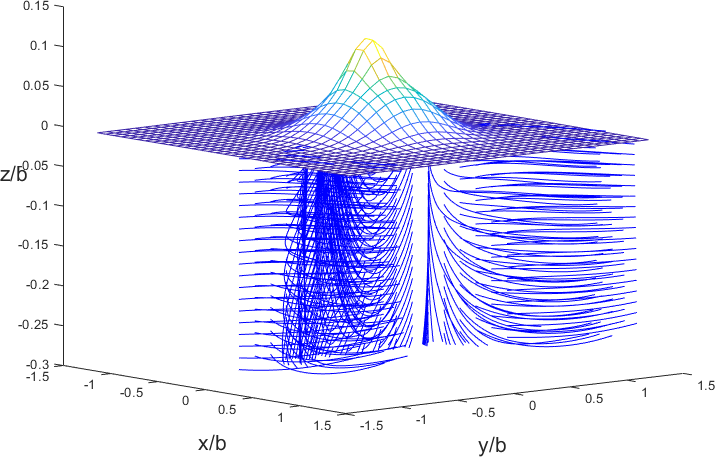} 
  &\includegraphics[width=.48\columnwidth]{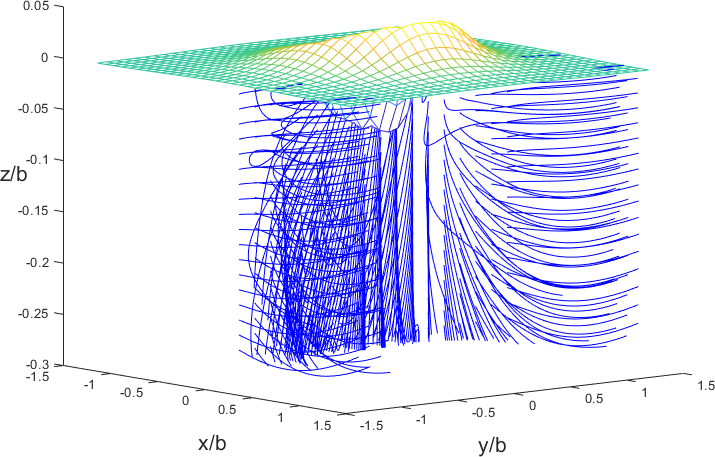}\\
  $T=0.5$ & $T=1.0$ \\
  \includegraphics[width=.48\columnwidth]{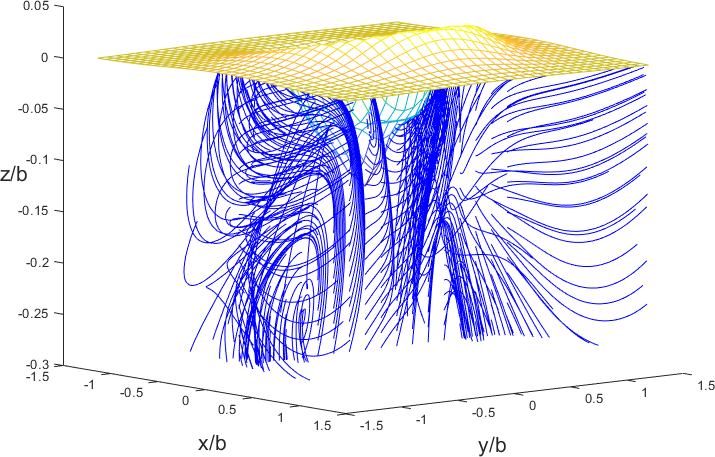}
  &\includegraphics[width=.48\columnwidth]{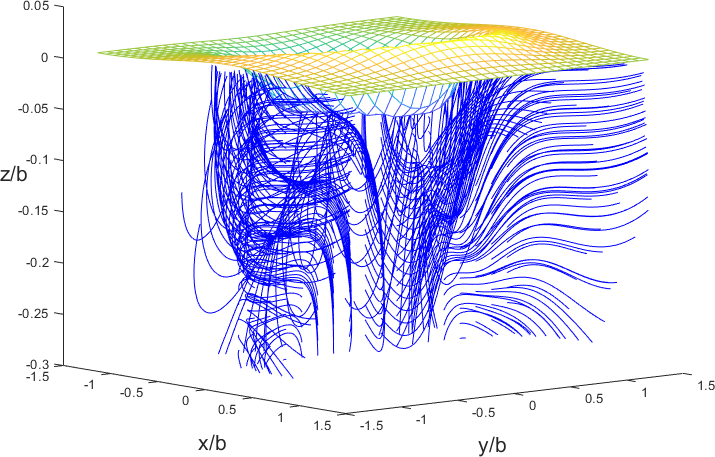} \\
  $T=1.5$ & $T=2.0$
\end{tabular*}
\caption{Motion from second order oblique wave interaction 
(originating from the $  w_\tpart\ott$-terms first appearing in \eqref{eq:w_sol})
showing a swirling contribution to the motion beneath the surface of a ring wave on a shear current.
%\adl{beneath a the ring wave in Fig.~\ref{fig:ringwave_vel:A}.}
This, together with the homogeneous motion of Fig.~\ref{fig:ringwave_vel:A}, comprise the net second order motion.
Initial parameters 
and surfaces 
 are 
%$\FrS=1.0$ and $H=0.2$, 
as in Fig.~\ref{fig:ringwave_vel:A}.
}
\label{fig:ringwave_vel:cross}
\end{figure}

Finally, we will consider the velocity fields below the ring wave surface. 
For visualization we use streamline plots where each streamline starts from $x=0$ in the $yz$-plane.
Above these are drawn the net surface elevation to second order.
Similar to the surface waves, we split the internal velocity field into two groups based on our physical interpretation of the solution of the Rayleigh equation \eqref{eq:w_sol}.
Fig.~\ref{fig:ringwave_vel:A} shows the homogeneous part of the velocity field, 
involving the eigenfunction $A\ot\of{t}$ in our solution. 
This motion, characterized by a purely exponential decay in $z$, is 
analogous to the first order solution with the time dependency closely related to the free surface and the dispersion it engenders. 
The remaining second order contribution to the velocity field, shown in Fig.~\ref{fig:ringwave_vel:cross}, is derived from the cross term $\f w_\tpart\ott$ --- the particular solution of the second--order Rayleigh equation. % \eqref{eq:Rayleigh_dt}.
These kinematics arise from oblique nonlinear wave interactions and their interaction with the shear.

The plots are seen to exhibit swirling motion and a helix twisting of the streamlines in a manner reminiscent to that seen in the Langmuir circulation examples.
This swirling motion is particular for the near field as dispersion separates wavelengths in the far field, leaving only self-interactions, parallel in nature. 
The sub--surface motion exemplifies how the second--order interactions between surface waves of \emph{any} shape and a sub--surface shear currents may create some amount of vortical or swirling motions. 
The indication could be that such motions contribute to mixing in the upper oceans
in the same way that long regular structures known as conventional Langmuir vortices do \citep{Belcher_2012_global_perspective_on_Langmuir_turbulence}, even if the particular conditions for Langmuir circulation are absent.
Whether and under what conditions this is important in realistic scenarios, however, is an open question.

%%%%%%%%%%%%%%%%%%%%%%%%%%%%%%%%%%%%%%%%%%%%%%%%%%%%%%%%	
%%%%%%%%%%%%%%%%%%% S E C T I O N %%%%%%%%%%%%%%%%%%%%%%	
%%%%%%%%%%%%%%%%%%%%%%%%%%%%%%%%%%%%%%%%%%%%%%%%%%%%%%%%	

\section{Summary}

\label{sec:conclusions}

A mode coupling solution to second order in Stokes perturbation has been constructed for the Cauchy--Poisson boundary value problem in the presence of a uniform vorticity field.
The solution captures the oblique wave--shear flow resonance mechanism that sets up large accelerating vortex motion close to the surface. 
It also allows for similar asymmetric cases to be investigated; a shear current acting upon a wave field at an angle to the mean wave direction is seen to skew the structure of the vortex cells and limit inviscid flow acceleration.
%\arev{
The rate of vortex acceleration diminishes as the misalignment between current and mean wave direction increases, but does not depend notably on
%whether the current and wave field directions are similar or opposing. 
whether the waves follow the current or oppose it.
%}

In a full Cauchy--Poisson boundary value problem four types of second order surface waves are identified, distinct in their dispersive properties. 
Their physical interpretation has been given as follows:
%\begin{itemize}
	(i) Second order harmonics that account for nonlinearities from convection and from surface boundary conditions.
	(ii) An initial velocity field that initially cancels the added second order harmonics in the internal flow. 
	This initial velocity field is then advected with the shear current.
	(iii \& iv) Additional first order harmonics that cancel the alteration that the two aforementioned waves make on the initial surface.
%\end{itemize}
%
These types of waves will with time disperse in distinguishable groups.
The free waves decay slower than the bound dispersive waves and will therefore dominate the second order far-field.

The second order internal flow consists of additional linear-type motion and motion from oblique nonlinear wave interactions. 
The latter, combined with action from the shear, serves to curve the streamlines of the internal near field of a ring wave in a spiralling manner.

\section*{Acknowledgements}
The research here reported was funded by the Research Council of Norway (FRINATEK), project number 249740. We have benefited from discussions with Dr.\ Ivan Savelyev.

%%%%%%%%%%%%%%%%%%%%%%%%%%%%%%%%%%%%%%%%%%%%%%%%%%%%	
%%%%%%%%%%%%%%%% A P P E N D I X %%%%%%%%%%%%%%%%%%%	
%%%%%%%%%%%%%%%%%%%%%%%%%%%%%%%%%%%%%%%%%%%%%%%%%%%%	
\appendix

%%%%%%%%%%%%%%%% A P P E N D I X %%%%%%%%%%%%%%%%%%%	
%\section{Velocities components}
\section{\arev{Components of the second order solution}}
\label{sec:uvw}

\renewcommand{\ot}{}

\arev{
The Euler equations \eqref{eq:problem:Euler} can be manipulated to yield explicit expressions for the leading order horizontal velocity components
\begin{equation}
	 \pdtbar(\p\nabla\_h^2 \p\bu\_h +\p\nabla\_h \pp_z \p w)  = 
	U' \pp_y \p\nabla\_h\!\times\! \p w 
	- \p\nabla\_h\!\times\!\p\nabla\_h\!\times\!(\p\bu\!\cdot\!\p\nabla)\p\bu\_h.
	\label{eq:huh}
\end{equation}%
}%
\arev{
In Fourier space, 
$\p\nabla\_h\times \rightarrow \rmi \bk\bb \cdot$ with $\bk\bb =  (-k_y,k_x)$.
The solution of \eqref{eq:huh} contains an integration coefficient representing horizontal motion present at $t=0$.
Similar to the first order components \eqref{eq:O1_flowfeeld}, 
we impose an initially irrotational horizontal velocity field, leaving only the irrotational term $\rmi A \bk/k$ initially active. 
The continuity equation, which dictates $\bk \cdot \f \bu\_{h} =0$, is upheld in this configuration.
As seen at first order, this choice removes the horizontal singularities associated with critical layers from our initial value problem and replaces them with levels of unbounded velocity growth.
}

\arev{Integrals of the advective $w\_{\tpart}$ term are in the below evaluated by inserting the exponential integral definition \eqref{eq:Ej} and then applying partial integration on the outer integral. 
We also take advantage of the  recurrence property $j \tilde \E_{j+1}\of{\mu}=1-\mu \tilde\E_{j}\of{\mu}$ \citep{abramowitz64}.
The result is
}
\arev{
\begin{align}
\f\bu\_h\ot &=
\br{
 \rmi \frac{\bk}{k} A\ot\of t 
+S\frac{\bk\bb}{k} \frac{k_y}{k} \int_0^t \!\dd t' \, A\ot\of {t'} \rme ^{\rmi k_x S z (t'-t)}
} \rme ^{k z} 
\nonumber\\&%
+ \rmi\frac{\bk\bb}{k} \wigbrac{ \frac{\bk\bb}{k}\!\cdot\! [(\f\bu\!\cdot\!\f\nabla)\f \bu\_h]\Big|_{t=0}\ott} \frac{\rme ^{\rmi \oms t}-1}{\oms}\rme ^{-\rmi \om t}
\nonumber\\&%
-\frac{1}{2 k} \frac{A_1\oo A_2\oo}{S k_x} \frac{\ktimesk}{k_1 k_2}
\sumij \sumpm  \frac{\pms b_{ij}}{(\xi_i-z)^{j-1}}
\wigbrac{
\frac{\bk}{k} G_{1ij\pms}  + \frac{k_y}{k}\frac{\bk\bb}{k}G_{2ij\pms}  }\rme ^{\ks z-\rmi \omsum t}
%
%+\frac{1}{2 k} \frac{A_1\oo A_2\oo}{S k_x} \frac{\ktimesk}{k_1 k_2}
%\br{\frac{\bk}{k} G_1 - \bm e_x S  \frac{\rme ^{-\rmi \oms t}-1}{\oms} G_2 -  \frac{\bk\bb}{k} \frac{k_y}{k_x}  G_3  } \rme ^{\ks z-\rmi \omsum t}
\end{align}
with
\newcommand{\mmut}{\kpmt (\xi_i-z)}
\newcommand{\mmu}{\kpm (\xi_i-z)}
\begin{align*}
G_{1ij\pm} &= \br{ \frac{j-1}{k(\xi_i-z)}\mp1}\wigbrac{\tilde \E_j\sqbrac\mmut\rme ^{\rmi \oms t}  -\tilde \E_j\sqbrac{\mmu}}
\\
&+ \frac{\kpmt}{k}\tilde \E_{j-1}\sqbrac\mmut\rme ^{\rmi \oms t}  -  \frac{\kpm}{k}\tilde \E_{j-1}\sqbrac{\mmu}
\\
G_{2ij\pm} &= S  \frac{\rme ^{\rmi \oms t}-1}{\oms} \tilde \E_j\sqbrac{\mmu}
\\
& + \frac{\rme ^{\rmi \oms t}}{k_x(\xi_i-z)}\Bigg\{ \sum_{\m=0}^{j-2} \frac{(j-1)!}{\m!}  \br{ \frac{\tilde\E_j\sqbrac\mmut - \frac{1}{j-l-1}}{[-\mmut]^{j-l-1}}
-  \frac{\tilde\E_j\sqbrac{\mmu} - \frac{1}{j-l-1}}{[-\mmu]^{j-l-1}}
}
\\& 
+\tilde\E_j\sqbrac\mmut+\ln\sqbrac\mmut
-\tilde\E_j\sqbrac{\mmu}-\ln\sqbrac{\mmu}
\Bigg\}.
\end{align*}
%\mut
}
Horizontal convection terms (in Fourier space) read
\begin{align}
[(\f\bu\cdot\f\nabla)\f \bm u\_h]\ott &=
\frac{\rmi }{2} A_1\oo A_2\oo \br{
\bk\,\frac{k_1k_2-\kdk}{k_1k_2}  + \sum_{i,j=1}^2 \frac{\bm c_{ij}}{(\xi_i-z)^j}
}\rme ^{\ks z - \rmi \omsum t}
\end{align}
with
\begin{align*}
\bm c_{i1} &= \frac{ \tanthetai}{k_i}
\sqbrac{ \bk_i\bb\frac{k_1k_2-\kdk}{k_1k_2}  
+ (-1)^i \frac{\ktimesk}{k_1k_2} \br{\bk_m + \frac{\tanthetam }{k_m}\frac{\bk_1\bb- \bk_2\bb }{\xi_1-\xi_2}  } },
\\
\bm c_{i2} &=  \frac{ \tanthetai}{k_i}  \frac{ \bk_i\bb}{k_i},
\end{align*}
where
%$ \bk\bb = (-k_y,k_x,0)^T$,
 $i\neq m$ and $\tanthetai = \frac{k_{iy}}{k_{ix}}$.

\arev{
Pressure is obtained by evaluating the $z$-momentum equation directly to find
}
\begin{equation}
p_\tpart(z,t)\ot  = \f p_\wave\ot\of z \rme ^{-\rmi \omsum t} + \f p_\bg\ot\of {z,t} 
	\label{eq:p_cross}
\end{equation}
with
\begin{align}
p_\wave\ot&=
 -\frac12\sumpmpm A_1\oo A_2\oo \bigg(
  \frac{ k_1 k_2 -\kdk }{k_1 k_2}
		\nonumber\\&
    - \frac{\ktimesk }{k_1 k_2} \bigg\{
		\frac{k_2}{k_1}\frac{k_{1y}}{k_{1x}}\tilde\E_1\sqbrac{\ks(\xi_1-z)}
		-\frac{k_1}{k_2}\frac{k_{2y}}{k_{2x}}\tilde\E_1\sqbrac{\ks(\xi_2-z)}
		\bigg\}
		\nonumber\\&
	+ \frac{1}{k^3}  \frac{\ktimesk}{k_1 k_2}     
	\sumij \sumpm \frac{b_{ij}}{\dxiz^{j-1} } 
	\bigg\{
	[k(\xi_i-z)\mps1] \tilde\E_j\sqbrac{\kpms \dxiz}
	\nonumber\\&
	+ \sqbrac{k(\xi_i-\xi_3) + (j-1)\frac{k}{\ks}}  \tilde\E_j\sqbrac{\ks \dxiz}- \frac{k}{\ks}
	\bigg\}
	\bigg) \rme ^{\ks z},
	\label{eq:p_wave}
	\\
	p_\bg\ot&=
	 \frac{1}{2k^3}  \sumpmpm A_1\oo A_2\oo 
	   \frac{\ktimesk}{k_1 k_2}     
	\sumij \sumpm \frac{b_{ij}}{\dxiz^{j-1} } 
	\nonumber\\&
		\times\wigbrac{ 
		\sqbrac{k(\xi_i-z)\mps 1  + (j-1) \frac{k}{\kpmst}} \tilde\E_j\sqbrac{\kpmst \dxiz}- \frac{k}{\kpmst}}\rme ^{\ks z-\rmi  k_x S z t} .
	\label{eq:p_bg}
\end{align}

As for the second order boundary conditions, the Taylor expanded free surface dynamic and kinematic boundary conditions in \eqref{eq:disp_eq_zeta_RHS} respectively read 
\begin{subequations}
\begin{align}
\Ddyn\ot &= 
p\ot_\tpart
+\frac12\br{\f\zeta_1 \pp_z\f p_2+\f\zeta_2 \pp_z \f p_1}, %&z=&0,
\label{eq:Tp:p}
\\
\Dkin\ot &= \f w_\tpart\ot+\frac12\sqbrac{
  \f\zeta_1 \pp_z \f w_2+\f\zeta_2 \pp_z \f w_1
-(\f\bu_1\! \cdot \! \f\nabla_2)\f\zeta_2
-(\f\bu_2 \!\cdot\!  \f\nabla_1)\f\zeta_1
-\rmi  k_x S \f\zeta_1 \f\zeta_2}, %&z=&0
\end{align}%
\label{eq:Tp}%
\end{subequations}%
evaluated at $z=0$. Here, $\psi_i = \psi\oo\of{\bk_i}$.
The dispersive component of $\tDdisp$, appearing in \eqref{eq:zeta_bound_wave}, is
\begin{align}
		\tDdisp=&
- \frac12 \f\zeta\oo_1 \f\zeta\oo_2 
				\br{\omsum+S \frac{k_x}{k}}
				\bigg[
				k_1  \om_1 + k_2 \om_2  + \kdk \br{  \frac{\om_1}{k_1} + \frac{\om_2}{k_2}}
				\nonumber\\&
				+ S k_x + S \ktimesk \br{\frac{ k_{1y}}{k_1^2} - \frac{ k_{2y}}{k_2^2}}
				\bigg] 
				\nonumber\\&
- \frac12  A_1\oo A_2\oo  k  \br{ \frac{\om_1}{\om_2} +\frac{\om_2}{\om_1}  }
+k \f p_\wave\ot\Big|_{z=0}
							\nonumber\\&
- \frac{k \xi_3 +1}{2k^2} A_1\oo A_2\oo  	\frac{\ktimesk}{k_1 k_2}     
	\sumij 
	\sumpm \pms\frac{b_{ij}}{\xi_i^{j-1} } \tilde\E_j\br{\kpms \xi_i}
	%\frac{b_{ij}}{\xi_i^{j-1} } ( \tilde\E_j\br{\kp \xi_i}- \tilde\E_j\br{\km \xi_i})
	.
	\label{eq:D_pmpm}
\end{align}

Finally, the $A\ot$-coefficient of the second order free modes, governed by \eqref{eq:A_n}, is found to be 
\begin{equation}
A\ot\of t = \sumpm A_{\free,\pms}\ot \rme ^{-\rmi \om_\pms t} + A_\wave\ot  \rme ^{-\rmi \omsum t} + A_\bg\ot\of t,
\label{eq:Aot}
\end{equation}
where
\begin{subequations}
\begin{align}
A_{\free,\pm}\ot &=
 -\rmi \, \om_\pm \f\zeta_{\free,\pm}\ot,
\\
A_\wave\ot&=
-\rmi \, \omsum\f\zeta_{\bound,\wave}\ot \Big|_{t=0}
\nonumber\\&\quad
+\frac{\rmi }{2} \f\zeta_1\oo\f\zeta_2\oo \bigg[
				k_1  \om_1 + k_2 \om_2  + \kdk \br{  \frac{\om_1}{k_1} + \frac{\om_2}{k_2}}
				+ S k_x + S \ktimesk \br{\frac{ k_{1y}}{k_1^2} - \frac{ k_{2y}}{k_2^2}}
				\bigg]
\nonumber\\&\quad
+\frac{\rmi }{2k} \frac{A_1 A_2}{k_x S}
	\frac{\ktimesk}{k_1 k_2} \sumij \sumpm \pms \frac{b_{ij}}{\xi_i^{j-1} }
		 \tilde\E_j\br{\kpms \xi_i},
\label{eq:A_disp}
\\
 A_\bg\ot &=
- \sumpm
\sqbrac{
\rmi\, \om_\pms \f\zeta_{\bound,\bg,\pms}
+
\frac{\rmi }{2k}\frac{A_1 A_2}{k_x S} \frac{\ktimesk}{k_1 k_2} \sumij \pms \frac{b_{ij}}{\xi_i^{j-1} }
		 \tilde\E_j\br{\kpmst \xi_i}	
}
\end{align}
\end{subequations}
and its advective integral is
\begin{align}
\int \dd t & A\ot\of t \rme ^{\rmi k_x S z t} =
\sumpm \frac{1}{\om_\pms-k_x S z}\br{
\rmi A_{\free,\pms}\ot \rme ^{-\rmi \om_\pms t} + \om_\pms \f\zeta_{\bound,\bg,\pms}
} \rme ^{\rmi  k_x S z t}
\nonumber\\&
+\rmi  A_\wave\ot\frac{\rme ^{-\rmi\oms t}}{\oms}  
%\nonumber\\&
-\frac{z}{2k}\frac{A_1\oo A_2\oo \rme ^{\rmi  k_x S z t}}{(\om_+-k_x S z)(\om_--k_x S z)}
\frac{\ktimesk}{k_1 k_2} \sumij \sumpm \frac{b_{ij}}{\xi_i^{j-1} }
\nonumber \bigg[
\tilde\E_j\br{\kpt \xi_i}	\\
&+ \kpmst z 
\br{1+{\frac{\pms g k}{(k_x S z)^2}-\frac{\pms1}{k z}}} \sum_{\m=0}^{j-1} \frac{(j-1)!}{\m!}
	\frac{ \tilde\E_{j-\m}\br{\kpmst z} - \tilde\E_{j}\br{\kpmst \xi_i}}{[-\kpmst(\xi_i-z)]^{j-\m}}
\bigg].
\end{align}
$\f\zeta_{\free,\pm}\ot$, $\f\zeta_{\bound,\wave}\ot$ and $\f\zeta_{\bound,\bg,\pm}\ot$ are given in \eqref{eq:zeta_ICn_evaluated} and \eqref{eq:zeta_bound_wave}--\eqref{eq:zeta_bound_background_pm}.
%}
The dispersion relation yielding \eqref{eq:omega_pm} has also been applied in the simplification.

%%%%%%%%%%%%%%%% A P P E N D I X %%%%%%%%%%%%%%%%%%%	
\section{Special cases}
\label{sec:special_cases}

%\acomm{I merged to two subsections the were here before, describing case A-C before case D.
 %I also removed the special case $k_{xl} = 0, l=1,2$ (in my code I started dealing with that case by perturbing $k_{xl}$ instead of the cumbersome analytical alterations.)  }\\

Special wave interaction of symmetry were encountered in the two-wave example of Sec.~\ref{sec:Craik_case}, where two monochromatic wave trains, propagating at equal but oppositely directer angles to the current, gave rise to the four types of second order interaction listed in Table.~\ref{tab:Craik_cases}.

Interaction between parallel wave vectors, designated case A, have in common that $\ktimesk \equiv 0$, removing all expressions containing exponential integral functions or $b_{ij}$-coefficients.

Equal but directly opposing wave vectors (case B) generate infinite wavelength interactions whose amplitude must be zero --- such interactions may be ignored.

In the case of $k_{1x}=k_{2x}$ and $k_{1y}=-k_{2y}$, where also $\om_1=\om_2$ (case C), a solution is obtained by letting $j$-inidces run form $2$ to $4$, letting $i=1$ only and replacing $b_{i,j}$-coefficients with $a_{i,j-1}$. % from \eqref{eq:a_coef_xi1eqxi2}.
The $a$-coefficients, \eqref{eq:a_coef}, also change under the present circumstance (since $\xi_1=\xi_2$) to
\begin{subequations}
\begin{align}
a_{11}&= -\br{\frac{k_{1y}}{k_{1x}} -\frac{k_{2y}}{k_{2x}}}  (k_1 k_2-\kdk) ,\\
a_{12}&= -\br{\frac{k_{1y}}{k_{1x}} \frac1{k_1} -\frac{k_{2y}}{k_{2x}}\frac1{k_2}} (k_1 k_2-\kdk)
-\frac{k_{1y}}{k_{1x}}\frac{k_{2y}}{k_{2x}}\ks \frac{\ktimesk}{k_1 k_2},\\
a_{13}&= -\frac{k_{1y}}{k_{1x}} \frac{k_2}{k_1} + \frac{k_{2y}}{k_{2x}}\frac{k_1}{k_2}-2\frac{k_{1y}}{k_{1x}}\frac{k_{2y}}{k_{2x}} \frac{\ktimesk}{k_1 k_2},\\
a_{21}&= a_{22}=a_{23}=0.
\end{align}
\label{eq:a_coef_xi1eqxi2}
\end{subequations}

Finally, and most important in regard to the discussions of this paper, are interactions of the type $k_{1x}=-k_{2x}$ and $k_{1y}=k_{2y}$, where also $\om_1=-\om_2$ (case D).
This is a standing wave with $\omsum\equiv0$.
%
%
%
%%%%%%%%%%%%%%%% A P P E N D I X %%%%%%%%%%%%%%%%%%%	
%\subsection{Second order solution for special case D}
%
%\adl{
%Consider a pair of wave components $\{\bk_1,\bk_2\}$ where $k_{1x}=-k_{2x}$ and $k_{1y}=k_{2y}$ such as will arise if plane waves propagate symmetrically at oblique angles to the direction of the shear. We also pick out the frequency pair $\pm_1=\mp_2$ such that $\omsum=\Om_{\pm_1}\of{\bk_1}+\Om_{\pm_2}\of{\bk_2}=0$.
%}
%
With this resonance active the vertical cross velocity becomes
\begin{equation}
 w\_{\tpart}\ott =-
 \frac{ t }{ 2k } A_1\oo A_2\oo  \frac{\ktimesk}{k_1 k_2}
 \sum_{j=1}^3 \sumpm \pms
\frac{a_{1j}}{ (\xi_1-z)^{j-1} }
\tilde\E_j\sqbrac{\kpms (\xi_1-z)}
\rme ^{\ks z}
\end{equation}
where $a_{ij}$ are here the coefficients in \eqref{eq:a_coef_xi1eqxi2} for the special case $\xi_1=\xi_2$.
%
%\bs
%\begin{align}
%a_{11}&= -\br{\frac{k_{1y}}{k_{1x}} -\frac{k_{2y}}{k_{2x}}}  k_1 k_2 (k_1 k_2-\kdk) ,\\
%a_{12}&= -\br{\frac{k_{1y}}{k_{1x}}k_2 -\frac{k_{2y}}{k_{2x}}k_1}  (k_1 k_2-\kdk)
%-\frac{k_{1y}}{k_{1x}}\frac{k_{2y}}{k_{2x}}\ks\ktimesk,\\
%a_{13}&= -\frac{k_{1y}}{k_{1x}}k_2^2+ \frac{k_{2y}}{k_{2x}}k_1^2-2\frac{k_{1y}}{k_{1x}}\frac{k_{2y}}{k_{2x}}\ktimesk,\\
%a_{21}&= a_{22}=a_{23}=0.
%\label{eq:a_coef_xi1eqxi2}
%\end{align}
%\es
%
As for the surface elevation, this can be expressed by setting $\f\zeta\ot_{\bound,\bg}=0$ and replacing the $ij$ summation in $\tDdisp$ with
$
	\sum_{j=1}^3 \frac{a_{1,j}}{\xi_1^{j-1}}\br{\tilde E_j[\kp\xi_1]-\tilde E_j[\ks\xi_1]}.
$
We further have
$
A\ot = \sumpm A_{\free,\pms}\ot \rme ^{-\rmi \om_\pms t}  -w\_{\tpart}\ott\of{0,t}/k.
$
$[(\f\bu\cdot\f\nabla)\f u]\ott$ vanishes under the present conditions and
\begin{subequations}
\begin{align}
\f u\ot &= - S \int\!\dd t\,  \f w\of t  = - \frac{S t} {2} \f w
\\
\f v\ot &= - \{[(\f\bu\cdot\f\nabla)\f v]\ott  +2 \rmi k_{1y} \f p\} \, t 
= \ldots
\end{align}%
\end{subequations}
\arev{
Because this interaction is two-dimensional we can express its stream function in the $yz$-plane with $\rmi w_\tpart/k_y$.
}
As noted by \cite{Craik_1970_Langmuir_myidea}, the velocity in the $yz$-directions increases linearly with time while it increases quadratically in the $x$-direction.
\\

As a comment, we point out \arev{the} error in the reference \cite{Craik_1970_Langmuir_myidea} \arev{relevant for the comparison presented here.} It originates from a sign error in two of the exponents in equation 3.11 of said reference \arev{and is easily confirmed by inserting the solution}. 
%This is easily verified by evaluating the left-hand side of his equation 3.10.
The consequences of this error are
\arev{apparent only for strong shear currents and has little bearing on the findings of that paper. }
%, in terms of streamline graphs, perceptible only for shear currents considerably stronger than those investigated,
%and they seem unlikely to have any bearing on the findings of that paper. 
For completeness, we here quote the correct expression for `$f(z)$' in equation 3.12 of \cite{Craik_1970_Langmuir_myidea}: % when solving equation 3.10:
\begin{subequations}
\begin{align}
\tilde f\of z &= 
\frac{k_{1x}^2- k_{1y}^2}{\xi_1-z} + \sumpm\br{\tfrac12 k_1^3\mps 2 k_{1x}^2 k_{1y}} \tilde E_1[\kpms(\xi_1-z)],
\\
 f\of z &= \tilde f\of z \rme ^{\ks z} - \tilde f\of 0 \rme ^{k z},
\label{eq:Craik_stream_function}
\end{align}
\end{subequations}
streamlines given by $ f\of z \cos(k y)=\const$.
We have here used the notation and coordinates of the present paper (with the $z$-axis pointing upwards).
%\\\acomm{About comment to Craik: is Craik still alive?}
%%%%%%

%%%%%%%%%%%%%%%% A P P E N D I X %%%%%%%%%%%%%%%%%%%	
\section{Stationary phase approximation in 2D}
\label{sec:statphase}

\renewcommand{\ot}{^{(2)}}

\arev{The method of stationary phase is a powerful technique
for approximating the asymptotic far-field of
 wave integrals. 
Although commonplace in use with one-dimensional wave integrals, the method's  extension to the present problem warrants a brief summary.
%A more comprehensive presentation is given by \cite{wong01}.
A more comprehensive mathematical derivation is given by \cite{wong01}.
}

\arev{
In integral form, the problem of inverse transformation consists of evaluating
\begin{equation}
  \p\func\of{\bmr,t} = \int \frac{\dd \bk}{(2\pi)^2} \f\func\of\bk\rme ^{\rmi\expfunc\of\bk t}
\end{equation}
for large values of $t$.
}
\arev{
The kernal $\expfunc$  varies slowly in the neighbourhood of a stationary point $\bk_0$ --- a point satisfying 
\begin{equation}
	(\nabla_{\bk} \expfunc)_{\bk=\bk_0} =0
\label{eq:stationary_phase_def}
\end{equation}
--- 
and this is where the asymptotic leading-order integral contribution resides.
}
%
%
%
%Integrals approximated by the method of stationary phase (Eq.~\eqref{eq:asymptote_formula}) are assumed to get their leading-order  contribution from a small integration neighbourhood about a stationary point $\bk_0$, within which $\expfunc$ 
%varies slowly by virtue of \eqref{eq:stationary_phase_def}.
\arev{Assuming $\det(\Hessian\expfunc)_{\bk=\bk_0}\neq 0$ and $\p\func$ to be smooth near the stationary point, we expand}
 $\expfunc$ in a Taylor series around the stationary point and
substituting $\bdk=\bk-\bk_0$
and write %\citep{wong01}
\begin{equation}
  \p\func\of{\bmr,t} \sim \frac{\f\func\of{\bk_0}}{(2\pi)^2} \rme ^{\rmi\expfunc_0  t} \int
  \dd \bdk \exp{ \br{\frac{\rmi t} 2\bdk^T  (\Hessian \expfunc)_0  \bdk }}, \qquad t\rightarrow\infty,
\label{eq:asymptote_integral_generic}
\end{equation}
where  $\Hessian$ being the Hessian operator and subscript $0$ means evaluation is performed at $\bk=\bk_0$.
The integral can now be decomposed with a linear substitution which transforms the integral to the form 
  $ \textstyle \int \dd \bm \kcomplex\, \rme ^{ - \kcomplex_x^2-\kcomplex_y^2 }$:
	\begin{equation}
    \begin{pmatrix}
  	\kcomplex_x\\[1.5ex]  \kcomplex_y
    \end{pmatrix}
    = \frac{\rmi\sqrt{\rmi t}}{\sqrt{2\pp_{k_x}^2 \expfunc_{0}} }
		%= \rmi \sqrt{ \frac{\rmi t}{2 \pp_{k_x}^2 \expfunc_{0} }}
		    \begin{pmatrix}
		 \pp_{k_x}^2 \expfunc_{0} & \pp_{k_x} \pp_{k_y} \expfunc_{0} \\[1.5ex]
		0&\det(\Hessian\expfunc_0)
		\end{pmatrix}
			\begin{pmatrix}
		 \dk_x \\[1.5ex] \dk_y
		\end{pmatrix}.
  \end{equation}
	The new variables	$\kcomplex_x$ and $\kcomplex_y$ can now be evaluated separately.
	Our transformation rotates 	the two integration paths, which in \eqref{eq:asymptote_integral_generic} run from $-\infty<(k_x',k_y')<\infty$,
	either an angle $\pm \pi/4$ or an angle $\pm 3\pi/4$ in the complex plane, depending on the signs of the second derivatives of $\phi$ at the stationary point. 
	Evaluation yields $\sqrt{\pi}$ in the former case and $-\sqrt{\pi}$ in the latter.
	The double integral thus evaluates to $\pm \pi$ and the approximation
	\begin{equation}
  \p\func\of{\bmr,t} 
  \sim \frac{\rmi \f\func\of{\bk_0}\rme ^{\rmi\expfunc(\bk_0) t} }{2\pi t\sqrt{\det(\Hessian\expfunc)_{\bk=\bk_0}} } 
   ; \qquad \expfunc\in\mathbb R,\; t\rightarrow\infty,
  \label{eq:asymptote_formula}
\end{equation}
 results.
The expression changes sign if the stationary point is a local maximum, though we find that stationary points in our problem are saddle points.
\\

%%%%%%%%%%%%%%%%%%%% S E C T I O N %%%%%%%%%%%%%%%%%%%%%%	
%\subsection{An asymptotic approximation of the initial value problem}
%\label{sec:asymptotic:IVP}

\arev{In terms of the initial value problem of this paper we will have a}
%Taking the first order solution \eqref{eq:O1_sol_form},
%we have
%the 
phase function
\begin{equation}
\phi_\pm(\bk) = \bk \cdot \bmr/t - \om_\pm (\bk),
%\label{eq:}
\end{equation}
\arev{$\bmr = (x,y)$.}
Consider large times and also large $|\bmr|$ so that $\bmr/t$ remains bounded.
Two frequencies $\om_\pm=\Om_\pm\of\bk$ exist for each wavelength and so there are two stationary wave vectors $\bk_{0\pm}$.
From the stationary phase definition \eqref{eq:stationary_phase_def} we have
\begin{equation}
\nabla_{\!\bk}\, \Om_\pm\of{\bk}\big|_{\bk=\bk_{0\pm}} = \bmr/t
\label{eq:stationary_phase_om}
\end{equation}
which tells us that the dominating wave is the wave
 %that will have reached the point $\bmr$ at time $t$ when travelling 
%\arev{at its linear group velocity.}
\arev{whose presence, travelling with the group velocity, will have reached the point $\bmr$ at time $t$.}
%with group velocity of a linear plane wave $\bk_{0\pm}$.
%
In order to establish the relationship between the two critical wave vectors $\bk_{0+}$ and $\bk_{0-}$ we use the calculus result
\arev{
$
\nabla_{\!\bk}  { f\of{-\bk}}\big|_{\bk=\bk_0} 
= -\nabla_{\!\bk}\,   f\of{\bk}\big|_{\bk=-\bk_0}
$
together with \eqref{eq:symmetry_prop_Omage}
to find that the complimentary of \eqref{eq:stationary_phase_om} is $\bk_{0-}=-\bk_{0+}$.
}
It also follows that 
$\expfunc_{+}\of{\bk_{0+}}=-\expfunc_{-}\of{\bk_{0-}}$.

Inserting the asymptotic approximation \eqref{eq:asymptote_formula} for the two stationary phases, we find, after considering \eqref{eq:psi_symmetry_prop}, that 
\begin{equation}
\det\sqbrac{\Hessian \Om_+\of{\bk_{0+}}}
=\det\sqbrac{\Hessian \Om_-\of{\bk_{0-}}} < 0,
%\label{eq:}
\end{equation}
is a requirement  for 
\arev{$\p \func$}
%$\p\zeta\oo$ 
to be real, 
\ie, the critical point is a saddle point (in $\expfunc$),
\arev{as is the case for the present problem.}
For the first order surface elevation, the result is then
\begin{equation}
\p\zeta\oo\of{\bmr,t} \sim \frac{
\re
\wigbrac{  \f\zeta\oo_+\of{\bk_{0+}}\rme ^{\rmi\br{\bk_{0+}\!\cdot\bmr-\Om_+\of{\bk_{0+}}t}}}
}{\pi t \sqrt{-\det\br{\Hessian \Om_+\of{\bk_{0+}}}}}.
  \label{eq:asymptote:O1}
\end{equation}
Second order frequency derivatives are found to be
\newcommand{\krx}{\K_{x}}
\newcommand{\kry}{\K_{y}}
\newcommand{\krxsq}{\krx^2}
\newcommand{\krysq}{\kry^2}
\begin{align*}
\pp_{k_x}^2
\Om_{\pm}  &= 
\frac{1}{2 k^2}
\br{
3\krx \krysq {S}
\pm \krysq\frac{ 2 g k+S^2 \left(1-4\krxsq\right)}{ \sqrt{4 g k+S^2\krxsq}}
\mp \krxsq \frac{ \left(2 g k+S^2\krysq\right)^2}{ \left(4 g k+S^2\krxsq\right)^{3/2}}
},
\\
\pp_{k_x}\pp_{k_y}
\Om_{\pm } &= 
\frac{\kry}{2 k^2}\left(
S \left(1-3\krxsq\right)
\mp 2\krx \frac{g k+S^2 \left(1-2\krxsq\right)}{ \sqrt{4 g k+S^2\krxsq}}
\pm\krx \frac{ \left(-2 g k+S^2\krxsq\right) \left(2 g k+S^2\krysq\right)}{\left(4 g k+S^2\krxsq\right)^{3/2}}
\right),
\\
\pp_{k_y}^2
\Om_{\pm } &= 
\frac{1}{2 k^2}
\br{
\krx S \left(1-3\krysq\right)
\pm\krxsq\frac{  2 g k-S^2 \left(1-4\kry\right)}{ \sqrt{4 g k+S^2\krxsq}}
\mp \krysq \frac{  \left(2 g k-S^2\krxsq\right)^2}{ \left(4 g k+S^2\krxsq\right)^{3/2}}
},
\end{align*}
with
$\bK=\bk/k$.
%\\

Next we look at asymptotic expressions for the higher order components.
The Fourier kernels are decoupled in the $n$th order purely bound mode;
$\rme ^{\rmi (\bk\cdot r-\omsum t)}=\textstyle\prod_{m=1}^n \rme ^{\rmi (\bk_m\cdot r-\om_m t)}$,
 and so the asymptote formula \eqref{eq:asymptote_formula} may be applied recursively.
Then, to second order, using \eqref{eq:psi_symmetry_prop},
\begin{equation}
\p\zeta_{\bound,\wave}\ot\of{\bmr,t} 
\sim  \frac{
\re
\wigbrac{  \f\zeta\ot_{\bound,\wave,++}\of{t=0;\bk_{0+},\bk_{0+}}\rme ^{2\rmi\br{\bk_{0+}\!\cdot\bmr-\Om_+\of{\bk_{0+}}t}}}
}{2\pi^2 t^2 \wigbrac{-\det\sqbrac{\Hessian \Om_+\of{\bk_{0+}}}}}.
\label{eq:asymptote:bound_wave_O2}
\end{equation}
Sign subscripts in $\f\zeta$ indicate the frequency branch of the interacting free waves.
For self--interaction of a single branch 
the dispersive mode equation \eqref{eq:zeta_bound_wave} simplifies to
\begin{equation}
  \f\zeta\ot_{\bound,\wave,++}\of{t=0;\bk_0,\bk_0} 
  =   k_0\br{\f\zeta\oo_+\of{\bk_{0+}}}^2 \frac{2\br{\om_{0+} + \frac{S}{2}\frac{k_{0x}}{k_0}}^2-\om_{0+}^2}
  {2\om_{0+}^2+\om_{0+}  S\frac{k_{0x}}{k_0}-g k_0}.
\end{equation}
We see that the asymptotic approximation 
contains only self-interaction terms of the same value of $\bk$. 
This is expected because 
wave components of different 
wavelengths will with time disperse across the physical plane, allowing a far point in $\bmr$ to be approximately represented by a single wave vector. 
The approximation fails at points of resonance where the 
integrand is no longer smooth.

The method of stationary phase is less powerful for the free waves, where it can be used to evaluate the outer integral in form of \eqref{eq:convolution_int_O2_inner_form},
but where one 
still has to compute a convolution integral at $t=0$.
\arev{This is because the free mode amplitudes are chosen to suit higher order terms to the initial conditions. This is a near-field correction. }
\\

\begin{figure}
\begin{tabular*}{\textwidth}{cccc}
	  \includegraphics[width=.24\columnwidth]{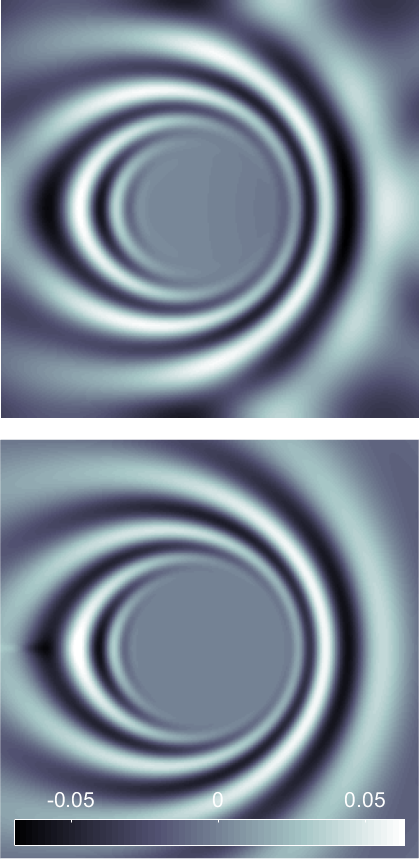} 
  &\includegraphics[width=.24\columnwidth]{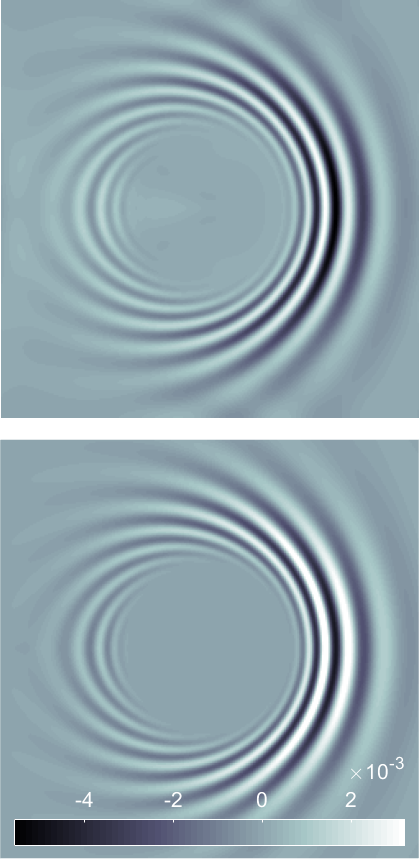}
  &\includegraphics[width=.24\columnwidth]{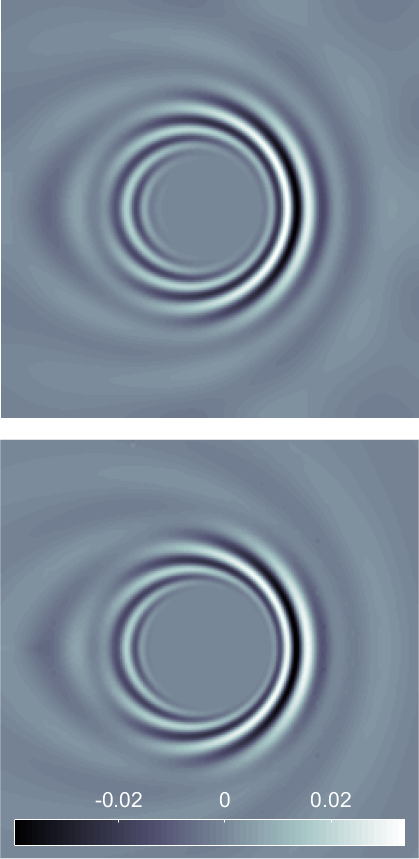}
  &\includegraphics[width=.24\columnwidth]{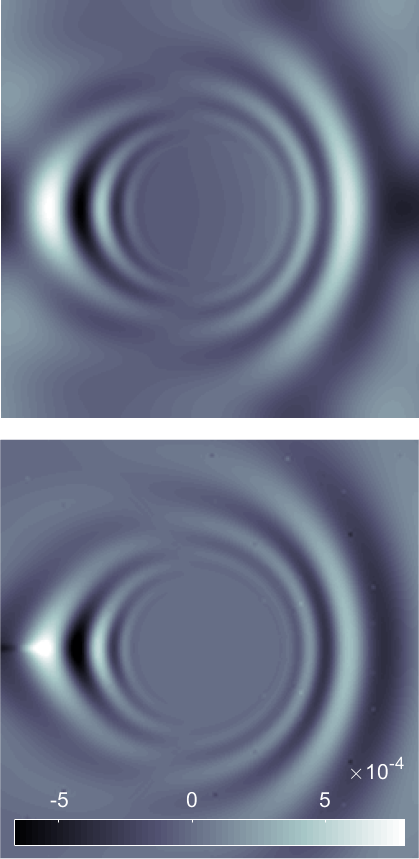} \\
  First order & Bound, dispersive & Free, dispersive & Free advective
\end{tabular*}

\caption{Comparison of Stationary phase approximation (top row%
%, theory from Sec.~\ref{sec:asymptotic:IVP}
) to solution obtained with discrete FFT in a restricted domain (bottom row).
$\FrS=1.0$,
$T =15$,
$H=0.2$.
Domain length: $20b/3 $.%$10\tfrac23\,b$.
}%
\label{fig:ringwave_stat_phase}%
\end{figure}

\arev{A decisive advantage with the method of stationary over FFT computation, apart from possible improvements in computational efficiency,  }
%Another advantage of the stationary phase approximation 
is that it is founded in a continuous spectrum and so does not impose periodicities around the computed domain.
In contrast, solutions computed with a discrete Fourier transform will be confined in a periodic domain where domain size requirements can be large if one is to avoid the influence of reflective boundaries. 
This effect can be seen in Fig.~\ref{fig:ringwave_stat_phase} where we compare the stationary phase approximation to solutions computed with a full FFT in a tight domain.
The approximation is seen to be good at this moderately large time $T=15$,
and it improves with increasing time. 
A Newton--Raphson method 
\arev{was here used for finding}
the stationary phases from \eqref{eq:stationary_phase_def}.
The stationary phase of the same problem without shear, 
$\bk_{0\pm}  \coloneqq  \pm g t^2 \bmr /(4 r^3)$,
was used as an initial condition for the root search.
%The convergence 
%in the ring-wave problems
\arev{Convergence for this problem}
was found to be quick, but could, if the shear was strong enough, diverge close to the abscissa in the far shear-assisted direction.

\bibliographystyle{jfm}
\bibliography{refs_wave}

\end{document}